\def\bea{\begin{eqnarray}}
\def\eea{\end{eqnarray}}

\documentclass[aps,pre,twocolumn,superscriptaddress,amsmath,amssymb,longbibliography]{revtex4-1}

\usepackage{graphicx}

\usepackage{dcolumn}

\usepackage{enumitem,kantlipsum}

\numberwithin{equation}{section}

\usepackage{bm}
\usepackage[normalem]{ulem}
\usepackage{color}
\usepackage{gensymb}

\usepackage{epstopdf}
\newcommand{\bew}{\begin{widetext}}
\newcommand{\ew}{\end{widetext}}
\newcommand{\nn}{\nonumber}

\newcommand{\bz}{\mathbf{z}}

\newcommand{\bq}{\mathbf{q}}
\newcommand{\bQ}{\mathbf{Q}}
\newcommand{\bv}{\mathbf{v}}

\newcommand{\br}{\mathbf{r}}
\newcommand{\brp}{\mathbf{r}_{_\parallel}}
\newcommand{\bR}{\mathbf{R}}
\newcommand{\bM}{\mathbf{M}}
\newcommand{\bff}{\mathbf{f}}

\newcommand{\bw}{\mathbf{w}}

\newcommand{\bh}{\mathbf{h}}
\newcommand{\hx}{\hat{{\bf x}}}
\newcommand{\hy}{\hat{{\bf y}}}
\newcommand{\hz}{\hat{{\bf z}}}
\newcommand{\hp}{\hat{{\bf p}}}

\newcommand{\sep}{ \ \ \ , \ \ \ }

\newcommand{\beq}{\begin{equation}}
\newcommand{\eeq}{\end{equation}}
\newcommand{\beqn}{\begin{eqnarray}}
\newcommand{\eeqn}{\end{eqnarray}}
\newcommand{\pp}{\partial}
\newcommand{\dd}{{\rm d}}
\newcommand{\ee}{{\rm e}}

\newcommand{\cO}{{\cal O}}

\usepackage{xcolor}
\usepackage{color}

\definecolor{green}{rgb}{0,0.5,0}

\begin{document}
\title{ Hydrodynamic theory of flocking at a solid-liquid interface: long range order and giant number fluctuations}
\author{Niladri Sarkar}\email{niladri2002in@gmail.com}
\affiliation{Instituut-Lorentz, Leiden University, P.O. Box 9506, 2300 RA Leiden, The Netherlands}
\author{Abhik Basu}\email{abhik.123@gmail.com, abhik.basu@saha.ac.in}
\affiliation{Condensed Matter Physics Division, Saha Institute of
Nuclear Physics, Calcutta 700064, West Bengal, India} 
\author{John Toner}\email{jjt@uoregon.edu}
\affiliation{Department of Physics and Institute of Theoretical Science, University of Oregon, Eugene, Oregon 97403, USA}

\date{\today}
\begin{abstract}
 We construct the hydrodynamic theory 
  of coherent collective motion (``flocking'') at a solid-liquid interface. The 
polar order parameter and concentration of a collection of ``active'' (self-propelled) particles at a planar interface between a passive, isotropic bulk fluid and a solid surface are dynamically coupled to the bulk fluid. We find that such systems are stable, and have long-range orientational order, over a wide range of parameters. When stable, these systems exhibit ``giant number fluctuations'', i.e., large fluctuations of the number of active particles in a fixed large area. Specifically, these number fluctuations grow as the $3/4$th power of the mean  number within the area.  Stable systems also exhibit anomalously rapid diffusion of tagged particles suspended in the passive fluid along any directions in a plane parallel to the solid-liquid interface,  whereas the diffusivity along the direction perpendicular to the plane is non-anomalous. 
In other parameter regimes, the system becomes unstable.
\end{abstract}

\maketitle

\section{Introduction}\label{intro}

Active fluids  are fluids containing  self-propelled  units  (``{\em active particles}"), 
each capable
of converting stored or ambient free energy into motion.    These intrinsically non-equilibrium systems often display orientationally ordered phases,  both in  
living~\cite{kruse04,kruse05,goldstein13,saintillan08,hatwalne04} and 
nonliving~\cite{saha14,cates15,narayan07,lubensky09,marchetti08}  systems. Such phases of active fluids exhibit many phenomena impossible in equilibrium orientationally ordered phases (e.g., nematics~\cite{deGennes}), among them spontaneous breaking of continuous symmetries in two dimensions~\cite{vicsek95, tonertu95, toner98, toner05}, instability in the extreme Stokesian limit~\cite{simha2002}, 
and giant number fluctuations~\cite{Chate+Giann, toner2019giant, ramaswamy03}.

As in  equilibrium systems, the  ``hydrodynamic" (i.e., long distance, long-time) behavior of active fluids depends crucially on whether or not momentum is conserved.
In what has become the standard nomenclature of this field, systems lacking momentum conservation due to, e.g.,  friction with a substrate, are referred to as ``dry"~\cite{wolgemuth2002, toner98,ramaswamy03,carlos2019,ricard2020}, while active fluids {\it with} momentum conservation are called ``wet"~\cite{lushi2014, lushi2014, act turb, Goldstein21, yeom}.  

In this paper, we treat a heretofore unconsidered  class of systems that is a natural hybrid of these two cases: a collection of 
polar active particles at a solid-liquid interface. Our study is inspired by experiments on high-density actin motility
assays~\cite{schaller13},
in which highly concentrated actin filaments on a solid-fluid interface are
propelled by  motor proteins. Another experimental realization is the work of Bricard {\em et al} ~\cite{bricard2013, Geyer17}, who studied the emergence of macroscopically directed motion in ``Quincke rotators": a dilute collection of motile colloids, in contact with a solid substrate, and immersed in a passive bulk electrolytic  fluid. The colloids spontaneously roll, making them self propelled, when a sufficiently strong electric field is applied to the electrolytic fluid.

 The hybrid system we consider here is in some ways reminiscent of  the phenomenon of ``wet to dry crossover"  studied by other authors \cite{Nejad, amin2016}. However, there are important differences. The systems studied by \cite{Nejad, amin2016} are fully two dimensional, but had both viscous and frictional damping. In such a problem, the asymptotic, longest wavelength behavior is that of dry active matter. In our case, as long as our bulk fluid is semi-infinite, the asymptotic long-distance behavior is completely different, even in its scaling, from dry active matter. The mixture of wet and dry behavior in our system is not a {\it crossover} phenomenon, only occurring over an intermediate range of length scales. Rather, it is the true, asymptotic,  long-wavelength (i.e., hydrodynamic) behavior (again, for a semi-infinite bulk passive fluid).

The systems that we consider here differ  from both dry and  wet active matter, as defined above, by having both 
friction from the underlying solid substrate and the long range 
hydrodynamic interactions due to the overlying bulk passive fluid. Very 
interestingly, the  experiments of ~\cite{schaller13} found giant fluctuations of the active particle 
number in a fixed volume,  with a standard deviation that scaled with $N^{0.8}$ in the 
ordered phase, where $N$ is the mean number in the volume. Although the exponent of $0.8$ is the value predicted~\cite{Chate+Giann, toner2019giant} for dry polar active fluids, we will argue here that this is a coincidence, and that in fact polar fluids at a solid-liquid interface belong to an entirely different and heretofore unstudied universality class, with number fluctuations scaling like $N^{0.75}$, which probably lies within experimental error of the $0.8$ exponent found in the experiments.

\begin{figure}[htb]
\includegraphics[height=6cm]{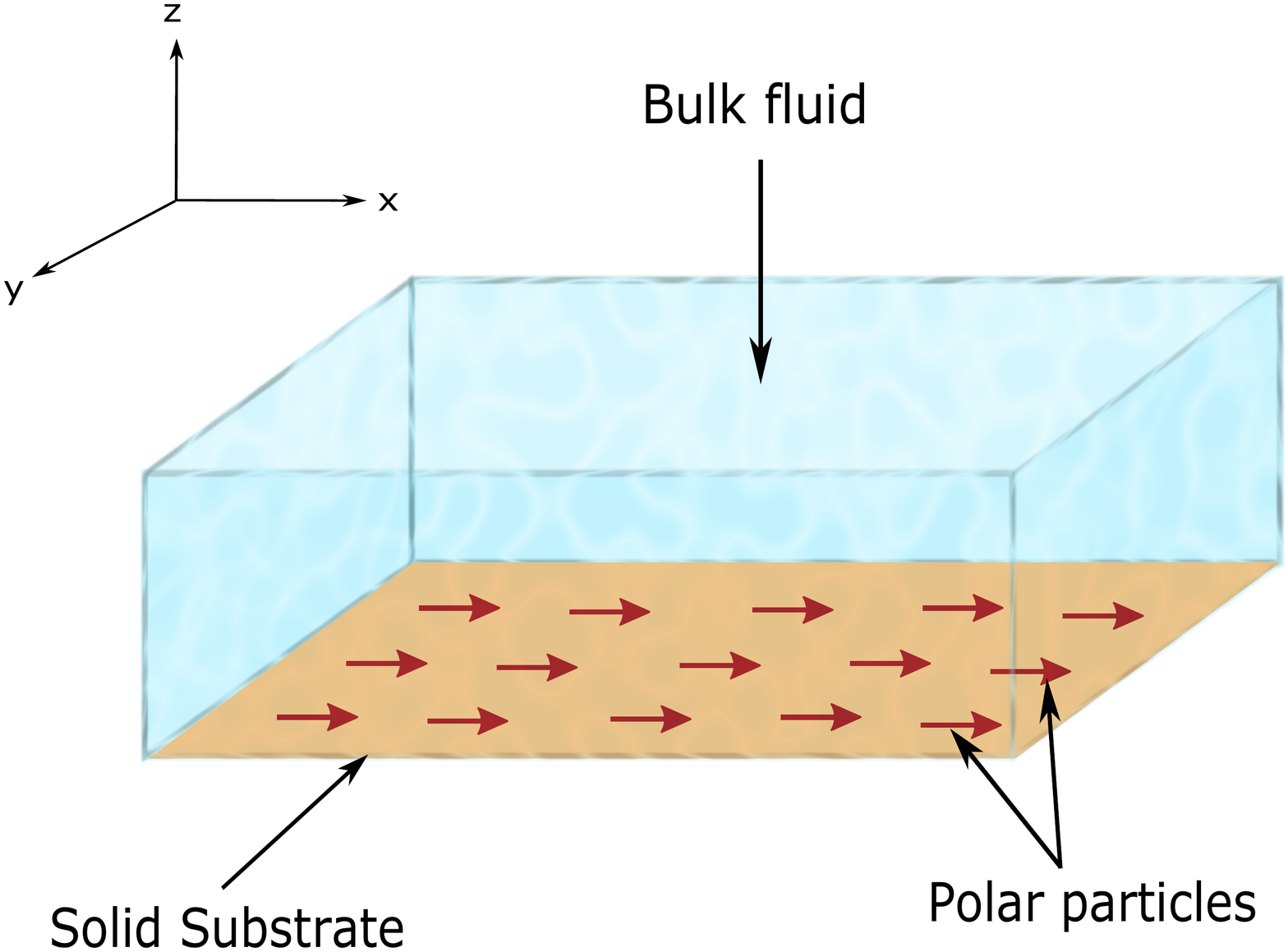}
\caption{(Color online) Schematic diagram of a layer of active polar particles 
embedded on a solid substrate surrounded by passive ambient fluid from the top}
\label{schem}
\end{figure}

In this  article, inspired by the experiments of Ref.~\cite{schaller13, bricard2013}, we formulate  
the hydrodynamic theory for a collection of polar, self propelled particles 
 at the interface between a solid 
substrate and a bulk isotropic passive liquid that is semi-infinite in the 
$z$-direction, as illustrated in Fig.~\ref{schem}. 
We focus on the extreme Stokesian limit, in which the momentum of both the active particles and the bulk fluid is entirely negligible relative to viscous drag. The theory is then used to study the stability and fluctuations of a uniformly ordered active polar state of the particles.

 The governing equations of the active particle number density and polarization fluctuations in this system, which are introduced here for the first time are very different from those of ``dry" and ``wet" active matter''.The breakdown of the Galilean invariance due to the friction from the substrate and the long ranged hydrodynamic interactions mediated by the passive bulk fluid lead to novel linear couplings between the density and polarization fluctuations, and  non-local damping. The resultant behavior constitutes  a new universality class, whose properties we describe in detail. 

Our most surprising result is that this system can exhibit a long range ordered polar state, even in the presence of noise. This should be contrasted with both ``wet" active systems, which are generically unstable\cite{simha2002} at very low Reynolds number, and with equilibrium systems, which cannot develop long ranged orientational order in two dimensions~\cite{MW, xtalfoot, 2dxtal,teth}.  Further, the number fluctuations in these systems are giant;  i.e., 
the 
standard deviation $\sqrt{\langle(N-\langle N\rangle)^2\rangle}$ of the number $N$ of the active particles  contained in a fixed open area grows  with its average $\langle N\rangle$ much {\em faster} than its equilibrium off-critical counterpart. Specifically, we find
\beq
\sqrt{\langle(N-\langle N\rangle)^2\rangle}\propto\langle N\rangle^{3/4} \,.
\label{gnf}
\eeq

We also find that the diffusivities of tagged particle in any direction parallel to the active fluid layer is {\em logarithmically  anomalous},  by which we mean the mean squared displacement $\langle|\brp(t)|^2\rangle$ grows according to  $<|r_\parallel(t)|^2>\propto t\ln (t)$, in contrast to the usual linear in $t$ scaling of normal diffusion. Diffusion normal to the plane remains conventional, with  $\langle|r_\perp(t)|^2\rangle\propto t$.

Unlike dry polar active matter, which can also exhibit long range order and giant number fluctuations~\cite{vicsek95, tonertu95, toner98, toner05}, these systems do so {\it even in a linear theory}. Indeed, such a linear theory proves to provide an asymptotically exact long wavelength description of these systems, again in contrast to dry active matter.

The principal condition for stability of our systems  is similar to the stability criterion for a simple compressible bulk fluid: the analog in our system of the bulk compressibility must be positive, as must our analog of the viscosities. 

Thus, in contrast to ``wet" active matter in the ``Stokesian" limit\cite{simha2002, toner05}, our ``mixed" system can be generically stable. Indeed, the requirements for stability are as easily met for these systems as for an equilibrium fluid.

 These results are testable in standard experiments, e.g., in systems in which the active particles can be imaged, like those of \cite{schaller13,bricard2013}.

{ The remainder of this paper is organized as follows. We begin with a short summary of the technical results, which include the governing hydrodynamic equations themselves, in section~\ref{tech-sum}.
In section (\ref{EOM}), we formulate the linear hydrodynamic theory, including the equations of motion, for our system, and calculate the eigenfrequencies of small fluctuations about the perfectly ordered state. We thereby demonstrate that the ordered state is dynamically stable. In section (\ref{corr}), we calculate the correlation functions implied by our equations of motion, and use them to show that long ranged polar order is robust against noise in these systems. In section (\ref{section:GNF}), we show that the density-density correlation functions found in section (\ref{corr}) imply giant number fluctuations (equation (\ref{gnf})). We also show that those giant number fluctuations depend on the shape of the ``counting volume", as well as the mean number of particles it contains, in contrast to equilibrium systems, in which the mean number alone determines the variance. In section (\ref{bulk}), we calculate the velocity fields induced in the passive fluid by the active particles, and the correlations of those fields. We also consider the motion of tracer particles in the passive fluid, and thereby derive the anomalous diffusion (equations (\ref{anomdifshort})-(\ref{anomsedxy}).)  Finally, in section (\ref{RG}), we use a dynamical renormalization group analysis   to show that these linear results are asymptotically exact at long wavelengths. That is, we show that all non-linear terms allowed in the hydrodynamic equations of motion are irrelevant, in the renormalization group sense, in the long-wavelength limit.
In appendix  (\ref{appa}), we solve the bulk Navier-Stokes equation to relate the passive fluid velocity throughout the bulk to the fluid velocity on the liquid-solid interface. In (\ref{appb}), we derive the stability limit for the ordered state. In appendix~(\ref{appc}), we evaluate an integral that arises in the calculation of the real-space correlation function. }

\section{Summary of the principal results}\label{tech-sum}

The variables of our hydrodynamic theory are two time-dependent fields that are confined to the plane of a flat interface between a solid substrate and a three-dimensional (3D) 
passive, isotropic, bulk fluid.  We choose our coordinates $(x,y,z)$ so that the interface sits at $z=0$, with the solid substrate underneath  (i.e., $z<0$), and the bulk fluid above (i.e., $z>0$). We will denote two dimensional position within that plane by a two component vector $\brp\equiv(x,y)$. One of the hydrodynamic fields is the conserved density fluctuation $\delta\rho(\brp,t)\equiv\rho(\brp,t)-\rho_0$ of the areal density $\rho(\brp,t)$ about its mean value $\rho_0$.

 Our second hydrodynamic variable reflects the polar order. By ``polar", we mean each particle (labeled by an index $\alpha$, where $\alpha$ runs from $1$ to the number of particles $N$), can have associated with it a unique unit vector $\hp_\alpha$ (imagine, e.g., a collection of arrows, with  $\hp_\alpha$ pointing from the tail to the head of arrow $\alpha$).

To formulate hydrodynamics, we coarse grain this particle-based unit vector  in to a position dependent vector {\it field} $\hp(\brp,t)$ defined on the surface at $z=0$.

Our theory studies small fluctuations around a uniform reference state with polarization ${\bf p}=\hx$ and density $\rho=\rho_0$.

 The linear hydrodynamic equations for the transverse orientation fluctuation $p_y$ and density fluctuation $\delta\rho$  for a polar-ordered flock at a solid-liquid interface will be  systematically derived in section~\ref{EOM} below. They are obtained by by linearizing around a uniform reference state with polarization ${\bf p}=\hat x$ and density $\rho=\rho_0$.    In Fourier space, these   equations  are
\bea
\partial_t\delta\rho =- iv_\rho [q_x\delta\rho+\rho_c q_yp_y] + i\bq\cdot{\bf f}_\rho \,, \label{rholin-intro}
\eea

\bew
\bea
\partial_tp_y =- iv_pq_xp_y -\gamma\left({q^2+q_y^2 \over q}\right)p_y -\left({\gamma_\rho\over\rho_c}\right)\left({q_xq_y \over q}\right)
\delta\rho
-i\sigma_t q_y\delta\rho + f_y \,.
\nonumber\\
\label{pyfin-intro}
\eea
\ew
 Here, $v_p$, $v_\rho$, $\gamma$, $\gamma_\rho$, and $\rho_c$ are phenomenological parameters of our model, the first four of which have the dimensions of speed, while $\rho_c$ has the dimensions of (areal) density. The parameters $\gamma$ and $\gamma_\rho$ are damping 
coefficients that play roughly the same role in our problem as the viscosities play in a simple fluid\cite{funny speed}. The speeds $v_\rho$ and $v_p$ are fundamentally active parameters, with no analog in equilibrium fluids, which arise from the self-propulsion of the active particles.

The Gaussian noises ${\bf f}_\rho$ and $f_y$ have zero mean and variances
\begin{eqnarray}
 &&\langle f_{\rho i} ({\bf r}_{_\perp},t)f_{\rho j} ({\bf r}_{_\perp}',t')\rangle=2 D_\rho\delta_{ij}^s\delta(
{\bf r}_{_\perp} - {\bf r}_{_\perp}') \delta(t-t')\,\nonumber\\\\
&&\langle f_{ y} ({\bf r}_{_\perp},t)f_{ y} ({\bf r}_{_\perp}',t')\rangle=2 D_p\delta(
 {\bf r}_{_\perp}-{\bf r}_{_\perp}') 
\delta(t-t')\,,
 \end{eqnarray}
where ${\bf r}_{_\perp}=(x,y)$ denotes the position of a point on the plane of the active fluid layer, and the parameters $D_{p,\rho}$ are constants with the dimensions of diffusion coefficients.

Fluctuations about  the uniform ordered state are underdamped; that is, they both propagate and decay. The propagation is non-dispersive; that is, the wavespeed is independent of the wavenumber $q$, as in a simple compressible fluid. However, the wavespeed is direction dependent, as will be discussed in more detail below.

Even more strangely, the decay rate of these fluctuations scales completely differently from that of dry active matter; indeed, it scales linearly with $q$, just like the real part. This means that, unlike almost every other physical system that exhibits non-dispersive propagating modes \cite{smecticfoot, smectic}, the ``quality factor" $Q$ of our system, defined as the ratio of the real part of $\omega(\bq)$ to its imaginary part, does {\it not} diverge as wavenumber $q$ goes to zero. Instead, it approaches a finite, system-dependent constant, independent of $q$, as $q\to0$.

This unusual damping 
in turn leads to  the aforementioned  giant number fluctuations of the active particles,  as given by equation~(\ref{gnf}).



We also find that, when the active particles at the interface are in their ordered state, they ``stir" the passive fluid above them. 
The components $\langle v_x^2(\brp,z)\rangle$ and $\langle v_y^2(\brp,z)\rangle$ of the mean squared velocity thereby induced in the passive fluid  are inversely proportional to the distance $z$ from the solid-fluid interface, whereas $\langle v_z^2(\brp,z)\rangle$ is inversely proportional to the cube of the distance $z$, i.e, it falls of as $1/z^3$.  Here, the $x$, $y$, and $z$ axes are respectively the direction of the polarization of the active particles, the in-plane direction orthogonal to that, and the normal to the interface, as illustrated in figure (\ref{schem}). 

These predictions for the passive fluid velocity correlations could be tested experimentally by tracking neutrally buoyant passive tracer particles in the passive fluid. Such particle tracking thus provides a probe of the fluctuations of the active particles in the ordered state.

These velocity fluctuations in the passive fluid also exhibit long ranged spatio-temporal correlations, which decay as $1/t$  for the projection of the velocity on the interface. The projection of the passive fluid velocity  along $z$ (perpendicular to the surface) decay as $1/t^3$.

 These correlations in turn lead to anomalous diffusion of neutrally buoyant passive particles in the $x$- and $y$-direction, with variances of the displacements obeying
\begin{eqnarray}
&&\langle(x(t)-x(0))^2\rangle =2D_xt\left[\ln\left({v_0t\over z_0}\right)+O(1)\right]\,\,, \,\,\,\,\,\, t\ll{z_0^2\over D_z} \,,\nonumber\\
&&\langle(y(t)-y(0))^2\rangle =2D_yt\left[\ln\left({v_0t\over z_0}\right)+O(1)\right]\,\,, \,\,\,\,\,\, t\ll{z_0^2\over D_z} \,,\nonumber\\
&&\langle(x(t)-x(0))(y(t)-y(0)\rangle = 0\,\,, \,\,\,\,\,\, t\ll{z_0^2\over D_z} \,.
\label{anomdifshort}
\end{eqnarray}
\vspace{.1in}
 
\noindent The cross-correlation function in the third line of (\ref{anomdifshort}) vanishes because the system is symmetric under inversion of $y$. Here, $v_0$ is a system-dependent characteristic speed associated with the active particles (roughly speaking, it is the speed at which they move over the surface due to their self-propulsion),  $z_0$ is the initial distance of the neutrally buoyant particle from the surface, and $D_x,\,D_y$ are   diffusion constants of the same order of magnitude as the diffusion constant in the $z$ direction.

The results (\ref{anomdifshort}) hold for times $t$ short compared to the time it takes for the 
neutrally buoyant particle to diffuse in the $z$ direction a distance comparable to $z_0$; that is, 
for $t\ll{z_0^2\over D_z}$, where $D_z$ is the diffusion constant in the $z$-direction. This is a 
genuine diffusion constant; that is, the motion along $z$ is simple diffusion, with a $z$-
independent diffusion constant. For $t\gg{z_0^2\over D_z}$, after which the neutrally 
buoyant particle will have diffused in the $z$ direction a distance much larger than $z_0$, the 
behavior of $\langle(x(t)-x(0))^2\rangle$, and $\langle(y(t)-y(0))^2\rangle$ can be obtained by replacing $z_0$ in (\ref{anomdifshort}) with  $\sqrt{D_zt}$,
which  is  just the rms distance the particle will have diffused away from the surface in that time. 
Doing so, we find that the motion on these much longer time scales is still superdiffusive, but with 
precisely half the ``superdiffusion" constant. That is, 
\begin{eqnarray}
&&\langle(x(t)-x(0))^2\rangle =D_xt\left[\ln\left({v_0^2t\over D_z}\right)+O(1)\right]\,\,, \,\,\,\,\,\, t\gg{z_0^2\over D_z} \,,\label{anomdiflongx}\nonumber\\
\\
&&\langle(y(t)-y(0))^2\rangle =D_yt\left[\ln\left({v_0^2t\over D_z}\right)+O(1)\right]\,\,, \,\,\,\,\,\, t\gg{z_0^2\over D_z} \,,\label{anomdiflongy}\nonumber\\\\
&&\langle(x(t)-x(0))(y(t)-y(0)\rangle = 0 \,\,, \,\,\,\,\,\, t\gg{z_0^2\over D_z} \,,
\label{anomdiflongxy}
\end{eqnarray}
\vspace{.1in}
Again, as in (\ref{anomdifshort}) above, the cross-correlation in the third line of (\ref{anomdiflongxy}) vanishes because of symmetry under inversion of $y$. Diffusion in the $z$-direction remains conventional. Furthermore, the diffusion  constant for diffusion in the $z$-direction is independent of height $z$. This set of predictions could also be tested experimentally by particle tracking of neutrally buoyant    tracer particles in the passive fluid.
 
Particles denser than the passive fluid, which therefore sediment, will also be affected by this activity induced flow. We find that particles sedimenting from an initial height $z_0$ will, when they reach the surface, be spread over a region of dimensions $\sqrt{\langle (x(z=0)-x(z=z_0))^2\rangle}$ and $\sqrt{\langle (y(z=0)-y(z=z_0))^2\rangle}$ in the $x$ and $y$ directions, respectively, with
\bew
\beq
\langle (x(z=0)-x(z=z_0))^2\rangle=2D_x\left({z_0\over v_{\rm sed}}\right)\ln\left({v_0\over v_{\rm sed}}\right) \sep v_{\rm sed}\ll v_0 \,, 
\label{anomsedx}
\eeq
\beq
\langle (y(z=0)-y(z=z_0))^2\rangle=2D_y\left({z_0\over v_{\rm sed}}\right)\ln\left({v_0\over v_{\rm sed}}\right) \sep v_{\rm sed}\ll v_0 \,,
\label{anomsedy}
\eeq
\beq
\langle(x(z=0)-x(z=z_0))(y(z=0)-y(z=z_0)\rangle = 0 \sep v_{\rm sed}\ll v_0  \,,
\label{anomsedxy}
\eeq
\ew
\vspace{.1in}
 
\noindent where $v_0$ is roughly the mean speed of the active particles, and $v_{\rm sed}$ is  the speed at which the sedimenting particles sink. These results only hold when the sedimenting speed $v_{\rm sed}\ll v_0$; in the opposite limit, the lateral diffusion of a sedimenting particle becomes conventional.


Once again, these predictions should be readily testable in particle tracking experiments.

We turn now to correlations of the active particles themselves. These are determined by the  characteristic frequencies of the fluctuations. We find that these eigenfrequencies $\omega(\bq)$  are complex, and given by
\beq
\omega(\bq)=\Upsilon(\theta_\bq) q \,.
\label{omega lin q intro}
\eeq
Here $\Upsilon$ is a complex   (and complicated!) function of the angle $\theta_\bq$ between the wavevector $\bq$ and the direction of spontaneous polarization of the active particles, and is the solution of 
a parameter and direction dependent quadratic equation   given in section~\ref{stab}. 

 Allowing the fluctuations of the stable modes to be driven by an additive white noise (which will arise from both noisy self-propulsion of the active particles, and thermal fluctuations), we find the orientational correlation function

\beq
C_{pp}({\bf q},\omega)\equiv\langle |p_y(\bq, \omega)|^2\rangle=\left({1\over q^2}\right)F_{pp}\bigg(\left({\omega\over q}\right), \theta_\bq\bigg)\, ,
\label{pycorscale}
\eeq
\noindent where  $F_{pp}(\omega/q, \theta_\bq)$ is a scaling function whose precise form is given in Section~\ref{corr}. We also find that the peak positions of the scaling function, which we give in detail in Section~\ref{corr}, agree with those found in  the experiments of \cite{Geyer17} on Quinke rotators.

The density-density correlation function and the orientation-density correlation function obey  similar scaling laws:
\bea
&&C_{\rho\rho}({\bf q},\omega)\equiv\langle |p_y(\bq, \omega)|^2\rangle={1\over q^2}F_{\rho\rho}\bigg(\left({\omega\over q}\right), \theta_\bq\bigg)\, ,\nonumber\\
\label{rhocorscale}\\
&&C_{p\rho}({\bf q},\omega)\equiv\langle  \rho({\bf q},\omega) p_y(-{\bf q},-\omega)  \rangle
={1\over q^2}F_{p\rho}\bigg(\left({\omega\over q}\right), \theta_\bq\bigg)\, ,\nonumber\\
\label{crossprhocorscale}
\eea
with slightly different scaling functions $F_{\rho\rho}(\omega/q, \theta_\bq)$ and $ F_{p\rho}(\omega/q, \theta_\bq)$, whose precise forms   are  given by equations (\ref{rhoscalefn}) and (\ref{crossprhoscalefn}) of section~\ref{corr}.


 {

Integrating these spatio-temporally Fourier-transformed correlation functions over all frequencies 
$\omega$ gives  surprisingly simple expressions for the equal-time correlation functions.   Specifically, 
\beq
C_{pp}(\bq)\equiv\langle |p_y(\bq, t)|^2\rangle=\int_{-\infty}^\infty{d\omega\over2\pi}C_{pp}(\bq,\omega)={f_{pp}(\theta_\bq)\over q} \,,
\label{CppETq}
\eeq
where the function $f_{pp}$ depends only on $\theta_\bq$ and not $\bf q$, whose precise form is discussed later in Section~\ref{corr}.

For the equal-time density autocorrelation function, we obtain
\beq
C_{\rho\rho}(\bq)\equiv\langle |\rho(\bq, t)|^2\rangle={f_{\rho\rho}(\theta_\bq)\over q}
\label{CrhorhoETq}
\eeq
where $f_{\rho\rho}$ is another function of $\theta_\bq$, to be given later.

Finally, the equal-time cross-correlation function is given by
\beq
C_{p\rho}(\bq)\equiv\langle p_y(\bq, t)\rho(-\bq,t)\rangle={f_{p\rho}(\theta_\bq)\over q}
\label{crossETq}
\eeq
where the direction dependence is given by the function $f_{p\rho}$ that depends only on $\theta_\bq$.
}

Fourier transforming these in space leads to equally simple expressions for the real space, equal-time correlation functions,  all of which scale as $1/r$ with anisotropic amplitudes.

That is:
\beqn
C_{pp}(\br)&\equiv&\langle \delta\hp(\br+\bR,t)\delta\hp(\bR,t)\rangle\approx\langle p_y(\br+\bR,t)p_y(\bR,t)\rangle\nonumber \\&=&{f_{pp}(\theta_\br)\over r} \,,
\label{CppETqr}
\eeqn
where, the function $f_{pp}(\theta_\br)$ depends only on $\theta_\br$. the angle between $\br$ and the direction of the mean polarization, and not on the magnitude $r\equiv|\br|$. The  precise form of $f_{pp}(\theta_\br)$ is given in equation blah  of section blah.

  For the real space equal-time density autocorrelation, we obtain
\beq
C_{\rho\rho}(\br)\equiv\langle \delta\rho(\br+\bR,t)\delta\rho(\bR,t)\rangle={f_{\rho\rho}(\theta_\br)\over r}
\label{CrhorhoETqr}
\eeq
where $f_{\rho\rho}$ is another function of $\theta_\br$, to be given later.

Finally, the equal-time polarization-density cross-correlation function is given by
\beq
C_{p\rho}(\br)\equiv\langle p_y(\br+\bR,t)\rho(\bR,t)\rangle={f_{p\rho}(\theta_\bq)\over q}
\label{crossETqr}
\eeq
where the direction dependence is given by the function $f_{p\rho}$ that depends only on $\theta_\br$.




These predictions could also be tested experimentally in systems in which the active particles can be imaged, like those of \cite{schaller13,bricard2013}.

\section{Derivation of the  hydrodynamic theory\label{EOM}}

\subsection{Identifying the hydrodynamic variables}

We consider active polar particles of  areal density $\rho(\brp,t)$ 
confined to { a flat interface between a solid substrate and a three-dimensional (3D) 
passive, isotropic, bulk fluid.  We choose our coordinates $(x,y,z)$ so that the interface sits at $z=0$, with the} solid substrate underneath  (i.e., $z<0$), and the bulk fluid above (i.e., $z>0$). By ``polar", we mean each particle (labeled by an index $\alpha$, where $\alpha$ runs from $1$ to the number of particles $N$), can have associated with it a unique unit vector $\hp_\alpha$. (Imagine, e.g., a collection of arrows, with  $\hp_\alpha$ pointing from the tail to the head of arrow $\alpha$.) 

To formulate hydrodynamics, we coarse grain this particle-based unit vector  in to a position dependent vector {\it field} $\hp(\brp,t)$ defined on the surface at $z=0$.
In principle, this coarse graining can lead to fluctuations in the magnitude of the coarse grained 
$\hp(\brp,t)$ as well as its direction. 
However, since  we are considering  a system   in an orientationally ordered state,  {\it not} at a continuous transition from an ordered to a disordered state, fluctuations of the  magnitude $|\hp|$ of $\hp$ relax {\em fast}; i.e., the relaxation time-scale remains finite in the long wavelength limit.  This is because fluctuations in the magnitude of $|\hp|$, unlike fluctuations in its direction, are {\it not} Goldstone modes associated with the spontaneous breaking of continuous rotation invariance. Therefore, there is no symmetry reason for them to be slow, so we  expect them to be fast generically. Hence,  in the spirit of hydrodynamics, we ignore these magnitude fluctuations, and set $\hp^2=1$ without any loss of generality.  A renormalization group analysis\cite{JT thesis} confirms this heuristic argument.

 We choose   our coordinate system such that $p_x=1$ is the reference polar
ordered state (see figure (\ref{schem}). 

In the presence of friction from the substrate, there is no momentum 
conservation  on the surface, so  the only conserved 
variable on the surface of relevance for the present problem is the active particle 
number. The hydrodynamic theory that we develop below thus includes 
the active particle number density $\rho (\brp,t)$ and the orientational order parameter ${\hat{\bf p}}(\brp,t)$ fluctuations 
as the relevant slow  or ``hydrodynamic" variables.  We also include the {\it bulk} fluid velocity $\bv(\br, t)$, which is defined throughout the semi-infinite three dimensional space above the surface, since in that space momentum (which is equivalent to velocity in the limit of an incompressible bulk fluid) is conserved. However, we will work in the Stokesian limit, in which viscous forces dominate inertial ones; this has the effect of ``enslaving" the bulk velocity to the hydrodynamic variables $\rho (\brp,t)$ and  ${\hat{\bf p}}(\brp,t)$ on the surface, as we will show below.

\beq
\nonumber\\
\eeq
\subsection{Hydrodynamic equations of motion}

 We will formulate the hydrodynamic equations for these variables by expanding their equations of motion phenomenologically in powers of fluctuations of both fields $\hp$ and $\rho$ from their mean values, and in spatio-temporal gradients. In so doing, we must respect all symmetries and conservation laws of the underlying dynamics. Because this is a non-equilibrium system, however, additional constraints that would apply in equilibrium, such as detailed balance, do not hold here, and {\it any} term consistent with the symmetries and conservation laws is allowed, and therefore will, in general, be present.

We will consider systems with underlying rotational invariance in the plane of the surface. That is, we will focus on the case in which there is no {\it a priori} preferred direction in that plane; the active particles are free to {\it spontaneously} break this symmetry, by choosing a direction in which to spontaneously point.

 With these considerations in mind, we will now derive the hydrodynamic equations of motion for the fields $\rho(\brp,t)$, $\hp(\brp,t)$, and the bulk velocity velocity $\bv(\brp, z, t)$.
 
 We begin with the active particle density $\rho(\brp,t)$.
 
\subsubsection{Equation of motion for the active particle density $\rho$}
 
 The conservation of the active  particles implies that they obey a continuity equation: 
\bea
\partial_t\rho + {\boldsymbol\nabla}_s\cdot {\bf J}_\rho &=& 0\,, \label{cont}
\eea
where ${\boldsymbol \nabla}_s\equiv {\hat{\bf x}}\partial/\partial x +
{\hat{\bf y}}\partial/\partial y$ is the 2D gradient operator, with ${\hat{\bf x}}$ and
${\hat{\bf y}}$ the unit vectors along the $x$ and $y$ axis respectively. The active particle 
current ${\bf J}_\rho$ is generated by  friction with the  bulk fluid 
velocity ${\bf v}(\br=\brp, z=0)$, plus additional diffusive and active currents.  We will phenomenologically expand all of these effects to leading order in powers of the bulk velocity  evaluated at the surface ${\bf v}(\brp, z=0)$, and gradients, keeping all terms consistent with the underlying rotation invariance. In practice, this means we can make the vector ${\bf J}_\rho$ only out of vectors the active particle configuration itself chooses, that is, out of gradients, the surface velocity $\bv_s(\brp, t)$,  by which we mean the bulk velocity  evaluated at the surface; i.e.,  $\bv_s(\brp, t)\equiv{\bf v}(\brp, z=0,t)$, and the polarization $\hp(\brp, t)$. These constraints force ${\bf J}_\rho $ to take the form:
\begin{widetext}
\bea
{\bf J_\rho}(\brp) &=& \rho_e(\rho, |\bv_s|)  {\bf v}_s(x,y,z=0) + \kappa (\rho , { |\bv_s|} ) \hp-D_\rho\boldsymbol\nabla_s \rho-D_{\rho2}\hp(\hp\cdot\nabla_s)\rho -{\boldsymbol f}_\rho(x,y) \,. 
\nonumber\\ 
\label{currrho}
\eea
\end{widetext}
to leading order in gradients, where $D_\rho>0$ is the diffusion constant, and $D_{\rho 2}$ is an anisotropic diffusion coefficient.  Any other symmetry-permitted vector terms involving additional gradients, e.g., $\nabla_s^2 {\bf \hat p}$ would be irrelevant in the long wavelength limit. Note also that the first two terms on the rhs of (\ref{currrho}) reflect the fact that the  the ``active particle velocity'' ${\bf J}_\rho/\rho$ will, in general, be different from the velocity ${\bf v}_s$  of the bulk fluid at the surface, due to the activity.  The existence of a non-zero $D_{\rho 2}$  reflects the fact that, once the underlying rotation invariance is broken by the development of a spontaneous polarization $\hp$, diffusion along the direction of $\hp$ need not, and, in general, will not, proceed at the same speed as diffusion perpendicular to that direction.  The  ``effective density'' $\rho_e(\rho, |\bv_s|)$ would simply be $\rho$ itself in a Galilean invariant model, but since the solid surface breaks Galilean invariance, it can in general depend non-linearly on $\rho$, and even on the magnitude $|\bv_s|$ of the surface velocity. All of these dependences disappear in the linear model; furthermore, our analysis in section (\ref{RG}) of the non-linear model   shows that these dependences do not affect any of our predictions for the long distance behavior of the system. More precisely, they can all be absorbed into a suitable, finite ``renormalization'' of the parameters of the linear theory.  In renormalization group jargon, they are ``irrelevant''. We have only included them in the current (\ref{currrho}) to make our starting model  more general.

The factor   $\kappa (\rho, |\bv_s|)$ is an active parameter reflecting the self-propulsion of the particles through interaction with the solid substrate. It can also in principle depend  on $\rho$ and  the magnitude $|\bv_s|$ of the surface velocity. As for $\rho_e$, all of these dependences disappear in the linear model, and prove to also be irrelevant in the RG sense just described.

For completeness, 
we have included in this current a zero-mean Gaussian white noise  ${\boldsymbol f}_\rho$ that we take to be delta-correlated in space and time:
\begin{equation}
 \langle f_{\rho i} (x,y,t)f_{\rho j} (x',y',t')\rangle=2 D_\rho\delta_{ij}^s\delta(
x- x') \delta(y- y') \delta(t-t')\,.
\label{rho noise}
\end{equation}
This noise is added to model the stochastic nature of the density dynamics. In the equilibrium limit, this would have been a thermal noise representing the effects of a finite temperature and with a variance proportional to the diffusivity, as required by the Fluctuation-Dissipation theorem~\cite{chaikin}. In an active system  such as ours,  there is no constraint like the fluctuation-dissipation theorem or the Einstein relations relating the noise variance  (\ref{rho noise})  to the diffusivity. The white noise (\ref{rho noise}) in (\ref{currrho}) reappears in (\ref{cont}) as a conserved noise, consistent with the density being a conserved variable; see Eq.~(\ref{rho}) below.
However, as we will show later, this noise proves to be irrelevant in the long wavelength limit compared to the polarization noise we will introduce later. The diffusion constants $D_\rho$ and $D_{\rho2}$ prove to be irrelevant in the long-wavelength limit as well, as we shall see.

Using (\ref{currrho}) in (\ref{cont}), the continuity 
equation (\ref{cont}) 
can be written

\bea
\partial_t\rho =&-&\hp\cdot\nabla_s\kappa (\rho) -\kappa(\rho)\nabla_s\cdot\hp \nonumber \\&+& {\boldsymbol\nabla}_s\cdot\left[D_\rho\nabla_s\rho +D_{\rho 2}\hp (\hp\cdot{\boldsymbol\nabla}_s)\rho\right]
\nonumber\\
&-&\rho_e(\rho, |\bv_s|){\boldsymbol\nabla}_s\cdot \bv_s-\bv_s\cdot{\boldsymbol\nabla}_s\rho_e(\rho, |\bv_s|)+
{\boldsymbol\nabla}_s\cdot{\bf f}_\rho\,.\nonumber\\
\label{rho}
\eea


\vspace{.3in}

\subsubsection{The bulk velocity $\bv(\brp, z, t)$}

In calculating the bulk velocity $\bv(\brp,z,t)$, we will assume  the bulk fluid is  in the extreme ``Stokesian'' limit, in which inertia is negligible relative to viscous drag. This should be appropriate for most systems in  which the active particles are microscopic, since the Reynolds' number will be extremely low for such particles. 

The   3D incompressible bulk velocity field ${\bf 
v}=(v_i,v_z),\,i=x,y$ satisfies 
the 3D Stokes' equation
\begin{equation}
 \eta\nabla^2_3v_\alpha (\brp,z)= \partial_\alpha \Pi (\brp,z)  \,,
 \label{Stokes}
\end{equation}

\noindent  where the incompressibility constraint  
\beq
{\boldsymbol\nabla}_3\cdot {\bf v}=0
\label{incomp}
\eeq
determines 
$\Pi$, the hydrostatic pressure. 
In these expressions \eqref{Stokes}, \eqref{incomp}, ${\boldsymbol \nabla}_3\equiv {\hat{\bf x}} \partial/\partial x +
{\hat{\bf y}}\partial/\partial y+
{\hat{\bf z}}\partial/\partial z$ is the  full three-dimensional  gradient operator, with ${\hat{\bf x}}$, ${\hat{\bf y}}$,  and
${\hat{\bf z}}$ the unit vectors along the $x$, $y$, and $z$  axes respectively.


We could, in principle, add a random Brownian noise to this Stokesian force balance equation (\ref{Stokes}) to reflect thermal fluctuations in the non-zero temperature bulk fluid. However, since the bulk fluid is {\it passive} - that is, {\it equilibrium}, we know that the net effect of such a noise term on the surface polarization must be to introduce a purely equilibrium random noise in the polarization equation of motion. So we will instead take the simpler approach of simply adding such a thermal noise by hand to the active noises that will also appear in that equation.

 In the absence of such a noise term, the noiseless Stokes equation (\ref{Stokes}) can be solved exactly for the {\it bulk} velocity $\bv(\brp,z,t)$ in terms of the {\it surface} velocity $\bv_s(\brp,t)$. If we Fourier expand the surface velocity:
\beq
\bv_s(\brp,t)={1\over\sqrt{L_xL_y}}\sum_\bq \bv_s(\bq,t)e^{i\bq\cdot\brp}
\label{vs Fourier}
\eeq
where $(L_x,L_y)$ are the linear dimensions of our (presumed rectangular) surface, then, as we show in Appendix \ref{appa} , the bulk velocity $\bv(\brp,z,t)$  is given by
\beq
\bv(\brp,z,t)={1\over\sqrt{L_xL_y}}\sum_\bq [\bv_s-z(\bq\cdot\bv_s)({\hat \bq}+i{\hat \bz})]e^{-qz+i\bq\cdot\br_\perp}
\label{v bulk Fourier}
\eeq

This is to be 
supplemented by the boundary conditions at $z=\infty$, at which all stresses vanish, and at $z=0$, which we will now discuss. 
For a bulk passive fluid with a free surface, the boundary condition is given 
by the vanishing of the appropriate component of the shear stress.  On the other 
hand, for a bulk fluid resting on a solid surface, one imposes the ``no-slip'' 
boundary condition { that}  the relative velocity between the solid surface and 
the fluid layer in contact vanishes. More generally, if there is a slip 
velocity between the solid surface and the liquid layer in contact with the 
solid surface, then one must impose what in fluid dynamics is known as a {\em 
partial slip} condition on the fluid layer in contact with the solid substrate. 
For an active fluid layer at the solid-liquid interface, the fluid layer in 
contact is {\em active}; hence, there should be active forces acting on this 
layer. Generalizing the partial-slip boundary condition~\cite{partial-slip} for an active surface 
fluid, we impose the following {\em active} boundary condition at $z=0$:
\bea
&&\bv(\brp,z=0, t) \equiv  \bv_s(\brp,t) \nonumber \\ 
&& =   v_a(\rho) \hp(\brp,z=0,t)+ \mu{\bm \tau}_\parallel (\brp,z=0,t) \,,
\label{activebc2}
\eea
where $v_a(\rho)$ is the spontaneous self-propulsion speed of the active particles relative to the solid substrate  that can depend upon the local active particle density, and the force density ${\bm \tau}_\parallel$ parallel to the surface is given by

 
\beq
{\bm \tau}_\parallel(\brp,z=0,t)={\bf \cO}{\hat \bz}\cdot\bm{\sigma}^{{\rm bulk}} (\brp,z=0,t)+ {\bm \tau}^{{\rm surface}}(\brp, t) \,,
\label{tau}
\eeq 
with ${\bf \cO}$ the operator that projects any vector onto the plane of the surface, and  
${\bf \sigma}^{{\rm bulk}}$ the bulk stress tensor. The projection operator ${\bf \cO}$  is given by
\beq
\cO_{ij}=\delta_{ij}-\delta_{iz}\delta_{jz} \,,
\label{Odef}
\eeq
while the bulk stress tensor is
given by
\beq
\sigma^{{\rm bulk}}_{ij}=\eta(\partial_iv_j+\partial_jv_i) -\Pi\delta_{ij} \,.
\label{sigbulk}
\eeq
The $\eta$ term in \eqref{sigbulk}  is simply the viscous stress in the bulk fluid, while the pressure term $\Pi\delta_{ij}$ drops out of the force density ${\bm \tau}_\parallel$ parallel to the surface due to the projection operator ${\bf \cO}$.

The form of this boundary 
condition is dictated by symmetry. Essentially, {\it any} vector in the two dimensional plane of the solid surface that is {\it spontaneously} chosen by the system (as opposed to {\it specified a priori}, is allowed, and, therefore, will be present. Because the projection  $\cO{\hat \bz}\cdot\bm{\sigma}^{{\rm bulk}}$ is such a vector, a term proportional to it is allowed, and, hence, will generically be present in the boundary condition \eqref{activebc2}.

  Physically, we expect contributions to the surface stress from the bulk fluid stress since this stress is also experienced at the interface where the active particles are situated. So we would expect this stress in general to influence the active particles, and, therefore, the boundary condition on the bulk fluid velocity at the interface on which these particles sit.
Note that if we set both the active surface stress $\tau^{\rm surface}$ and the self-propulsion speed $v_a$ to zero, equations II.8 to II.11 reduce to the conventional
partial slip boundary condition of {\it passive} fluid dynamics\cite{partial-slip}.

Requiring that the bulk stress tensor \eqref{sigbulk} be divergenceless (as it must be in the Stokesian limit, since, by definition, in that limit the viscous stresses dominate the inertial ones) leads to 
 the Stokes' equation \eqref{Stokes} for the bulk flow. Note that $\tau^{\rm surface}$ exists only at the interface at $z=0$ and hence does not appear in the Stokes' equation for the bulk flow.  The  surface force ${\bm \tau}^{{\rm surface}} (\brp, t)$, when expanded to leading order in the polarization $\hp$ and gradients, must, by symmetry,  take the form\cite{maitra comment, maitra2018}
\bew
\bea
\tau^{{\rm surface}}_{i}=\zeta_1(\rho)\hp\cdot\nabla_Sp_i +\zeta_2(\rho)p_i\nabla_S\cdot\hp +p_i\hp\cdot\nabla_S\zeta(\rho) - \partial_iP'_s(\rho) \,.
\label{sigsurf}
\eea
\ew

In (\ref{sigsurf}), the $\zeta_{1,2}$ and $\zeta$ represent active stresses, and $P'_s$ is a surface osmotic pressure (not to be confused with the bulk pressure $\Pi$ in the Stokes equation (\ref{Stokes})), both of which depend on the density of the active particles.

Inserting (\ref{sigbulk}) and (\ref{sigsurf}) into our expression (\ref{tau}) for the force density ${\bf \tau}_\parallel$ parallel to the surface, and using the result in our  active boundary condition (\ref{activebc2}), gives, 
for $i=(x,y)$,
\bew
\begin{equation}
v_{si}(\brp,t)=v_a(\rho) p_i(\brp,t) + \zeta_1(\rho)\hp\cdot\nabla_Sp_i +\zeta_2(\rho)p_i\nabla_S\cdot\hp +p_i\hp\cdot\nabla_S\zeta(\rho) + \mu\eta\bigg(\frac{\partial 
v_i(\brp,z,t)}{\partial z}\bigg)_{z=0}- \partial_iP_s  \,,
\label{activebc}
\end{equation}
\ew
where we have defined $P_s\equiv\mu P'_s$, and used the fact that the velocity normal to the surface vanishes at the surface (i.e.,  $v_z(z=0)=0$)~\cite{maitra comment}.  Note that the velocity $v_i(\brp,z,t)$ is $i$'th component  the full {\it bulk} velocity.

The term proportional to $\partial_zv_i$ can be given another interpretation: if we define a length $a\equiv\mu\eta$, $ \mu\eta\bigg(\frac{\partial 
v_i}{\partial z}\bigg)_{z=0}=a\bigg(\frac{\partial 
v_i}{\partial z}\bigg)_{z=0}$ is simply the velocity of a fluid element a distance $a$ above the surface. Hence, we can alternatively interpret this term as modeling  active particles of finite thickness $a$ on the surface   which are passively convected by the local fluid velocity at that height. 
 For a system in thermal  equilibrium, $v_a=0=\zeta_{1,2}(\rho) = \zeta(\rho)$, and (\ref{activebc}) reduces to the well-known equilibrium partial slip boundary condition.

\subsubsection{Equation of motion for the polarization $\hp$}

We now turn to the equation of motion for $\hp$. As the active particles are polar, the system is 
not invariant under $\hp\rightarrow
-\hp$ symmetry. This means that terms even in $\hp$ are allowed in our phenomenological expression for $\partial_t\hp$. 

The most general equation of motion for $p_k$  allowed by symmetry is therefore:
\bew
\bea
\partial_tp_k=
T_{ki}\bigg(\alpha v_{si} -\lambda_{pv}(\bv_s\cdot\nabla_s)p_i +{h_i \over \gamma_1}+
\left({\nu_1-1 \over 2}\right)p_j\partial_iv_{sj} 
+\left({\nu_1+1 \over 2}\right)(\hp\cdot\nabla_s)v_{si}
-\lambda(\hp\cdot\nabla_s)p_i
-\partial_iP_p(\rho)+f_{i}\bigg) ,\nonumber\\ \label{pi}
\eea
\ew
where we have defined the transverse projection operator
\beq
T_{ki}\equiv\delta^s_{ki}-p_kp_i
\label{Tdef}
\eeq
which projects any vector orthogonal to $\hp$. Its presence in (\ref{pi}) insures that the fixed length condition $|\hp|=1$ on $\hp$ is preserved.

 In (\ref{pi}), the ``molecular field'' $\bh$ is given by
\beq
\bh=-{\delta F\over\delta\hp} \,,
\label{hdef}
\eeq
where the Frank free energy is 
\beq
F={1\over2}\int d^2r_\perp\,[K_1(\nabla_S\cdot\hp)^2+K_3|\nabla_S\times\hp|^2] \,.\label{frank-free1}
\eeq
Here, $K_1$ and $K_3$ are respectively the Frank splay and bend elastic modulii.  This $h_i/\gamma_1$-term, together with the Frank free energy $F$, gives the damping terms in (\ref{pilin}) below. Notice that the more familiar   classical alignment-free energy term $K|\nabla_S\hp|^2$ is already implicitly included 
in (\ref{frank-free1}) separately,  as can be seen by using the vector calculus identity
\beq
|\nabla_S\times\hp|^2=|\nabla_S\hp|^2-(\nabla_S\cdot\hp)^2-\nabla_S\cdot {\bf W} \,,
\label{vecid}
\eeq
where ${\bf W}\equiv(\hp\cdot\nabla_S)\hp-\hp(\nabla_S\cdot\hp)$, in (\ref{frank-free1}). Doing so, and dropping the surface term arising from the total divergence in \eqref{vecid}, we can rewrite (\ref{frank-free1}) in the form
\beq
F={1\over2}\int d^2r_\perp\,[(K_1-K_3)(\nabla_S\cdot\hp)^2+K_3|\nabla_S\hp|^2] \,.\label{frank-free2}
\eeq
In this form, it is {\it explicit} that we have included the usual alignment-free energy term proportional to $|\nabla_S\hp|^2$ in (\ref{frank-free1}).

The reasoning that leads to \eqref{pi} is essentially identical to that leading to our earlier expression \eqref{activebc} for the active boundary condition: we want to include {\it every} vector {\it spontaneously picked by the system}, provided that vector is also leading order in an expansion in powers of spatial gradients. It is straightforward to see that \eqref{pi} does so. It contains all of the vectors contained in \eqref{activebc}, except for an explicit term proportional to $\hp$ itself. We need not include such a term in \eqref{pi}, since it would be ``killed" by he projection operator $T_{ki}$. All the other terms in \eqref{pi} are vectors spontaneously picked by the system. Furthermore, they all have independent coefficients, since there is no symetry argument that would lock them ito any particular relation.

Note that terms proportional to some {\it a priori} specified vector, say $\hx$, are not allowed in \eqref{pi}, since they violate rotation invariance by singling out a special direction that is not spontaneously chosen by the system (i.e., by any of the vectors $\hp$, $\bv$, etc., or by spatial gradients).

The fact that we are performing a gradient expansion allows us to drop terms proportional to, e.g., $\nabla_S^2 v_{si}$, $\nabla_S^4 v_{si}$, and so on, in \eqref{pi}. While such terms are certainly allowed by symmetry, they are irrelevant at long distances relative to the $\alpha v_{si}$ and $p_j\partial_iv_{sj}$ terms already present  in \eqref{pi}. 

When all such symmetry forbidden and irrelevant in the gradient expansion terms are dropped, one is left with \eqref{pi} as the only possible equation of motion for $\hp$.


To summarize, the equation of motion (\ref{pi})
includes all of the relevant nonequilibrium terms allowed by symmetry, in addition to the usual
equilibrium terms and the convective covariant derivative of $\hp$. Here, 
the presence of the solid substrate underneath breaks any Galilean 
invariance, and hence a term directly proportional to $v_i$ in (\ref{pi}) is 
permissible: indeed the 
$\alpha$-term in (\ref{pi}) clearly breaks the Galilean invariance. The presence 
of this terms implies {\em flow alignment} or {\em antialignment} of the polarization $\hp$ depending 
upon the sign of $\alpha$  in 
Eq.~(\ref{pi}). The ``polarization pressure'' $P_p(\rho)$ is an additional function of the density, 
independent of the ``osmotic pressure'' $P_s(\rho)$ introduced earlier.
Furthermore,  the $\lambda$ term represents active self advection. Finally, the terms proportional to $\nu_1$ are ``flow alignment terms'', identical in form to those found in nematic liquid crystals\cite{martin1972}.

We have also added to the equation of motion (\ref{pi}) a white noise ${\bf f}$ with statistics
\begin{equation}
 \langle f_{ i} ({\bf r}_{_\perp},t)f_{ j} ({\bf r}_{_\perp}',t')\rangle=2 D_p\delta_{ij}\delta(
 {\bf r}_{_\perp}-{\bf r}_{_\perp}') 
\delta(t-t')\,.
\label{p noise}
\end{equation}
As discussed earlier, this noise also incorporates equilibrium contributions from thermal fluctuations of the bulk fluid. These must, by the fluctuation-dissipation theorem, be spatiotemporally white, as we have assumed in (\ref{p noise}), since the equilibrium dynamics of $\hp$ is local in space and time. The actual noise strength $D_p$ in (\ref{p noise}) is larger than the equilibrium value (which is proportional to $k_BT$)  due to active contributions to the noise.

\vspace{.3in}

\subsubsection{Summary of the equations of motion}

 Our hydrodynamic model, then, is summarized by the equations of motion (\ref{rho}) and (\ref{pi}) for $\rho$  and $\hp$, respectively, and the solution (\ref{v bulk Fourier}) of the Stokes equation (\ref{Stokes}) for the bulk velocity field $\bv(x,y,z,t)$obtained with the boundary condition (\ref{activebc}). When considering fluctuations, we will also need the noise correlations (\ref{rho noise})  and (\ref{p noise}) .

\subsection{Linearization of the equations of motion}

We now show that linearizing Eqs.~(\ref{rho}) and (\ref{pi}) about a uniform reference state with polarization ${\bf p}=\hat x$ and density $\rho=\rho_0$ produces Eqs.~(\ref{rholin-intro}) and (\ref{pyfin-intro}) in section~\ref{tech-sum}.

The equations of motion and boundary conditions found in the previous section have an obvious spatially uniform, steady state solution:
\bea
\rho(\brp,t)=\rho_0
\label{rho steady}\\
\hp(\brp,t)=\hx
\label{p steady}\\
\bv_s(\brp,t)=v_0\hx
\label{surface steady}\\
\bv(\brp,z,t)=v_0\hx
\label{bulk steady}
\eea
where we have defined
\beq
v_0\equiv v_a(\rho_0)
\label{v0def}
\eeq
and have chosen the $\hx$ axis of our coordinate system to be along the (spontaneously chosen) direction of 
polarization, as illustrated in figure (\ref{schem}).

As a first step towards understanding fluctuations about this steady state, we will  write 
\bea
&&\rho(\brp,t)=\rho_0+\delta\rho(\brp,t) \sep
\label{delrhodef}\\
&&\hp(\brp,t)=\hx\sqrt{1-p_y^2(\brp,t)}+p_y(\brp,t)\hy \sep
\label{p lin}\\
&&\bv_s(\brp,t)=(v_0+\delta v_{sx}(\brp,t))\hx+v_{sy}(\brp,t)\hy \,\,,\nonumber\\
\label{surface lin}
\eea
and expand the equations of motion \ref{rho}) and (\ref{pi}) for $\rho$  and $\hp$, and the boundary condition (\ref{activebc}), to linear order in $\delta\rho$ and $p_y$. 
We will obtain the bulk velocity $\bv(\brp,z,t)$ from the surface velocity $\bv_s(\brp,t)$ using our solution (\ref{v bulk Fourier}) of the Stokes equation.


We will expand the surface osmotic pressure of the particles $P_s(\rho)$  to linear order in $\delta\rho$: 
\beq
P_s(\rho)=\sigma \delta\rho \,.
\label{Ps exp}
\eeq
We will also expand the ``surface polarization  
pressure'' $P_p(\rho)$ to linear order in $\delta\rho$ (and will show later that higher powers of $\delta\rho$ are irrelevant at long wavelengths):
\beq
P_p(\rho)= \sigma_p \delta\rho \,,
\label{Pp exp}
\eeq
and likewise expand the $\zeta$'s in equation (\ref{activebc}) . To linear order, it is sufficient to take
\bea
&&\zeta_1(\rho)=\zeta_1(\rho_0)\equiv \zeta_{10} \sep
\label{zeta1lin}\\
&&\zeta_2(\rho)=\zeta_2(\rho_0)\equiv \zeta_{20} \sep
\label{zeta2lin}\\
&&\zeta(\rho)=\zeta(\rho_0)+\bar\zeta\delta\rho \,\,\,\,\,\,\,\,.
\label{zeta}
\eea
Finally, we expand $v_a(\rho)$, $\kappa(\rho)$ and $\rho_e(\rho)$ to linear order in $\delta\rho$:
\bea
&&v_a(\rho)\equiv v_0+v'_a\delta\rho \sep
\label{valin}\\
&&\kappa(\rho)\equiv \kappa_0+\bar\kappa\delta\rho \sep
\label{kappalin}\\
&&\rho_e(\rho)=\rho_1+\rho_e'\delta\rho\,\,\,\,\,\,\,\,,
\label{rhoelin}
\eea
where $\kappa_0$, $\bar\kappa$, $\rho_1$, and $\rho_e'$ are all constants to linear order. All of the other parameters (i.e., the diffusion constants $D_\rho$ and $D_{\rho2}$, $v_a$, $\alpha$, $\lambda$, $\lambda_{pv}$, $\gamma_1$, and $\nu_1$ can to linear order be replaced by their values at $\rho=\rho_0$, $|\bv_s|=v_0$, and treated as constants. 


 Now using our active boundary condition Eq.~(\ref{activebc}), the fluctuating velocity components $\delta v_x$ and 
$v_y$ can be expressed up to linear order as
\bea
\delta v_{sx} &=& v'_a\delta\rho+\bar\zeta\partial_x\delta\rho +\mu\eta\left({\partial_zv_x}\right)_{z=0} +\zeta_{20}\partial_yp_y 
-\sigma\partial_x \delta\rho\,, \label{vsx} \nonumber\\
\\
v_{sy} &=& v_0 p_y +\mu\eta \left({\partial_zv_y}\right)_{z=0}  +\zeta_{10}\partial_xp_y 
-\sigma\partial_y\delta\rho \,. \nonumber\\
 \label{vsy}
\eea

 These expressions for $\delta v_{sx}$ and $v_{sy}$ are implicit equations, since the bulk velocities $v_x$ and $v_y$ on the right hand side also
depend on the surface velocity through (\ref{v bulk Fourier}) . 

We can make them explicit by solving them iteratively. This is most conveniently done in Fourier space, because we can then obtain the iterative solution to lowest order in powers of the wavenumber $q$. 

Performing a {\it two-dimensional} Fourier transform on  (\ref{vsx}) and (\ref{vsy}) - that is, Fourier transforming over $\brp$, and using our solution (\ref{v bulk Fourier})  for the bulk velocity in terms of the surface velocity, we obtain
\bew
\bea
&&\delta v_{sx}(\bq,t) =v'_a\delta\rho(\bq,t)+i \bar\zeta q_x\delta\rho(\bq,t) +\mu\eta W_x(\bq,t) +i\zeta_{20}q_yp_y(\bq,t)
-i\sigma q_x \delta\rho(\bq,t)\,, \label{vsxFT} \nonumber\\
\\
&&\delta v_{sy}(\bq,t) = v_0 p_y(\bq,t) +\mu\eta W_y(\bq,t)  +i\zeta_{10}q_xp_y(\bq,t)
-i\sigma q_y\delta\rho(\bq,t) \,, \nonumber\\
 \label{vsyFT}\nonumber\\
\eea
\ew
where we have defined
\bea
{\bf W}(\bq,t) &\equiv & \left(\partial_z\left\{[\bv_s(\bq,t)-z(\bq\cdot\bv_s(\bq,t))({\hat \bq}+i{\hat \bz})]e^{-qz}\right\}\right)_{z=0} \nonumber \\ 
&=& -q\bv_s(\bq,t)-(\bq\cdot\bv_s(\bq,t)){\hat \bq}+W_z(\bq,t)\hz\,\,
\label{Wdef}
\eea
In the second equality of (\ref{Wdef}), we have not bothered to evaluate $W_z(\bq,t)$, since it does not appear in (\ref{vsxFT}) or (\ref{vsyFT}).

If we now insert our expressions  (\ref{vsxFT}) and (\ref{vsyFT}) for $\bv_s(\bq,t)$ after the second equality in (\ref{Wdef}), we see that the leading order in $q$ terms (which are  $O(q)$)  come from the $v'_a\delta\rho$ term   in  (\ref{vsxFT}) and the $v_0p_y$ term in  (\ref{vsyFT}). Keeping only those terms implies 
\beq
\bv_s(\bq,t)\approx v'_a\delta\rho\hx+v_0p_y(\bq,t)\hy \,.
\label{vsq0}
\eeq
 Inserting this into (\ref{Wdef}) gives
\bea
W_x &=& -\left({q^2+q_x^2 \over q}\right)v'_a\delta\rho -v_0\left({q_xq_y\over q}\right)p_y 
 \,\,, \label{vxqz} \\
W_y &=&  -v'_a\left({q_xq_y\over q}\right)\delta\rho-v_0\left({q^2+q_y^2 \over q}\right) p_y 
\,. \label{vyqz}
\eea

Inserting these expressions back into  (\ref{vsxFT}) and (\ref{vsyFT}) gives us our final closed form expression for the two components of the surface velocity, written entirely in terms of $p_y$ and $\delta\rho$:
\bew
\bea
\delta v_{sx}(\bq,t) &=&\left[v'_a+i (\bar\zeta-\sigma) q_x -\mu\eta v'_a\left({q^2+q_x^2 \over q}\right)\right]\delta\rho(\bq,t) +\left[i\zeta_{20}q_y-\mu\eta v_0 \left({q_xq_y\over q}\right)\right]p_y(\bq,t)
\,, \label{vsxFTclosed} 
\nonumber\\\\
v_{sy}(\bq,t) &=&\left[ v_0  -\mu\eta\left({q^2+q_y^2 \over q}\right)   +i\zeta_{10}q_x\right]p_y(\bq,t)
-\left[i\sigma q_y +\mu\eta v'_a\left({q_xq_y\over q}\right)\right]\delta\rho \,, \nonumber\\
 \label{vsyFTclosed}
\eea
\ew
Note the non-analytic character of the $\mu\eta$ terms in these expressions; this reflects the long-ranged hydrodynamic interaction between active particles on the surface mediated by the bulk passive fluid. Indeed, it is only through these terms that the presence of the bulk fluid makes itself felt. Note also that these terms are real; we'll see in a moment that this makes them {\it damping} terms. They are also the {\it same} order in $q$ as the imaginary $\zeta$ and $\sigma$ terms, which are associated with propagation. This is the origin of the peculiar property of our system that damping and propagation are the same order in wavevector (or, equivalently, that the quality factor $Q$ of the normal modes of this system is finite and independent of wavenumber $q$ at small $q$). This should be contrasted with, e.g., a simple bulk equilibrium fluid, for which the propagating terms are $O(q)$, while the damping terms are $O(q^2)$.

Note that the parameter $\zeta_{20}$ has dropped out at this point in our calculation, because it only leads to terms of higher order in $q$ than we have kept here.

With these expressions (\ref{vsxFTclosed}) and (\ref{vsyFTclosed}) for the surface velocity  in terms of $p_y$ and $\delta\rho$ 
in hand, we can now derive closed equations of motion for $p_y$ and $\delta\rho$. First, we must linearize the general equations of motion (\ref{rho}) and (\ref{pi}), using the linearizations (\ref{Pp exp})- (\ref{rhoelin}) for the pressures and $\zeta$'s, and    (\ref{delrhodef}) and (\ref{p lin}) for the density and the polarization. Doing so for the 
continuity equation (\ref{rho}), we obtain

\bew
\beq
\partial_t\delta\rho =-\kappa_0\partial_yp_y -(\bar\kappa+v_0\rho'_e)\partial_x\delta\rho -\rho_1\nabla_s\cdot\bv_s+ [D_\rho\nabla_s^2\rho +D_{\rho 2} \partial_x^2]\delta\rho+
\nabla_s\cdot{\bf f}_\rho\,.
\label{rholin1}
\eeq
\ew
Fourier transforming equation (\ref{rholin1}), and keeping only terms to leading order in $q$, we find
\beq
\partial_t\delta\rho(\bq,t) =-i\kappa_0q_yp_y -i(\bar\kappa+v_0\rho'_e)q_x\delta\rho -i\rho_1\bq\cdot\bv_s+ 
i\bq\cdot{\bf f}_\rho\,.
\label{rholinq}
\eeq

Setting $k=y$ in the equation of motion (\ref{pi}) for $\hp$, and linearizing the resulting equation gives
\bew
\beqn
\partial_tp_y=\alpha(v_{sy}-v_0p_y)-(\lambda_{pv}v_0+\lambda)\partial_xp_y+\left({\nu_1-1 \over 2}\right)\partial_y\delta v_{sx} 
+\left({\nu_1+1 \over 2}\right)\partial_xv_{sy}
-\sigma_p\partial_y\delta\rho+{1\over\gamma_1}(K_1\partial_y^2+K_3\partial_x^2)p_y+f_{y} \,.
\nonumber\\ 
\label{pilin}
\eeqn
\ew
Note the appearance of the combination $v_{sy}-v_0p_y$ in the $\alpha$ term on the RHS of this equation. The fact that this particular combination appears is not an accident, but, rather, a consequence of rotation invariance: if we work to {\it zeroeth} order in gradients in our solution   for the surface velocity, this term vanishes, as it must, since $p_y$ is a Goldstone mode associated with the spontaneous breaking of the continuous rotational symmetry. It is because of this exact cancellation that we needed to evaluate $\bv_s$ to higher order in gradients of $p_y$ in (\ref{vsxFT}) and (\ref{vsyFT}), which is why we had to keep higher order gradient terms in that expression.  Note that no such cancellation happened when we calculated the vector ${\bf W}$ eqn. (\ref{Wdef}) earlier, so {\it there} we {\it could} truncate the expansion for the surface velocity at zeroeth order in  the gradient term. 

Fourier transforming equation (\ref{pilin}), and keeping only terms to leading order in $q$, we find that  the molecular field terms proportional to the Frank constants $K_{1,3}$ are higher order in $q$, and so can be dropped, leaving us with
\bew
\beq
\partial_tp_y=\alpha(v_{sy}-v_0p_y)-i(\lambda_{pv}v_0+\lambda)q_xp_y+i\left({\nu_1-1 \over 2}\right)q_y\delta v_{sx} 
+i\left({\nu_1+1 \over 2}\right)q_xv_{sy}
-i\sigma_pq_y\delta\rho+f_{y}\bigg) \,.
\nonumber\\ 
\label{pilinq}
\eeq
\ew
Now inserting our expressions (\ref{vsxFTclosed}) and (\ref{vsyFTclosed}) for the surface velocity in terms of $p_y$ and $\delta\rho$ into the Fourier transformed equations of motion (\ref{rholinq}) and (\ref{pilinq}), and keeping only terms to leading order in $q$,  gives our final closed form for the linearized equations of motion:

\bea
\partial_t\delta\rho =- iv_\rho [q_x\delta\rho+\rho_c q_yp_y] + i\bq\cdot{\bf f}_\rho \,, \label{rholin}
\eea

\bew
\bea
\partial_tp_y =- iv_pq_xp_y -\gamma\left({q^2+q_y^2 \over q}\right)p_y -\left({\gamma_\rho\over\rho_c}\right)\left({q_xq_y \over q}\right)
\delta\rho
-i\sigma_t q_y\delta\rho + f_y \,, 
\nonumber\\
\label{pyfin}
\eea
\ew
where we have defined the characteristic velocities 
\beq
v_p\equiv\lambda_{pv}  v_0+\lambda-\zeta_{10}\alpha-\left({\nu_1+1\over2}\right) v_0
\label{vpdef}
\eeq
 and  
 \beq
 v_\rho\equiv \bar\kappa+\rho_e'v_0+\rho_1v'_a\,, 
  \label{vrhodef}
\eeq

the total inverse compressibility $\sigma_t\equiv\sigma_p+\sigma\alpha - \bigg(\frac{\nu_1-1}{2}\bigg)
v_a'$, 
 the characteristic density 
\beq
\rho_c\equiv {\rho_1v_0+\kappa_0\over v_\rho} \sep
\label{rhocdef}
\eeq
\noindent and the bulk fluid damping coefficients  $\gamma\equiv\alpha  v_0\mu\eta$ and $\gamma_\rho\equiv\alpha\mu\eta v'_a\rho_c$. Both $\gamma$ and $\gamma_\rho$ have the dimensions of speed. Further  more, all the contributions from $\delta v_x({\bf q},t)$ in (\ref{vsxFTclosed}) in (\ref{pilinq}) are subleading and hence do not appear in (\ref{pyfin}), where only the terms leading order in $\bf q$ are kept.


We thus obtain the linearized hydrodynamic equations  (\ref{rholin}) and (\ref{pyfin}),  which are nothing but  equations (\ref{rholin-intro}) and (\ref{pyfin-intro}) mentioned earlier. These
are invariant under $q_y\rightarrow -q_y,\,p_y\rightarrow 
-p_y,\rho\rightarrow\rho$, but not invariant under $q_x\rightarrow -q_x$. This 
is equivalent to invariance in real space under  $y\rightarrow -y, 
p_y\rightarrow -p_y,\rho\rightarrow \rho$, but  with no analogous invariance under 
$x\rightarrow -x$.  In these two equations $v_\rho\neq v_p$ in general, due to the abscence of Galilean invariance because of
friction with the solid substrate.
The effective coefficient $\gamma$, when positive (which can be achieved by tuning the signs of the various original model parameters), serves as the effective damping coefficient in the model. Interestingly, the effective damping here is ${\cal O}(q)$, which is far stronger than the ${\cal O}(q^2)$ damping in the linearized Toner-Tu model for flocking~\cite{tonertu95,toner98} in the hydrodynamic limit. Hydrodynamic interactions mediated by the passive, bulk fluid above are responsible for this ${\cal O}(q)$ damping. 

An alert reader might wonder how hydrodynamic interactions mediated by the bulk fluid  can dominate friction from the underlying solid substrate,  which, after all, is $O(q^0)$. The reason is that friction with the substrate, while playing a very important role (in particular, it is the principal mechanism limiting the speed $v_0$ of the active particles), does not act to suppress fluctuations in the {\it directions} of motion of those particles (or, similarly, the polarization. The leading order damping of such fluctuations, which are the Goldstone modes of our problem, come from the hydrodynamic interactions.

\vspace{.2in}

\section{Mode structure of the linearized equations and stability of the uniform state\label{stab}}
Having written down the equations of motion, we will now analyze the linear stability of the 
system.
 We  work  in polar coordinates ${\bf 
q}=(q\cos\theta_\bq,q\sin\theta_\bq)$,  and
set the noises to zero.
Assuming a time-dependence of the
form $\rho, p_y\sim  \exp (-i\omega t)$, this   leads to
the eigenvalue condition on $\omega$:

\bew
\beq
\omega^2+i\omega[i(v_p+v_\rho)q\cos\theta_\bq+\gamma q (1+\sin^2\theta_\bq)]-i[iv_pq\cos\theta_\bq+\gamma q (1+  \varphi\sin^2\theta_\bq)]v_\rho q\cos\theta_\bq-c_0^2q^2\sin^2\theta_\bq=0 \,,
\label{ev1}
\eeq
\ew
where we have defined 
\beq
c_0^2\equiv \sigma_t\rho_c v_\rho
\label{c0def}
\eeq
and 
\beq
\varphi\equiv1-{\gamma_\rho\over\gamma} \,.
\label{phidef1}
\eeq.

The eigenfrequencies always scale as $q$, independent of all other parameters, as can be seen by inserting the ansatz
\beq
\omega=\Upsilon(\theta_\bq) q 
\label{omega lin q}
\eeq
into (\ref{ev1}). This leads to a $q$ independent condition on $\Upsilon$:
\bew
\beq
\Upsilon^2+i\Upsilon[i(v_p+v_\rho)\cos\theta_\bq+\gamma  (1+\sin^2\theta_\bq)]-i[iv_p\cos\theta_\bq+\gamma  (1+ \varphi\sin^2\theta_\bq)]v_\rho \cos\theta_\bq-c_0^2\sin^2\theta_\bq=0 \,,
\label{ev2}
\eeq
\ew
thereby proving that $\omega$ scales like $q$.

For stability, we must have ${\rm Im}(\omega)<0$ for all $\bq$, which means ${\rm Im}(\Upsilon(\theta_\bq))<0$ for all $\theta_\bq$. We show in appendix (\ref{appb})  that this condition is satisfied for sufficiently small $\gamma_\rho$   (which appears in (\ref{ev2}) through $\varphi\equiv1-{\gamma_\rho\over\gamma}$), provided $\gamma$ and $c_0^2$ are both $>0$. Note that this condition is similar to the stability criterion for a simple compressible bulk fluid: $c_0^2>0$ simply means the bulk compressibility is positive, while $\gamma>0$ is analogous to requiring positive shear and bulk viscosities.  The stability condition on $\gamma_\rho$ is
\bew
\beq
-|\varpi-1|-\sqrt{(\varpi-1)^2+{1\over m^2}}<{\gamma_\rho\over\gamma}<-|\varpi-1|+\sqrt{(\varpi-1)^2+{1\over m^2}} \,,
\label{stability}
\eeq
\ew
where we have defined the ``Mach number''
\beq
m\equiv {v_\rho\over c_0} \,,
\label{mach1}
\eeq
and the speed ratio 
\beq
\varpi\equiv
{v_p\over v_\rho} \,.
\label{ratio1}
\eeq
It is easy to see that this condition can always be satisfied for sufficiently small $\gamma_\rho$; in particular, the allowed region always includes $\gamma_\rho=0$. That a sufficiently large $\gamma_\rho$ relative to the effective damping $\gamma$ can lead to instabilities is not surprising. A non-zero $\gamma_\rho$ (or, equivalently, $v_a'$) implies that different patches of the system move at different speeds. Of course, damping tends to homogenize  the density by exchanging particles,  thereby reducing the  speed  differences. However, if $\gamma_\rho$ is too large, damping may not be sufficient to suppress these speed  fluctuations.  The eventual steady state  may be nonuniform, leading to a patterned state, though its actual nature cannot be ascertained from the  linearized equations of motion. Nonlinear amplitude equations would be necessary for a full-fledged analysis of the steady state. It will be interesting to explore the relation between these instabilities here and the banding instabilities reported in ~\cite{bertin,mishra}.



Thus 
generic underdamped propagating waves with anisotropic, $\theta_\bq$-dependent 
wavespeed proportional to $q$ are expected for the wide range of parameters satisfying the stability condition derived in appendix B.

\section{Correlation functions and robustness of long-ranged order against noise\label{corr}}

\begin{figure}[htb]
\includegraphics[width=4cm]{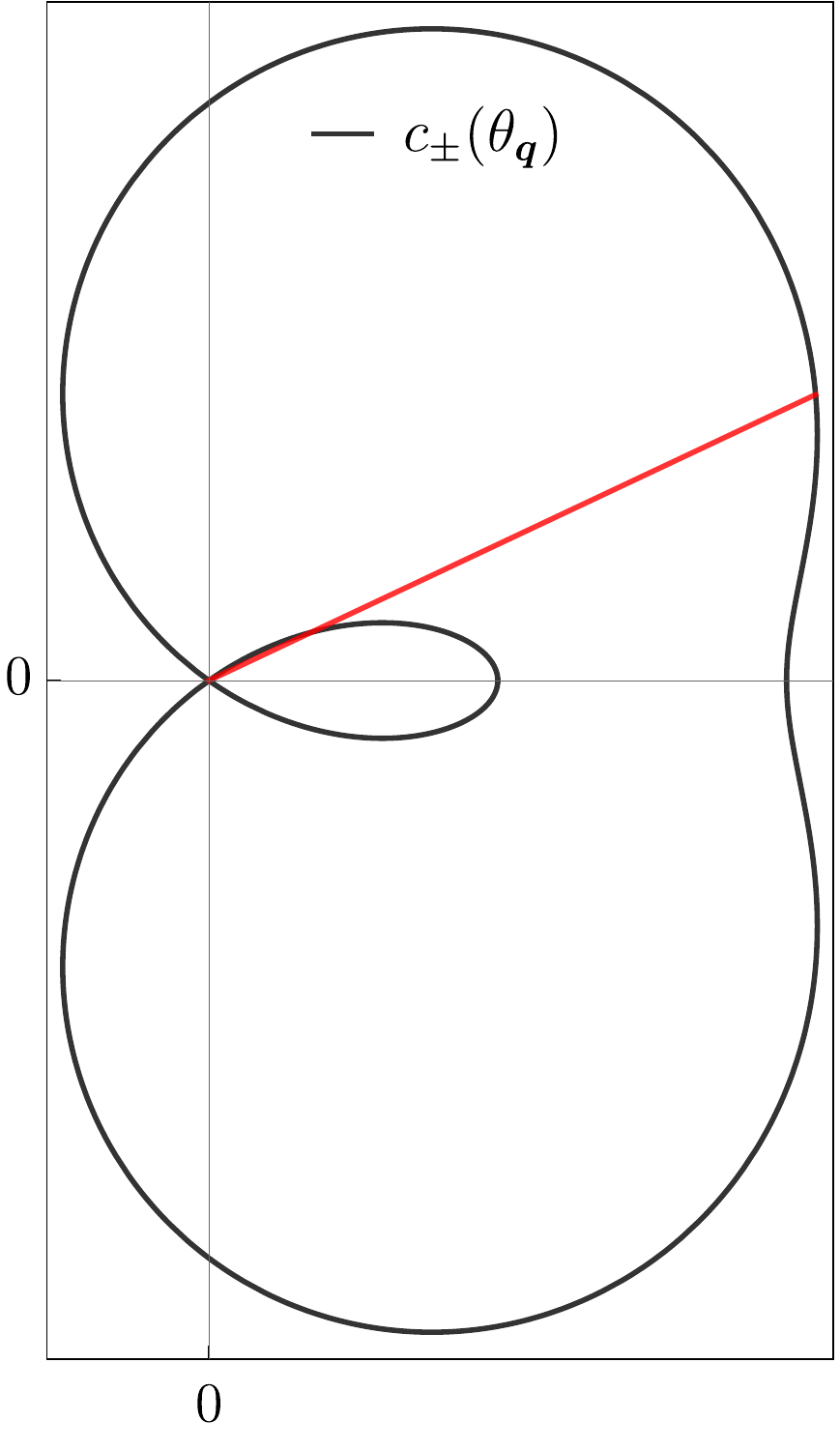}
\caption{(Color online) Polar plot of the sound speeds (\ref{cplusminus}) (black curve). Here we have taken $q=1$, $v_\rho=1$, $v_p=c_0=2$, and  
$\gamma=.3$ (all in arbitrary units).The polarization points directly  to the right. The   red straight line indicates the direction of $\bq$ for which the scaling function $F_{pp}$ is plotted in figure (\ref{scaling functions}).}
\label{sound speeds}
\end{figure}

In the stable region of the parameter space, 
the 
correlation functions for the system in the steady state may be calculated
from the noise-driven equations of motion. Dropping for now the density force ${\bf f}_\rho$ (we will show later that it is irrelevant in the long-wavelength limit), and solving the linear equations of motion for the spatio-temporally Fourier transformed fields $p_y(\bq, \omega)$ and $\delta\rho(\bq, \omega)$ gives
\bew
\beq
p_y(\bq, \omega)={i(\omega-v_\rho q\cos\theta_\bq)f_y(\bq,\omega)\over(\omega-c_+(\theta_\bq)q)(\omega-c_-(\theta_\bq)q)+ i(\omega\Psi(1,\theta_\bq)-v_\rho  q \Psi(\varphi,\theta_\bq)\cos\theta_\bq)q}\,,
\label{pysol1}
\eeq
\ew
\bew
\beq
\delta\rho(\bq, \omega)={i\rho_cv_\rho q\sin\theta_\bq f_y(\bq,\omega)\over(\omega-c_+(\theta_\bq)q)(\omega-c_-(\theta_\bq)q)+ i(\omega\Psi(1,\theta_\bq)-v_\rho { q} \Psi(\varphi,\theta_\bq)\cos\theta_\bq)q}\,,\label{rhosol1}
\eeq
\ew
where we have defined 
\beq
\Psi(\varphi, \theta)\equiv\gamma(1+\varphi\sin^2\theta)\,,
\label{Psidef}
\eeq
and with the ``sound speeds'' $c_\pm(\theta_\bq)$, defined by the positions of the peaks in the scaling function versus $u$ ~\cite{unreal},  precisely those found for dry active matter in \cite{tonertu95, toner98, toner05}; i.e., 
\bew
\begin{eqnarray}
c_{\pm}\left(\theta_\bq \right) =
\left({v_\rho + v_p \over 2}\right)\cos\theta_\bq
\pm \sqrt{{1 \over 4}\left(v_\rho -v_p\right)^2 \cos^2
\theta_\bq + c^2_0
\sin^2 \theta_\bq} \quad ,
\label{cplusminus}
\end{eqnarray}
\ew 
The `sound speeds'' $c_\pm(\theta_\bq)$ in (\ref{cplusminus}), defined by the positions of the peaks in the scaling function versus $u$~\cite{unreal},  are precisely those found for dry active matter in~\cite{tonertu95, toner98, toner05}. Further, these peak positions agree with those found in  the experiments of \cite{Geyer17} on Quinke rotators. 
In Fig. (\ref{sound speeds})  we show a  polar plot of these sound speeds.

Autocorrelating these fields with themselves then gives their spatio-temporally Fourier transformed correlations:
\beq
\nonumber
\eeq

\bew
\beq
C_{pp}({\bf q},\omega)\equiv\langle |p_y(\bq, \omega)|^2\rangle={2D_p(\omega-v_\rho q\cos\theta_\bq)^2\over(\omega-c_+(\theta_\bq)q)^2(\omega-c_-(\theta_\bq)q)^2+ (\omega\Psi(1,\theta_\bq)-v_\rho { q} 
\Psi(\varphi,\theta_\bq)\cos\theta_\bq)^2q^2}\, ,
\label{pysol}
\eeq
\ew
\beq
\nonumber
\eeq
\beq
\nonumber
\eeq
\bew
\beq
C_{\rho\rho}({\bf q},\omega)\equiv\langle |\delta\rho(\bq, \omega)|^2 \rangle ={2D_p\rho_c^2v_\rho^2 q^2\sin^2\theta_\bq\over(\omega-c_+(\theta_\bq)q)^2(\omega-c_-(\theta_\bq)q)^2+ (\omega\Psi(1,\theta_\bq)-v_\rho 
{ q} \Psi(\varphi,\theta_\bq)\cos\theta_\bq)^2q^2}\,,
\label{rhosol}
\eeq
\ew
where we have used (\ref{p noise}) to obtain the autocorrelation $\langle|f_y(\bq,\omega)|^2\rangle=2D_p$.

Similar reasoning gives the cross-correlation function
\bew
\beq
C_{p\rho}({\bf q}, \omega)\equiv  \langle \rho({\bf q},\omega) p_y(-{\bf q},-\omega)\rangle 
= \frac{2D_p \rho_c v_\rho q\sin\theta_\bq (\omega - v_\rho q\cos\theta_\bq)} {(\omega-c_+(\theta_\bq)q)^2(\omega-c_-(\theta_\bq)q)^2+(\omega\Psi(1,\theta_\bq)-v_\rho q \Psi(\varphi,\theta_\bq)\cos\theta_\bq)^2q^2}\,.
\label{crossprhoFT}
\eeq
\ew

Pulling a factor of $q^2$ out of the numerator of each of these expressions, and a factor of $q^4$ out of their denominators, gives the scaling forms  (\ref{pycorscale}), (\ref{rhocorscale}),  and 
(\ref{crossprhocorscale}), with the scaling functions $F_{pp}(u_f)$, $F_{\rho\rho}(u_f)$ and $F_{p\rho}(u_f)$.


The scaling function $F_{pp}(u_f, \theta_\bq)$ is given by 
\bew
\beq
F_{pp}(u_f, \theta_\bq)={2D_p(u_f-v_\rho \cos\theta_\bq)^2\over(u_f-c_+(\theta_\bq))^2(u_f-c_-(\theta_\bq))^2+(u_f\Psi(1,\theta_{\bf q})-v_\rho \Psi(\varphi,\theta_{\bf q})\cos\theta_\bq)^2}\,,
\label{pyscalefn}
\eeq
\ew
where  $u_f\equiv \omega/q$.

Note that the existence of the scaling form (\ref{pycorscale}) implies that the ratio of the widths of the peaks
in  $C_{pp}(\bq, \omega)$, plotted versus $\omega$ for fixed $\bq$, to their positions does not change as $\bq\to{\bf 0}$; this is what we meant by our earlier cryptic comment that the ``quality factor $Q$'' becomes independent of $q$.

In Fig. (\ref{sound speeds}),  we show a  polar plot of the sound speeds $c_\pm(\theta_\bq)$, which are the positions of the peaks in the scaling function $F_{pp}$ when plotted versus the scaling argument $u_f$ for fixed direction of propagation $\theta_\bq$. As such, they are the analog in our system of the sound speeds in a simple compressible fluid.

The density-density correlation function obeys a similar scaling law:
\beq
C_{\rho\rho}({\bf q},\omega)\equiv\langle |p_y(\bq, \omega)|^2\rangle={1\over q^2}F_{\rho\rho}\bigg(\left({\omega\over q}\right), \theta_\bq\bigg)\, ,
\eeq
with a slightly different scaling function $F_{\rho\rho}(u_f, \theta_\bq)$  given by

\bew
\beq
F_{\rho\rho}(u_f, \theta_\bq)={2D_p  \rho_c^2v_\rho^2 \sin^2\theta_\bq\over (u_f-c_+(\theta_\bq))^2(u_f-c_-(\theta_\bq))^2+(u_f\Psi(1,\theta_{\bf q})-v_\rho \Psi(\varphi,\theta_{\bf q})\cos\theta_\bq)^2}\, .
\label{rhoscalefn}
\eeq
\ew

 The orientation-density correlation function also obeys a similar scaling law:
\beq
C_{p\rho}({\bf q},\omega)\equiv\langle  \rho({\bf q},\omega) p_y(-{\bf q},-\omega)  \rangle={1\over q^2}F_{p\rho}\bigg(\left({\omega\over q}\right), \theta_\bq\bigg)\, ,
\eeq
where the scaling function for this correlation function is
\bew
\beq
F_{p\rho}(u_f, \theta_\bq)
= \frac{2D_p \rho_c v_\rho \sin\theta_\bq (u_f - v_\rho \cos\theta_\bq)} {(u_f-c_+(\theta_\bq))^2(u_f-c_-(\theta_\bq))^2+(u_f\Psi(1,\theta_\bq)-v_\rho  \Psi(\varphi,\theta_\bq)\cos\theta_\bq)^2}\,.
\label{crossprhoscalefn}
\eeq
\ew

\begin{figure}[htb]
\includegraphics[height=5cm]{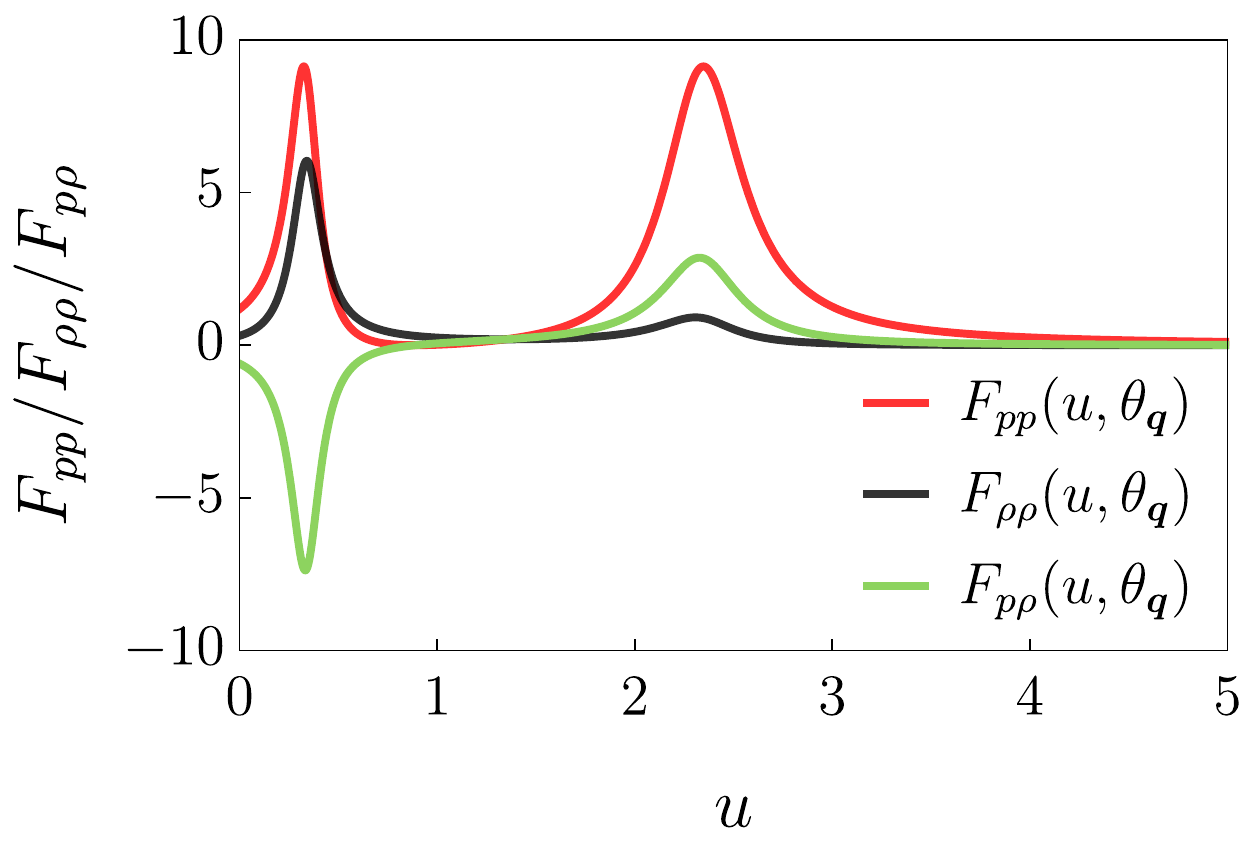}
\caption{(Color online) Plot of the scaling functions $F_{pp}(u_f,\theta_\bq)$  ( red curve), $F_{\rho\rho}((u_f,\theta_\bq)$ ( black curve), and  $F_{p\rho}(u_f,\theta_\bq)$ (green curve) versus scaling argument $u_f\equiv{\omega\over q}$  for fixed $\theta_\bq$. Here we have taken $v_\rho=1$, $v_p=c_0=2$, 
$\gamma=.3$, $D_p=.5$,  $\rho_c=1$, ${\gamma_\rho/\gamma}={1/2}$  (all in arbitrary units), and $\theta=27^\circ$ (the direction indicated by the straight line in figure (\ref{sound speeds}).}
\label{scaling functions}
\end{figure}

These three scaling functions are plotted versus the scaling argument $u$ for a fixed direction of propagation $\theta_\bq$ in Figure (\ref{scaling functions}). Note that the two peaks in the polarization scaling function $F_{pp}$ are exactly the same height, but have different widths. We have chosen to plot this figure for a fairly generic direction of propagation. In contrast, for $\theta=\pi/2$ (that is, propagation perpendicular to the polarization, which corresponds to a vertical line in figure (\ref{sound speeds})), the scaling functions $F_{pp}$ and $F_{\rho\rho}$ both become even functions of the scaling argument $u$,  with two symmetrically placed peaks at $\pm c_0$. This implies that  the correlation functions both become even functions of $\omega$, with two symmetrically placed peaks at $\pm c_0q$.
Furthermore, for this direction of propagation, the two correlations functions are precisely proportional to each other; that is, their ratio is a constant, independent of both $q$ and $\omega$.

Another special direction is $\theta_\bq=0$; i.e., propagation {\it along} the polarization. In this case, the asymmetry is maximized: the polarization scaling function has only {\it one} peak, at scaling argument $u=v_p$, meaning the correlation function has a single peak at $\omega=v_pq$. The density-density scaling function  on the other hand, vanishes identically in this limit (except at the singular point $\omega=v_\rho q$, where its value depends on precisely how the limit $\omega\to v_\rho q$, $\theta_\bq$ is approached).


Both of these special directions are, indeed, special: for generic directions of propagation $\theta\ne0,\pm \pi/2$, the scaling functions, and, hence, the $\omega$-dependence of the correlation functions, look like figure (\ref{scaling functions}). That is, each scaling function has two in general {\it asymmetrically} placed peaks, in the same positions for both correlation functions. 

The way the special limit $\theta=0$ is approached as $\theta\to0$ is that the first  (i.e., leftmost) peak of the polarization scaling function $F_{pp}$, while its height stays the same as that of the second (i.e., rightmost) peak, has its width continuously vanish as $\theta\to0$, while the density scaling function $F_{\rho\rho}$ has  the height of its first peak continuously vanish as $\theta\to0$. One can see the tendency towards this limit in the plot of figure (\ref{scaling functions}) at $\theta=27^\circ$.

We can use these expressions to calculate the equal time correlation functions; we find for the polarization $p_y$
\bew
\begin{eqnarray}
 &&C_{pp}(\bq)\equiv\langle |p_y(\bq,t)|^2\rangle=\int_{-\infty}^\infty  
\frac{d\omega}{2\pi}\langle |p_y(\bq, \omega)|^2\rangle 
={D_p\over\pi} \int_{-\infty}^\infty 
{(\omega-v_\rho q\cos\theta_\bq)^2d\omega\over(\omega-c_+(\theta_\bq)q)^2(\omega-c_-(\theta_\bq)q)^2+ (\omega\Psi(1,\theta_\bq)-v_\rho { q} \Psi(\varphi,\theta_\bq)\cos\theta_\bq)^2q^2}\nn\\
\label{pet1}
\end{eqnarray}
\ew
Introducing a new variable of integration 
\beq
s\equiv{\omega-v_\rho q\cos\theta_\bq \,{  \Xi(\varphi, \theta_\bq)}\over{ \Psi(1,\theta_\bq)q}}
\label{newvar}
\eeq
\beq
\nonumber
\eeq
where we have defined
\beq
\Xi(\Phi, \theta)\equiv{\Psi(\Phi, \theta)\over\Psi(1, \theta)}={1+\Phi\sin^2\theta\over1+\sin^2\theta}
\label{Xidef}
\eeq
gives
\bew
\begin{eqnarray}
 &&C_{pp}(\bq)={D_p\over\pi{ \Psi(1, \theta_\bq)}{ q}} \int_{-\infty}^\infty 
{{ (s+\psi(\varphi, \theta_\bq)}^2ds\over(s-\psi_+( \varphi,  \theta_\bq))^2(s-\psi_-({ \varphi}, \theta_\bq))^2+s^2} \,,
\label{pet2}
\end{eqnarray}
\ew
where we have defined
\beq
\psi_\pm(\varphi, \theta_\bq)\equiv{c_\pm(\theta_\bq)-v_\rho \cos\theta_\bq \, { \Xi(\varphi, \theta_\bq)}\over{\Psi(1, \theta_\bq)}} \,,
\label{psidef}
\eeq
 and
\beq
\psi(\varphi, \theta_\bq)\equiv{v_\rho \cos\theta_\bq \,( \Xi(\varphi, \theta_\bq)-1)\over \Psi(1,\theta_\bq)} \,.
\label{psi n def}
\eeq
We will show in appendix (\ref{appc})
 that,  whenever the stability condition is satisfied, $\psi_+>0$ and $\psi_-<0$; we will make use of these facts later.

 We also show in appendix \ref{appc} that this integral is equal to
{
\bew
\beq
\int_{-\infty}^\infty 
{(s+\psi(\varphi, \theta_\bq)^2ds\over(s-\psi_+(\varphi,  \theta_\bq))^2(s-\psi_-( \varphi, \theta_\bq))^2+s^2} = \pi\left[1+\left({\gamma_\rho\over\gamma}\right)^2\frac{m^2\sin^2\theta_\bq\cos^2\theta_\bq}{\left\{\left(1+\sin^2\theta_\bq\right)^2-m^2{\gamma_\rho \over \gamma}[\varpi-1+(\varpi-\varphi)\sin^2\theta_\bq]
\cos^2\theta_\bq\right\}}\right] \,.
\label{Icpp}
\eeq
\ew

Using (\ref{Icpp}), we obtain the expression for the equal-time correlation function:
\begin{align}
 C_{pp}(\bq) &={D_p\over q\Psi(1, \theta_\bq)}\left[1+\left({\gamma_\rho\over\gamma}\right)^2\frac{m^2\sin^2\theta_\bq\cos^2\theta_\bq}{D(m^2,\varpi,\varphi,\theta_\bq)}\right] \nonumber \\
 &={D_p\over\gamma q(1+\sin^2\theta_\bq)}\left[1+\left({\gamma_\rho\over\gamma}\right)^2\frac{m^2\sin^2\theta_\bq\cos^2\theta_\bq}{D(m^2,\varpi,\varphi,\theta_\bq)}\right]  \,,
\label{petfinal}
\end{align}
where we have defined
\bew
\beq
D(m^2,\varpi,\varphi,\theta_\bq)=\left[\left(1+\sin^2\theta_\bq\right)^2-m^2{\gamma_\rho \over \gamma}[\varpi-1+(\varpi-\varphi)\sin^2\theta_\bq]\cos^2\theta_\bq\right]
\eeq
\ew

This expression is easily Fourier transformed in space to give the real space correlations $C_{pp}(\br)$ of the polarization fluctuations: 
\beq
C_{pp}(\br)\equiv\langle p_y(\br+\bR,t)p_y(\bR,t)\rangle=\int  {d^2q\over(2\pi)^2} e^{i\bq\cdot\br } {
C_{pp}(\bq) } \,.
\label{Cppreal1}
\eeq
Defining $\theta_\br$ as the angle between $\br$ and the $x$-axis, we can rewrite this as

\bew
\bea
C_{pp}(\br) ={D_p\over(2\pi)^2\gamma} \int_0^{2\pi} {d\theta_\bq\over1+ \sin^2\theta_\bq}\left[1+\left({\gamma_\rho\over\gamma}\right)^2\frac{m^2\sin^2\theta_\bq\cos^2\theta_\bq}{D(m^2,\varpi,\varphi,\theta_\bq)}\right]\int_0^\infty dq ~e^{iqr\cos(\theta_\bq-\theta_\br)} \,.
\label{Cppreal2}
\eea
\ew

The integral over $q$ is given by
\beq
\int_0^\infty dq ~e^{iqr\cos(\theta_\bq-\theta_\br)}=\pi\delta(r\cos(\theta_\bq-\theta_\br))+\mathcal{P}\bigg[{1\over r\cos(\theta_\bq-\theta_\br)}\bigg]\nonumber\\
\label{qint}
\eeq
where $\mathcal{P}$ denotes the Cauchy principal value. 
Using this in (\ref{Cppreal2}), we have

\bew
\beq
C_{pp}(\br)={D_p\over2\pi\gamma} \int_0^{2\pi} {d\theta_\bq\over1+ \sin^2\theta_\bq}\left[1+\left({\gamma_\rho\over\gamma}\right)^2\frac{m^2\sin^2\theta_\bq\cos^2\theta_\bq}{D(m^2,\varpi,\varphi,\theta_\bq)}\right]\bigg(\delta(r\cos(\theta_\bq-\theta_\br))+\mathcal{P}\bigg[{1\over\pi r\cos(\theta_\bq-\theta_\br)}\bigg]\bigg) \,.
\label{Cppreal3}
\eeq
\ew


The Cauchy principal value term in this expression is odd under the operation $\theta_\bq\to\theta_\bq+\pi$, while ${1\over1+ \sin^2\theta_\bq}$ is even under this operation; hence, the contribution to integral from the Cauchy principal value term vanishes. We are therefore left with

\bew

\beq
C_{pp}(\br)={D_p\over 4\pi\gamma} \int_0^{2\pi} {d\theta_\bq\over1+ \sin^2\theta_\bq}\left[1+\left({\gamma_\rho\over\gamma}\right)^2\frac{m^2\sin^2\theta_\bq\cos^2\theta_\bq}{D(m^2,\varpi,\varphi,\theta_\bq)}\right]
\delta(r\cos(\theta_\bq-\theta_\br)) \,.
\label{Cppreal4}
\eeq

\ew

Using
\newpage

\bew

\beq
\delta(r\cos(\theta_\bq-\theta_\br))={\delta(\theta_\bq-\theta_\br+\pi/2)+\delta(\theta_\bq-\theta_\br-\pi/2)\over r\left|\left({\pp \cos(\theta_\bq-\theta_\br)\over\pp\theta_\bq}\right)_{\theta_\bq=\theta_\br\pm\pi/2}\right|}={\delta(\theta_\bq-\theta_\br+\pi/2)+\delta(\theta_\bq-\theta_\br-\pi/2)\over r} \,,
\label{deltas}
\eeq

\ew

where the first equality follows from the familiar identity 
\beq
 \nonumber
 \eeq
\beq
\delta(f(x))=\sum_{x_0}\delta(x-x_0)/|f'(x_0)| \,,
\eeq
where $\{x_0\}$ are the roots of $f(x)$,  
(\ref{Cppreal4}) becomes

\bew
\bea 
C_{pp}(\br) &=& {D_p\over 4\pi\gamma r} \int_0^{2\pi} {d\theta_\bq\over1+ \sin^2\theta_\bq}\left[1+\left({\gamma_\rho\over\gamma}\right)^2\frac{m^2\sin^2\theta_\bq\cos^2\theta_\bq}{D(m^2,\varpi,\varphi,\theta_\bq)}\right]
(\delta(\theta_\bq-\theta_\br+\pi/2)+\delta(\theta_\bq-\theta_\br-\pi/2)) \nonumber \\ 
&& ={D_pr\over 2\pi\gamma (r^2+ x^2)}\left[1+\left({\gamma_\rho\over\gamma}\right)^2\frac{m^2y^2 x^2}{\left\{(r^2+x^2)^2-m^2{\gamma_\rho \over \gamma}y^2[r^2(\varpi-1)+(\varpi-\varphi)x^2]\right\}}\right] \,.
\label{Cppreal5}
\eea
\ew


We now turn to the equal-time density correlations. Using (\ref{rhosol}), we have 
\bew
\begin{eqnarray}
 &&C_{\rho\rho}(\bq)\equiv\langle |\rho(\bq,t)|^2\rangle\nn\\&&=\int_{-\infty}^\infty  
\frac{d\omega}{2\pi}\langle |\rho(\bq, \omega)|^2\rangle 
={D_p\rho_c^2v_\rho^2q^2\sin^2\theta_\bq\over\pi} \int_{-\infty}^\infty 
{d\omega\over(\omega-c_+(\theta_\bq)q)^2(\omega-c_-(\theta_\bq)q)^2+(\omega\Psi(1,\theta_\bq)-v_\rho q \Psi(\varphi,\theta_\bq)\cos\theta_\bq)^2q^2} \,.\nn\\
\label{rhoet1}
\end{eqnarray}
\ew

}

Making the same change of variables of integration (\ref{newvar}) that we made earlier, and using partial fractions again, we find
\bew
\begin{eqnarray}
 C_{\rho\rho}(\bq)={D_p\rho_c^2v_\rho^2q^2\sin^2\theta_\bq\over2i\pi[\gamma q(1+ \sin^2\theta_\bq)]^3} (I_\rho^--I_\rho^+) \,,
\label{rhoet2}
\end{eqnarray}
\ew
where we have defined
\beq
\nonumber\\
\eeq
\bew
\begin{eqnarray}
 &&I_\rho^\pm\equiv \lim_{\epsilon\to0} \int_{-\infty}^\infty 
{ds\over(s-i\epsilon-\psi_+(\theta_\bq))(s-i\epsilon-\psi_-(\theta_\bq))\pm i(s-i\epsilon)}\frac{1}{s-i \epsilon} \,,
\label{irhopm}
\end{eqnarray}
\ew
where we have introduced the small parameter $\epsilon$ to regularize the integral. 

We can now do both integrals by the usual complex contour techniques. Note that, unlike the integral for the polarization correlation function $C_{pp}$, the integrand in (\ref{irhopm}) converges rapidly enough at infinity that the infinite semicircle needed to close the path of integration contributes nothing to the integral. Note also that the poles of $I^\pm_\rho$ are those of $I_\pm$ considered earlier, plus one more pole at $s=i\epsilon$.

If we choose $\epsilon>0$, then we can close the integral for $I_\rho^+$ in the upper half plane, picking up the pole at $s=i\epsilon$. This gives
\beq
I_\rho^+={2i\pi\over\psi_+\psi_-} \,,
\label{irhop}
\eeq
where we have taken the limit $\epsilon\to0$.

On the other hand, we can close the contour for $I_\rho^-$ in the lower half plane, finding (for $\epsilon>0$) no poles at all. This implies 
\beq
I_\rho^-=0 \,.
\label{i-0}
\eeq
Inserting (\ref{irhop}) and (\ref{i-0}) into (\ref{rhoet2}), we obtain 
\begin{eqnarray}
 C_{\rho\rho}(\bq)=-{D_p\rho_c^2v_\rho^2q^2\sin^2\theta_\bq\over[\gamma q(1+ \sin^2\theta_\bq)]^3\psi_+\psi_-}  \,.
\label{rhoet3}
\end{eqnarray}
It is straightforward to check that, had we chosen $\epsilon<0$, we would have gotten the same answer, although now with $I_\rho^+=0$ and $I_\rho^-=-{2i\pi\over\psi_+\psi_-}$.

It is an equally straightforward algebraic exercise, using the definitions (\ref{psidef}) of $\psi_\pm$ and our expressions (\ref{cplusminus})  for $c_\pm(\theta_\bq)$ to show that

\beq
\psi_+\psi_-=-\frac{D(m^2,\varpi,\varphi,\theta_\bq) v_\rho^2 \gamma^2\sin^2\theta_\bq }{m^2
[\Psi(1,\theta_\bq)]^4} \,.
\label{psiprod}
\eeq
Using this in (\ref{rhoet3}) gives
\begin{eqnarray}
 C_{\rho\rho}(\bq)={D_p\rho_c^2 m^2 (1+ \sin^2\theta_\bq)\over \gamma q D(m^2,\varpi,\varphi,\theta_\bq)} \,.
 \nonumber\\
\label{rhoet4}
\end{eqnarray}
The above expression can be easily Fourier transformed in space to obtain the real space density 
correlations $C_{\rho\rho}(\br)$ just as we did above to obtain $C_{pp}(\br)$.  We thereby obtain:
\bew
\beq
C_{\rho\rho}(\br)\equiv\langle \delta\rho(\br+\bR,t)\delta\rho(\bR,t)\rangle= 
\frac{D_p \rho_c^2m^2}{2\pi\gamma} \frac{r(r^2+x^2)}{\left\{(r^2+x^2)^2-m^2{\gamma_\rho \over \gamma}y^2[r^2(\varpi-1)+(\varpi-\varphi)x^2]\right\}} \,.
\label{Crhoreal5}
\eeq
\ew

Finally we work out the expression for the crosscorrelation function 
\begin{widetext}


\begin{eqnarray}
C_{p\rho}({\bf q})\equiv \langle \rho({\bf q})p_y(-{\bf q})\rangle&=&\int_{-\infty}^{\infty} \frac{d\omega}{2\pi} \langle \rho({\bf q},\omega) p_y(-{\bf q},-\omega)\rangle \nonumber \\
&=& \int_{-\infty}^{\infty} \frac{d\omega}{2\pi} \frac{2D_p \rho_c v_\rho q\sin\theta_\bq (\omega - v_\rho q\cos\theta_\bq)} {(\omega-c_+(\theta_\bq)q)^2(\omega-c_-(\theta_\bq)q)^2+(\omega\Psi(1,\theta_\bq)-v_\rho q \Psi(\varphi,\theta_\bq)\cos\theta_\bq)^2q^2}\nonumber \\
&=&\frac{D_p\rho_cv_\rho q\sin\theta_\bq} {\pi\gamma^2 q^2 (1+ \sin^2\theta_\bq)^2}\int_{-\infty}^{\infty}\frac{(s+ \psi(\varphi,\theta_\bq))ds}{(s-\psi_+)^2(s -\psi_-)^2 +s^2}. 
\label{crossprho}
\end{eqnarray}
From equations (\ref{I2}) and (\ref{I3}), we can easily  show that
\begin{eqnarray}
 \int_{-\infty}^{\infty}\frac{(s+ \psi(\varphi,\theta_\bq))ds}{(s-\psi_+)^2(s -\psi_-)^2 +s^2} = I_2 + 
 \psi(\varphi,\theta_\bq) I_3
\end{eqnarray}
where
\begin{equation}
 I_2=0, \; \mbox{and} \,\, I_3= -\frac{\pi}{\psi_+\psi_-}. 
 \label{i2i3} 
\end{equation}
\end{widetext}

Using eqs. (\ref{i2i3}) and (\ref{crossprho}), we obtain 

\beq
C_{p\rho}(\bq)= - \frac{D_p\rho_c \sin\theta_\bq\cos\theta_\bq m^2 \gamma_\rho}{\gamma^2 q D(m^2,\varpi,\varphi,\theta_\bq)} \,.
\label{crosscorrq}
\eeq

Fourier transformation of the above expression gives the real space cross correlation, given by

\bew
\bea
C_{p\rho}(\br) = - \frac{D_p\rho_c m^2 \gamma_\rho }{2\pi \gamma^2} \frac{xyr}{\left\{(r^2+x^2)^2-m^2{\gamma_\rho \over \gamma}y^2[r^2(\varpi-1)+(\varpi-\varphi)x^2]\right\}}.
\label{crosscorrr}
\eea
\ew

To summarize, we have shown that  over a finite range of the model parameters, the linearized model displays long range orientational order. We argue below that this linear theory prediction is asymptotically exact, in contrast to the well-known Toner-Tu model for dry active matter,  for which nonlinear effects are essential for the existence of the long range order~\cite{tonertu95,toner98}.

\section{Giant number fluctuations and their shape dependence\label{section:GNF}}

We will show in this section that the long-ranged density correlations found in equation (\ref{Crhoreal5})   of the previous section lead to ``giant number fluctuations''\cite{GNF}\cite{toner05,ramaswamy03, Chate+Giann}. These can be defined as follows: 

Consider a ``counting box'', defined as 
 a rectangular area $A=L_x\times L_y$, and define the aspect ratio $\alpha_A=L_x/L_y$. Our experiment will consist of counting the number of active particles in this box. The {\it mean} number of particles $\overline N$ in the box is, of course, given by
\begin{equation}
 \overline N= L_x L_y\rho_0 = \alpha_A L_y^2 \rho_0 \,.
\end{equation}

The total number fluctuation $\delta N\equiv N-\overline N$ in the area $A$ is
\begin{equation}
 \delta N = \int d^2 r\,\delta\rho(\br,t).
\end{equation}
Thence,

\begin{widetext}
\begin{eqnarray}
 &&\langle (\delta N)^2\rangle = \int_0^{L_x} dx\int_0^{L_y} dy \int_0^{L_x}dx'\int_0^{L_y} dy' \langle \rho ({\bf r},t)\rho({\bf r'},t)\rangle = \int_0^{L_x} dx\int_0^{L_y} dy \int_0^{L_x}dx'\int_0^{L_y} dy' C_{\rho\rho}({\bf r}- {\bf r'})\nonumber \\
 &&=\frac{D_p\rho_c^2 m^2}{2\pi \gamma } \int_0^{L_x} dx\int_0^{L_y} dy \int_0^{L_y} dy'\int_0^{L_x}dx'\,\frac{|{\bf r-r'}|}{({\bf r-r'})^2 + (x-x')^2}\nonumber \\
 &&\times \left[1-\frac{(v_p-v_\rho)(y-y')^2}{({\bf r-r'})^2 + (x-x')^2}\frac{\gamma_\rho v_\rho}{\gamma c_0^2} - \left(\frac{v_\rho\gamma_\rho}{c_0\gamma}\right)^2 \frac{(y-y')^2(x-x')^2} {\{({\bf r-r'})^2 + (x-x')^2\}^2}\right]^{-1}.
 \label{Nvar1}
\end{eqnarray}
\end{widetext}

We consider the limits $\alpha_A\gg 1$ and $\alpha_A \ll 1$. 

Consider first the limit $L_x\gg L_y$, i.e., $\alpha_A \gg 1$. Now consider the integral over $x'$  for fixed ${\bf r}=(x,y)$ in this limit. Note first that in this limit, for most of the range of integration over $x$, $x\gg L_y$. For such values of $x$, we can split  the integral over $x'$  into  three parts: 

\begin{widetext}
\begin{eqnarray}
&&\int_0^{L_x}dx'\frac{|{\bf r-r'}|}{({\bf r-r'})^2 + (x-x')^2}\nonumber \\
 &&\times \left[1-\frac{(v_p-v_\rho)(y-y')^2}{({\bf r-r'})^2 + (x-x')^2}\frac{\gamma_\rho v_\rho}{\gamma c_0^2} - \left(\frac{v_\rho\gamma_\rho}{c_0\gamma}\right)^2 \frac{(y-y')^2(x-x')^2} {\{({\bf r-r'})^2 + (x-x')^2\}^2}\right]^{-1}=I_1+I_2+I_3 \,,
\label{split1}
\end{eqnarray}
\end{widetext}

where we have defined
\bew
\begin{eqnarray}
I_1&\equiv&\int_0^{x-CL_y}dx'\frac{|{\bf r-r'}|}{({\bf r-r'})^2 + (x-x')^2}\nonumber \\
 &&\times \left[1-\frac{(v_p-v_\rho)(y-y')^2}{({\bf r-r'})^2 + (x-x')^2}\frac{\gamma_\rho v_\rho}{\gamma c_0^2} - \left(\frac{v_\rho\gamma_\rho}{c_0\gamma}\right)^2 \frac{(y-y')^2(x-x')^2} {\{({\bf r-r'})^2 + (x-x')^2\}^2}\right]^{-1} \,,
\label{I1def}
\end{eqnarray}
\begin{eqnarray}
I_2&\equiv&\int_{x-CL_y}^{x+CL_y}dx'\frac{|{\bf r-r'}|}{({\bf r-r'})^2 + (x-x')^2}\nonumber \\
 &&\times \left[1-\frac{(v_p-v_\rho)(y-y')^2}{({\bf r-r'})^2 + (x-x')^2}\frac{\gamma_\rho v_\rho}{\gamma c_0^2} - \left(\frac{v_\rho\gamma_\rho}{c_0\gamma}\right)^2 \frac{(y-y')^2(x-x')^2} {\{({\bf r-r'})^2 + (x-x')^2\}^2}\right]^{-1} \,,
\label{I2def}
\end{eqnarray}
\begin{eqnarray}
I_3&\equiv&\int_{x+CL_y}^{L_x}dx'\frac{|{\bf r-r'}|}{({\bf r-r'})^2 + (x-x')^2}\nonumber \\
 &&\times \left[1-\frac{(v_p-v_\rho)(y-y')^2}{({\bf r-r'})^2 + (x-x')^2}\frac{\gamma_\rho v_\rho}{\gamma c_0^2} - \left(\frac{v_\rho\gamma_\rho}{c_0\gamma}\right)^2 \frac{(y-y')^2(x-x')^2} {\{({\bf r-r'})^2 + (x-x')^2\}^2}\right]^{-1} \,,
\label{I3def}
\end{eqnarray}
\ew

\noindent where $C$ is an arbitrary constant enough larger than $1$ that we can safely neglect terms of ${\cal O}({1\over C})$. With this choice, throughout the region of integration for $I_1$, $x-x'\gg L_y$. Therefore, throughout this region, $|y-y'|\ll x-x'$ (since $|y-y'|$ is always $<L_y$). This gives $\left(\frac{y-y'}{x-x'}\right)^2 \ll 1$.
Making this approximation for $I_1$, our expression for it reduces to
\beq
I_1\approx {1\over 2}\int_0^{x-CL_y}{dx'\over x-x'} \,.
\label{I1approx}
\eeq
Performing this elementary integral gives
\beq
I_1={1\over2 }\ln\left({x\over CL_y}\right)={1\over2 }\ln\left({x\over L_y}\right) -{1\over2} \ln C \,.
\label{I1result1}
\eeq
Virtually identical reasoning can be applied to $I_3$, giving the result
\beq
I_3={1\over2 }\ln\left({L_x-x \over CL_y}\right) \,.
\label{I3result1}
\eeq

For $I_2$, a simple shift of variables of integration $x'=x+x''$ shows that $I_2$ is independent of $x$:
\bew
\begin{eqnarray}
I_2&=&\int_{-CL_y}^{CL_y}dx''\frac{\sqrt{x''^2 + (y-y')^2}}{(y-y')^2 + 2x''^2}\nonumber \\
&&\times \left[1-\frac{\gamma_\rho v_\rho}{\gamma c_0^2}\frac{(v_p-v_\rho)(y-y')^2} {(y-y')^2 + 2x''^2} - \left(\frac{v_\rho\gamma_\rho}{c_0\gamma}\right)^2 \frac{(y-y')^2{ x''}^2} {[(y-y')^2 + 2x''^2}]^2\right]^{-1}\,.
\label{I2result1}
\end{eqnarray}
\ew
Inserting these results (\ref{I1result1}),  (\ref{I2result1}), and  (\ref{I3result1}) into our expression (\ref{split1}), and using that result in our expression (\ref{Nvar1}) for $\langle (\delta N)^2\rangle$, we obtain

\bew
\begin{eqnarray}
 &&\langle (\delta N)^2\rangle  
\approx \frac{D_p\rho_c^2 m^2}{2\pi \gamma }\Bigg\{{1\over2 }\int_0^{L_x} dx\int_0^{L_y} dy \int_0^{L_y}dy'\left(\ln\left({x\over L_y}\right)+\ln\left({L_x-x\over L_y}\right) -2\ln C \right)\nonumber\\&&
+\int_0^{L_x} dx\int_0^{L_y} dy \int_0^{L_y}dy'\int_{-CL_y}^{CL_y}dx''\frac{ \sqrt{x''^2+(y-y')^2}}{ [(y-y')^2 +2x''^2]} 
\times \left[1-\frac{\gamma_\rho v_\rho}{\gamma c_0^2}\frac{(v_p-v_\rho)(y-y')^2}
{ { [(y-y')^2 +2x''^2]} }
- \left(\frac{v_\rho\gamma_\rho}{c_0\gamma}\right)^2 \frac{(y-y')^2{ x''^2}} { [(y-y')^2 +2x''^2]^2}\right]^{-1}\Bigg\} \,,\nonumber\\
\label{delNfinal1}
\end{eqnarray}
\ew
The integrals over $y$ and $y'$ in the (triple) integral in the first line of this expression trivially give a factor of $L_y^2$, 
since the integrand is independent of $y$ and $y'$. The remaining integral over $x$ is elementary. The net result is
\bew
\beq
{1\over2 }\int_0^{L_x} dx\int_0^{L_y} dy \int_0^{L_y}dy'\left(\ln\left({x\over L_y}\right)+\ln\left({L_x-x\over L_y}\right) -2\ln C \right)= L_xL_y^2\left(\ln\left({L_x\over CL_y}\right)-1\right)\,.
\label{1st line1}
\eeq
\ew
The $x$ integral in the remaining (quadruple) integral which appears on the second line in equation (\ref{delNfinal1}) can be done immediately, since the integrand is independent of $x$, yielding a factor of $L_x$. The remaining triple integral over $y$, $y'$, and $x''$ can be done by changing variables of integration to new, rescaled variables $u_x$, $u_y$, and $u'_y$ via

\beq
y\equiv L_yu_y
\sep
y'\equiv L_yu'_y
\sep
x''\equiv L_yu_x \,.
\label{varchange xyy'}
\eeq
This gives
\bew
\bea
\int_0^{L_x} dx\int_0^{L_y} dy \int_0^{L_y}dy'\int_{-CL_y}^{CL_y}dx''&&\frac{ \sqrt{x''^2+(y-y')^2}}{ [(y-y')^2 +2x''^2]}\times \left[1-\frac{\gamma_\rho v_\rho}{\gamma c_0^2}\frac{(v_p-v_\rho)(y-y')^2}
{ { [(y-y')^2 +2x''^2]} }
- \left(\frac{v_\rho\gamma_\rho}{c_0\gamma}\right)^2 \frac{(y-y')^2{ x''^2}} { [(y-y')^2 +2x''^2]^2}\right]^{-1}  \nonumber \\
&&=C' L_xL_y^2 \,,
\label{2nd line1}
\eea
\ew
where
\bew
\bea
C'&\equiv& \int_0^1 du_y\int_0^1 du_y'\int_{-C}^C\,du_x  \frac{\sqrt{u_x^2 + (u_y-u_y')^2}}{(u_y - u_y')^2 + 2 u_x^2}\nonumber \\&&
\left[1-\frac{\gamma_\rho v_\rho}{\gamma c_0^2}\frac{(v_p-v_\rho)(u_y-u_y')^2}{2u_x^2 + (u_y - u_y')^2} - \left(\frac{v_\rho\gamma_\rho}{c_0\gamma}\right)^2 \frac{(u_y-u_y')^2 u_x^2}{[2u_x^2 + (u_y - u_y')^2]^2}\right]^{-1} 
\label{C'def}
\eea
\ew
is an ${\cal O}(1)$ constant. Comparing  (\ref{1st line1}) and (\ref{2nd line1}), we see that the first line of 
(\ref{delNfinal1})  actually dominates the second in the large aspect ratio limit $L_x\gg L_y$ that we are considering here. Therefore, we obtain, in the limit of large aspect ratio ($\alpha_A\gg1$):
\bew
\begin{equation}
 \sqrt{\langle (\delta N)^2\rangle} \approx \frac{\rho_c m\sqrt{D_p}}{\sqrt{2\pi \gamma}} 
 \sqrt{L_xL_y^2\ln\left({L_x\over L_y}\right)}\approx\frac{\rho_c m\sqrt{D_p}}{\sqrt{2\pi \gamma}} L_y^{3/2}\sqrt{\alpha_A \ln \alpha_A}\,,
 \label{Nvarlar}
\end{equation}
\ew


This expression (\ref{Nvarlar}) can be rewritten in terms of 
the mean particle number $\overline N$ in the same area $A$,  which is given by 
\begin{equation}
 \overline N= L_x L_y\rho_0 = \alpha_A L_y^2 \rho_0 \,.
\end{equation}
This gives 
\beq
L_y=(\overline N/(\alpha_A\rho_0))^{1/2} \,.
\label{LyNb}
\eeq
 Using this result (\ref{LyNb}) in our expression (\ref{Nvarlar})  for the variance $(\Delta N)^2$ of the number fluctuations gives {
\begin{equation}
 \Delta N= \sqrt{\langle(\delta N)^2\rangle} = \left({\rho_c m\over \rho_0^{3/4}}\right)\left(\sqrt {D_p\over
 2\pi \gamma}\right) \left(\frac{{\overline N}^{3/4}\sqrt{\ln{\alpha_A}}}{\alpha_A^{1/4}}\right)    \,.
 \label{bigalpha}
\end{equation}

Three points should be noted about this result:

\noindent 1) The number fluctuations are {\it giant}; that is, they grow much more rapidly with $\bar N$ than the usual $\sqrt{\bar N}$ ``law of large numbers'' fluctuations, which are found in almost all systems, and, in particular, in most equilibrium systems away from fixed points\cite{SC}. { Specifically, we have $\Delta N \propto  \overline N^{3/4}$. Note also  that the exponent $3/4$ is very close to the exponent $.8$ found in the experiments of ~\cite{schaller13}.
}

\noindent 2) The size of the number fluctuations depend not only on the mean number $\bar N$, but also on the {\it shape} (i.e., on the aspect ratio $\alpha_A$). 

\noindent 3) The number fluctuations $\Delta N$ are a monotonically {\it decreasing} function of the aspect ratio $\alpha_A$ in this range of $\alpha_A\gg1$.

We analyze giant number fluctuations in the opposite regime of $L_y  \gg L_x$, i.e.,  small aspect ratio $\alpha_A\ll 1$ in Appendix~\ref{giant-app}. 
 As shown there, in this regime of small aspect ratio $\alpha_A$, $\Delta N$ is a monotonically {\it increasing} function of $\alpha_A$. Recall that in the opposite limit of  $\alpha_A\gg 1$, eqn (\ref{bigalpha})    , we found that $\Delta N$ is a monotonically {\it decreasing} function of $\alpha_A$. Hence, the maximum value of $\Delta N$ for a given mean number of particles $\overline N$ will occur when $\alpha_A\sim 1$; i.e., for a roughly square counting box.
}

  In contrast to our above results on giant number fluctuations, Ref.~\cite{bricard2013} did not report any giant number fluctuations in their experiment. We believe this is due to the fact that in their experiment, the density fluctuations were probed at length scales larger than the height of the passive fluid over the active layer, which is outside the regime of validity of our theory. 
  


\section{Bulk velocity fluctuations}\label{bulk}

We can use the relations (\ref{vsx}) and (\ref{vsy})  between the bulk fluid velocity field and the surface velocity, and our boundary condition (\ref{activebc}) for that surface velocity, to obtain expressions for the bulk velocity correlations  in terms of the polarization $\hp$ and density $\rho$ correlation functions. This gives, {\em to the lowest order in gradients}

\bew
\bea
&&\langle v_i(\brp,z)v_j(\brp, z)\rangle =v_0^2 \int  {d^2q\over(2\pi)^2} C_{pp}(\bq)e^{-2qz}\bigg[\delta_{iy}\delta_{jy}-{q_yz\over q}(\delta_{iy}q_j+\delta_{jy}q_i)+q_y^2z^2\left({q_iq_j\over q^2}+\delta_{iz}\delta_{iz}\right)\bigg] \nonumber \\ 
&& + v_0v_a' \int  {d^2q\over(2\pi)^2} C_{p\rho}(\bq)e^{-2qz}\bigg[\delta_{iy}\delta_{jx}+
\delta_{ix}\delta_{jy} - z({q_y\over q}\delta_{ix}q_j+ {q_y\over q}\delta_{jx}q_i + {q_x\over q}\delta_{jy}q_i 
+ {q_x\over q}\delta_{iy}q_j)+ 2q_xq_yz^2\left({q_iq_j\over q^2}+\delta_{iz}\delta_{iz}\right)\bigg] \nonumber \\ 
&& + v_a'^2 \int  {d^2q\over(2\pi)^2} C_{\rho\rho}(\bq)e^{-2qz}\bigg[\delta_{ix}\delta_{jx}-{q_xz\over q}(\delta_{ix}q_j+\delta_{jx}q_i)+q_x^2z^2\left({q_iq_j\over q^2}+\delta_{iz}\delta_{iz}\right)\bigg]
\label{vbulk1}
\eea
\ew
The second ($\delta_{iy}q_j$) term in the first integral is odd in $q_x$ if $j=x$. Since $C_{pp}(\bq)e^{-2qz}$ and $q_y$ are even in $q_x$, the integral of this term vanishes when $i=x$. Hence, we can replace this term by $\delta_{iy}\delta_{jy}q_y$. Similar arguments imply that we can also replace the third ($\delta_{jy}q_i$) term with $\delta_{iy}\delta_{jy}q_y$. Similarly we can replace ($\delta_{ix}q_j$) with 
$\delta_{ix}\delta_{jx}q_x$, and ($\delta_{jx}q_i$) with $\delta_{ix}\delta_{jx}q_x$ in the third integral. In the second integral ($\delta_{ix}q_j$) can be replaced with $\delta_{ix}\delta_{jy}q_y$, ($\delta_{jx}q_i$) 
with $\delta_{jx}\delta_{iy}q_y$, ($\delta_{jy}q_i$) with $\delta_{ix}\delta_{jy}q_x$, and ($\delta_{iy}q_j$) 
with $\delta_{iy}\delta_{jx}q_x$ respectively. Thus, we can rewrite (\ref{vbulk1}) as
\bew
\bea
\langle v_i(\brp,z)v_j(\brp, z)\rangle &=&v_0^2 \int  {d^2q\over(2\pi)^2} C_{pp}(\bq)e^{-2qz}\bigg[\delta_{iy}\delta_{jy}\left(1-{2q_y^2z\over q}\right)+q_y^2z^2\left({q_iq_j\over q^2}+\delta_{iz}\delta_{jz}\right)\bigg] 
\nonumber \\ 
&& + v_0v_a' \int  {d^2q\over(2\pi)^2} C_{p\rho}(\bq)e^{-2qz}\bigg[\bigg(\delta_{iy}\delta_{jx}+
\delta_{ix}\delta_{jy}\bigg)(1 - zq) + 2z^2q_xq_y\left({q_iq_j\over q^2}+\delta_{iz}\delta_{jz}\right)\bigg] \nonumber \\ 
&& + v_a'^2 \int  {d^2q\over(2\pi)^2} C_{\rho\rho}(\bq)e^{-2qz}\bigg[\delta_{ix}\delta_{jx}\left(1-{2q_x^2z\over q}\right)+q_x^2z^2\left({q_iq_j\over q^2}+\delta_{iz}\delta_{jz}\right)\bigg].
\label{vbulk2}
\eea
\ew
W{e also note that the $q_iq_j\over q^2$ term in the first integral is also odd in $q_x$ unless $i=j$, because, if $i\ne j$, then one, and only one, of the indices $(i,j)$ must be $x$. Hence, its integral also vanishes if $i\ne j$.  Identical reasoning implies that the third integral  also vanishes for $i \ne j$. This makes the first and third integrals diagonal.

The second integral is non-zero only when $i=x$, $ j=y$, or  $i=y$, $ j=x$. This is explicit for the first term, since that term is proportional to $\delta_{iy}\delta_{jx}+
\delta_{ix}\delta_{jy}$. To see that it is also true for the second term (i.e., the $z^2q_xq_y$ term), note that term is odd in at least one of $q_x$ or $q_y$
unless $i=x$, $ j=y$, or  $i=y$, $ j=x$, and will therefore integrate to zero otherwise. 

In light of these observations, we can rewrite our expression (\ref{vbulk2}) for the velocity correlations as
\beq
\langle v_i(\brp,z)v_j(\brp, z)\rangle = v_0^2 \varPi^{(1)}_{ij}(z)+v_0v_a'\varPi^{(2)}_{ij}(z)+v_a'^2\varPi^{(3)}_{ij}(z)  \,,
\eeq
where we have defined
\bew
\beq
\varPi^{(1)}_{ij}(z)\equiv \int  {d^2q\over(2\pi)^2} C_{pp}(\bq)e^{-2qz}\bigg[\delta_{iy}\delta_{jy}\left(1-{2q_y^2z\over q}\right)+q_y^2z^2\left({q_iq_j\over q^2}+\delta_{iz}\delta_{jz}\right)\bigg] 
\label{Pi1def}
\eeq
\beq
\varPi^{(2)}_{ij}(z)\equiv\int  {d^2q\over(2\pi)^2} C_{p\rho}(\bq)e^{-2qz}\bigg[1 - zq + {2z^2q_x^2q_y^2\over q^2}\bigg]\bigg(\delta_{iy}\delta_{jx}+
\delta_{ix}\delta_{jy}\bigg)
\label{Pi2def}
\eeq
\beq\varPi^{(3)}_{ij}(z)\equiv \int  {d^2q\over(2\pi)^2} C_{\rho\rho}(\bq)e^{-2qz}\bigg[\delta_{ix}\delta_{jx}\left(1-{2q_x^2z\over q}\right)+q_x^2z^2\left({q_iq_j\over q^2}+\delta_{iz}\delta_{jz}\right)\bigg].
\label{Pi3def}
\eeq
\ew
Using the fact that all three of the correlation functions $C_{\rho\rho}(\bq)$, $C_{p\rho}(\bq)$, and $C_{pp}(\bq)$ are proportional to ${1\over q}$ times functions that depend only on the direction $\theta_\bq$ of $\bq$, and making the  simple change of variables $\bq={\bQ\over z}$, shows that all three of these (tensor)  integrals $\varPi^{(1,2,3)}(z)$
are proportional to ${1\over z}$. 

Thus, we conclude that 
\beq
\langle v_i(\brp,z)v_j(\brp, z)\rangle = \left({1\over z}\right)\bM  \,,
\label{vrs scale}
\eeq
where we have defined the constant, parameter dependent matrix
\beq
\bM\equiv v_0^2 \bM^{(1)}_{ij}+v_0v_a'\bM^{(2)}_{ij}+v_a'^2\bM^{(3)}_{ij}  \,,
\eeq
with
\bew
\beq
\bM^{(1)}_{ij}(z)\equiv \int  {d^2Q\over(2\pi)^2} {f_{pp}(\theta_\bQ)\over Q}e^{-2Q}\bigg[\delta_{iy}\delta_{jy}\left(1-{2Q_y^2\over Q}\right)+Q_y^2\left({Q_iQ_j\over Q^2}+\delta_{iz}\delta_{jz}\right)\bigg], 
\label{Mi1def}
\eeq
\beq
\bM^{(2)}_{ij}(z)\equiv\int  {d^2Q\over(2\pi)^2} {f_{p\rho}(\theta_\bQ)\over Q}e^{-2Q}\bigg[1 - Q + {2Q_x^2Q_y^2\over Q^2}\bigg]\bigg(\delta_{iy}\delta_{jx}+
\delta_{ix}\delta_{jy}\bigg),
\label{M2def}
\eeq
\beq
\bM^{(3)}_{ij}(z)\equiv \int  {d^2Q\over(2\pi)^2} {f_{\rho\rho}(\theta_\bQ)\over Q}e^{-2Q}\bigg[\delta_{ix}\delta_{jx}\left(1-{2Q_x^2\over Q}\right)+Q_x^2\left({Q_iQ_j\over Q^2}+\delta_{iz}\delta_{jz}\right)\bigg] . 
\label{M3def}
\eeq
\ew
In light of the above discussion, $\bM$ is a symmetric matrix, whose only non-zero off-diagonal components are $M_{xy}=M_{yx}$. Its diagonal entries are all, in general, different from each other, and from $M_{xy}$.

The important point about these velocity correlations is that they scale like ${1\over z}$; that is, inversely proportional to the distance $z$ from the solid surface. This ${1\over z}$ scaling will break down once $z$ becomes microscopic, as can be seen as follows: 
our arguments above depended on our hydrodynamic theory, which breaks down for wavevectors $q$ comparable to an inverse microscopic length. Since the integrals over wavevector that we have done to derive (\ref{vrs scale}) were dominated by $q\sim{1\over z}$, the calculation clearly ceases to be valid once $z$ is a microscopic length, because then we'll need the correlation functions at wavevectors comparable to an inverse microscopic length, at which our hydrodynamic theory does not apply.

}

We now calculate the space and time-dependent velocity correlators 
\beq
\label{vcortimedef}
C^v_{ij}(\brp-\brp', z, z', t-t')\equiv\langle v_i(\brp, z,t) v_j(\brp', z',t')\rangle
\eeq 

This can also be written in terms of the  in terms of the polarization $\hp$ and density $\rho$ correlation functions using the relation (\ref{vbulk})  between the bulk fluid velocity field and the surface velocity, and our boundary condition (\ref{activebc}) for that surface velocity.
We find

\begin{widetext}

\begin{eqnarray}
 &&C^v_{ij}(\brp-\brp', z, z', t-t')=\int\frac{d\omega}{2\pi}\frac{d^2q}{(2\pi)^2} \exp [i\omega (t-t')-i{\bf q}\cdot ({\brp-\brp'})]\langle v_i({\bf q},z,\omega)v_j(-{\bf q},z',-\omega) \rangle\nonumber \\
 &&=v_0^2\int\frac{d\omega}{2\pi}\frac{d^2q}{(2\pi)^2} \exp [i\omega (t-t')-i{\bf q}\cdot ({\brp-\brp'})
-q(z+z')] C_{pp}({\bf q},\omega)\bigg[\delta_{iy}\delta_{jy}\left(1-{q_y^2( z+z') \over q}\right)+q_y^2z z'
\left({q_iq_j\over q^2}+\delta_{iz}\delta_{jz'}\right)\bigg] \nonumber \\
&&+ v_0 v_a'\int\frac{d\omega}{2\pi}\frac{d^2q}{(2\pi)^2} \exp [i\omega (t-t')-i{\bf q}\cdot ({\brp-\brp'})
-q(z+z')] C_{p\rho}({\bf q},\omega)\bigg[(\delta_{ix}\delta_{jy}+ \delta_{iy}\delta_{jx})\left(1-{q( z+z')
\over 2} + {2z z'q_x^2q_y^2 \over q^2} \right)\bigg] \nonumber \\
&&+v_a'^2\int\frac{d\omega}{2\pi}\frac{d^2q}{(2\pi)^2} \exp [i\omega (t-t')-i{\bf q}\cdot ({\brp-\brp'})
-q(z+z')] C_{\rho\rho}({\bf q},\omega)\bigg[\delta_{ix}\delta_{jx}\left(1-{q_x^2( z+z') \over q}\right)
+q_x^2z z'\left({q_iq_j\over q^2}+\delta_{iz}\delta_{jz'}\right)\bigg]
\label{vcorrreal}
 \nonumber\\
\end{eqnarray}
\ew
where $C_{pp} ({\bf q},\omega)$ , $C_{p\rho}({\bf q},\omega)$, and $C_{\rho\rho}({\bf q},\omega)$ are given by equations (\ref{pycorscale}) and (\ref{pyscalefn}), (\ref{rhocorscale}), (\ref{rhoscalefn}),   (\ref{crossprhocorscale}), and(\ref{crossprhoscalefn}),     of the introduction.

One of the best experimental probes of this correlation function is the diffusion of tracer particles. Consider first neutrally buoyant tracer particles, which, in the absence of diffusion, sit at a constant $z$, and are advected along $\hx$ at a speed $v_0$ by the mean motion of the fluid. If we wish to study the diffusion of these particles on a time scale small compared to the time required for them to diffuse a distance comparable to $z$, we therefore only need the velocity correlations (\ref{vcorrreal}) at  $\br-\br'=v_0(t-t')\hx$, $z=z'$. In this limit, (\ref{vcorrreal}) reduces to

\bew
\begin{eqnarray}
 C^v_{ij}(v_0(t-t')\hx, z, z, t-t') = C^{pp}_{ij}(v_0(t-t')\hx, z, z, t-t') + 
 C^{p\rho}_{ij}(v_0(t-t')\hx, z, z, t-t') + C^{\rho\rho}_{ij}(v_0(t-t')\hx, z, z, t-t') \,,
 \nonumber\\
 \label{cvij}
 \end{eqnarray}
 \ew
where
 \bew
\begin{eqnarray}
 && C^{pp}_{ij}(v_0(t-t')\hx, z, z, t-t') = \nonumber \\ 
 && v_0^2\int\frac{d\omega}{2\pi}\frac{d^2q}{(2\pi)^2} \exp [i(\omega-v_0q_x)(t-t')
-2qz] C_{pp}({\bf q},\omega)\bigg[\delta_{iy}\delta_{jy}\left(1-{2q_y^2z\over q}\right)+q_y^2z^2\left({q_iq_j\over q^2}+\delta_{iz}\delta_{jz}\right)\bigg] \nonumber \\
&&C^{p\rho}_{ij}(v_0(t-t')\hx, z, z, t-t') = \nonumber \\
&& v_a'v_0\int\frac{d\omega}{2\pi}\frac{d^2q}{(2\pi)^2} \exp [i(\omega-v_0q_x)(t-t')
-2qz] C_{p\rho}({\bf q},\omega)\bigg[(\delta_{ix}\delta_{jy}+ \delta_{iy}\delta_{jx})\left(1-qz + {2z^2 q_x^2q_y^2 \over q^2} \right)\bigg] \nonumber \\
&& C^{\rho\rho}_{ij}(v_0(t-t')\hx, z, z, t-t') = \nonumber \\
&& v_a'^2\int\frac{d\omega}{2\pi}\frac{d^2q}{(2\pi)^2} \exp [i(\omega-v_0q_x)(t-t')
-2qz] C_{\rho\rho}({\bf q},\omega)\bigg[\delta_{ix}\delta_{jx}\left(1-{2q_x^2z\over q}\right)+q_x^2z^2\left({q_iq_j\over q^2}+\delta_{iz}\delta_{jz}\right)\bigg]
\label{vcorrneutral}
 \nonumber\\
\end{eqnarray}
\ew

It is straightforward to see from this expression that the tensor $C^{pp}_{ij}(v_0(t-t')\hx, z, z, t-t')$ is purely diagonal. Begin by noting that the only term in the {\it integrand} in (\ref{vcorrneutral}) that has an off-diagonal component is the $q_iq_j$ term. Since $\bq$ is a vector in the $xy$ plane, this term vanishes if either index $i$ or $j$ is $z$. Hence, the only off-diagonal terms are $i=x$, $j=y$, or vis-versa. In either case, the integrand then becomes odd in $q_y$ (recall that  both $C_{pp}(\bq, \omega)$ and $q=|\bq|$ are even in $\bq$. These are  the only $q_y$ dependent pieces of the rest of the integrand (it appears in the $2qz$ term in the argument of the first exponential)).  

Hence, the integral in (\ref{vcorrneutral}), and, therefore, the correlation function $C^{pp}_{ij}(v_0(t-t')\hx, z, z, t-t')$, itself vanishes if $i=x$, $j=y$, or vice-versa. Since we have already established that the off-diagonal components of  $C^{pp}_{ij}(v_0(t-t')\hx, z, z, t-t')$ with one of the indices equal to $z$ also vanish, this completes the proof that $C^{pp}_{ij}(v_0(t-t')\hx, z, z, t-t')$ is, as claimed earlier,  purely diagonal. Similarly we can show from (\ref{vcorrneutral}) that $C^{\rho\rho}_{ij}(v_0(t-t')\hx, z, z, t-t')$ is also a 
diagonal tensor. 

However $C^{p\rho}_{ij}(v_0(t-t')\hx, z, z, t-t')$ is an off-diagonal symmetric tensor, as is easily seen from 
(\ref{vcorrneutral}). Only the components $i=x$, $j=y$, and $i=y$, $j=x$ are non-zero, and 
equal.

This shows that $C^v_{ij}(v_0(t-t')\hx, z, z, t-t')$ as a whole is symmetric tensor with four independent 
components. The long-time scaling behavior of the four non-zero independent components $C^v_{xx}(v_0(t-t')\hx, z, z, t-t')$, $C^v_{yy}(v_0(t-t')\hx, z, z, t-t')$ and 
$C^v_{xy}(v_0(t-t')\hx, z, z, t-t')$  are essentially same, as shown below .  However, $C^v_{zz}(v_0(t-t')\hx, z, z, t-t')$ has a different behavior in the long time limit. This is significant: diffusion along any direction in the $xy$-plane is anomalous, whereas it is normal (non-anomalous) along the vertical or $z$-direction.


To see this, let us consider each of these  four non-zero components in turn, starting with $C^v_{xx}(v_0(t-t')\hx, z, z, t-t')$. Using (\ref{cvij}), we see that this can be expressed as

\bew
\bea
C^v_{xx}(v_0(t-t')\hx, z, z, t-t') = C^{pp}_{xx}(v_0(t-t')\hx, z, z, t-t')+ C^{\rho\rho}_{xx}(v_0(t-t')\hx, z, z, t-t') 
\label{cvxx}
\eea
\ew
as $C^{p\rho}_{xx}(v_0(t-t')\hx, z, z, t-t')$ is zero, being a non-diagonal tensor itself.

We evaluate each of the two terms in (\ref{cvxx}). From (\ref{vcorrneutral}), we see that 
$C^{pp}_{xx}(v_0(t-t')\hx, z, z, t-t')$ is given by
\bew
\beq
 C^{pp}_{xx}(v_0(t-t')\hx, z, z, t-t') 
 =v_0^2\int\frac{d\omega}{2\pi}\frac{d^2q}{(2\pi)^2} \exp [i(\omega-v_0q_x)(t-t')
-2qz] C_{pp}({\bf q},\omega){q_x^2q_y^2z^2\over q^2}\,.
\label{vcorrxx}
\eeq
\ew

Making the change of variables of integration from $(\bq,\omega)$ to dimensionless variables ${\bf Q}, \Omega$ given by
\beq
\omega\equiv{v_0\Omega\over z}
\sep
\bq\equiv{{\bf Q}\over z} \,,
\label{varchange}
\eeq
and recalling that $C_{pp}({\bf q},\omega)$ obeys the scaling form (\ref{pycorscale}), we find that 
\bew
\beq
C^{ pp}_{xx}(v_0(t-t')\hx, z, z, t-t') 
=\left({v_0^3\over z}\right)\int\frac{d\Omega}{2\pi}\frac{d^2Q}{(2\pi)^2} \exp [i(\Omega-Q_x) 
v_0(t-t')/z-2Q] F_{pp}({v_0\Omega\over Q}, \theta_{\bf Q}){Q_x^2Q_y^2\over Q^4}\,,
\label{vcorrxx2}
\eeq
\ew
where $\theta_{\bf Q}$ is the angle between the rescaled, dimensionless vector ${\bf Q}$ and the direction of mean polarization $\hx$ (which is, of course, just the same as the angle $\theta_{\bf q}$between the original vector ${\bf q}$ and the direction of mean polarization $\hx$, since our rescaling (\ref{varchange}) was isotropic.

Using our expression (\ref{pyscalefn}) for $F_{pp}$, we have

\bew
\beqn
&&F_{pp}\bigg({v_0\Omega\over Q}, \theta_{\bf Q}\bigg)={2D_p(v_0{\Omega\over Q}-v_\rho \cos\theta_{\bf Q})^2\over(v_0{\Omega\over Q}-c_+(\theta_{\bf Q}))^2(v_0{\Omega\over Q}-c_-(\theta_{\bf Q}))^2+(v_0{\Omega\over Q}
\psi(1,\theta_{\bf Q})-v_\rho\psi(\varphi,\theta_{\bf Q}) \cos\theta_{\bf Q})^2} \,.
\label{fppscale}
\eeqn
\ew

Factoring out $D_p v_0^2$ from the numerator of this expression, and $v_0^4$ from the denominator, we see that this can be rewritten in terms of a completely dimensionless scaling function of dimensionless arguments as
\beq
F_{pp}\bigg({v_0\Omega\over Q}, \theta_{\bf Q}\bigg)=\left({D_p\over v_0^2}\right) H_{pp}\bigg({\Omega\over Q}, \theta_{\bf Q}; \bigg\{{v_\rho\over v_0},{v_p\over v_0},{\gamma\over v_0},{c_0\over v_0}\bigg\}\bigg) \,,
\label{Hdef}
\eeq
where we have defined

\bew
\beqn
&&H_{pp}\bigg({\Omega\over Q}, \theta_{\bf Q}; \bigg\{{v_\rho\over v_0},{v_p\over v_0},{\gamma\over v_0},{c_0\over v_0}\bigg\}\bigg)={2\bigg({\Omega\over Q}-{v_\rho\over v_0} \cos\theta_{\bf Q}\bigg)^2\over\bigg({\Omega\over Q}-{c_+(\theta_{\bf Q})\over v_0}\bigg)^2\bigg( {\Omega\over Q}-{c_-(\theta_{\bf Q})\over v_0}\bigg)^2+\bigg({\Omega\over Q} {\psi(1,\theta_{\bf Q}) \over v_0} -{v_\rho \over v_0}{\psi(\varphi,\theta_{\bf Q}) \over 
v_0} \cos\theta_{\bf Q}\bigg)^2} \,. \nonumber\\
\label{Hppscale}
\eeqn
\ew

Some of the dependence of this function on the dimensionless ratios ${v_\rho\over v_0}$, ${v_p\over v_0}$, ${\gamma\over v_0}$, and ${c_0\over v_0}$ is hidden in the ratios ${c_\pm(\theta_{\bf Q})\over v_0}$, since $c_\pm(\theta_{\bf Q})$ depend on $v_\rho$, $v_p$, $\gamma$, and $c_0$, as displayed in equation (\ref{cplusminus}).

Using (\ref{Hdef}) and (\ref{Hppscale}) in (\ref{vcorrxx2}), we find that 
$C^{pp}_{xx}(v_0(t-t')\hx, z, z, t-t')$
 itself obeys a simple scaling law:
\beq
C^{ pp}_{xx}(v_0(t-t')\hx, z, z, t-t')=\left({D_pv_0\over z}\right)F^{ pp}_{xx}\left({v_0|t-t'|\over z}\right)\, ,
\label{Cppxx}
\eeq
where the dimensionless scaling function 
\bew
\beq
F^{ pp}_{xx}(u_r) 
=\int\frac{d\Omega}{2\pi}\frac{d^2Q}{(2\pi)^2} \exp [i(\Omega-Q_x)u-2Q]H_{pp}\bigg({\Omega\over Q}, \theta_{\bf Q}; \bigg\{{v_\rho\over v_0},{v_p\over v_0},{\gamma\over v_0},{c_0\over v_0}\bigg\}\bigg){Q_x^2Q_y^2\over Q^4}\,,
\label{xscalefn}
\eeq
\ew
 where the dimensionless scaling variable $u_r\equiv v_0|t-t'|/ z$.

The limiting behaviors of $F^{ pp}_{xx}(u_r)$ for small and large $u_r$ are easy to obtain;, and useful: $u_r\ll 1$ and $u_r\gg 1$ give information about the velocity  correlations, and from those, the displacement correlations, in the ballistic and diffusive limits, respectively, as we'll soon show. Intermediate values of $u_r\sim {\cal O}(1)$  correspond to the crossover between the two regimes. For  $u\ll1$, which corresponds to time differences obeying $v_0|t-t'|\ll z$,  the $e^{-2Q}$ factor in the integral (\ref{xscalefn}) kills the integrand before (that is, at smaller $Q$ than) the $(\Omega-Q_x)u$ term becomes important. Hence, the integral, and $F^{ pp}_{xx}(u_r)$ itself, become independent of $u$ in this limit; that is, 
\beq
F^{ pp}_{xx}(u_r)=A_x\,\,\,\,\, {\rm for} \,\,\,\, u_r\ll1\,,
\label{xx small u}
\eeq
where the constant $A^{ pp}_{ xx}$ is given by
\bew
\beq
A^{pp}_{xx} \nonumber\\
=\int\frac{d\Omega}{2\pi}\frac{d^2Q}{(2\pi)^2} e^{-2Q}H_{pp}\bigg({\Omega\over Q}, \theta_{\bf Q}; \bigg\{{v_\rho\over v_0},{v_p\over v_0},{\gamma\over v_0},{c_0\over v_0}\bigg\}\bigg){Q_x^2Q_y^2\over Q^4}\,.
\label{Appxx}
\eeq
\ew
and is a function of all of the ratios ${v_\rho\over v_0}$, ${v_p\over v_0}$, ${\gamma\over v_0}$, and ${c_0\over v_0}$.  $A^{pp}_{xx}$  will be of ${\cal O}(1)$ when all of these ratios are of ${\cal O}(1)$.

In the opposite limit $u_r\gg1$, which corresponds to time differences obeying $v_0|t-t'|\gg z$,  the $e^{i(\Omega-Q_x)u}$ factor in the integral (\ref{xscalefn}) kills (by oscillation) all contributions to the integral coming from $Q\gtrsim{1\over u_r}\ll1$.  Hence, in the dominant region of the integral, $Q\ll1$, and so we can drop the $2Q$ term in the argument of the exponential. Doing so gives
\bew
\beq
F^{ pp}_{xx}(u_r) \nonumber\\
=\int\frac{d\Omega}{2\pi}\frac{d^2Q}{(2\pi)^2} \exp [i(\Omega-Q_x)u]H_{pp}\bigg({\Omega\over Q}, \theta_{\bf Q}; \bigg\{{v_\rho\over v_0},{v_p\over v_0},{\gamma\over v_0},{c_0\over v_0}\bigg\}\bigg){Q_x^2Q_y^2\over Q^4}\,.
\label{xscalefn1}
\eeq
\ew
This integral can easily be done with one further change of variables:
\beq
\Omega\equiv{\Omega'\over u_r}
\sep
{\bf Q}\equiv{{\bf Q}'\over u_r} \,,
\label{varchange Q}
\eeq
which gives
\beq
F^{ pp}_{xx}(u_r)={B^{pp}_{xx}\over u_r^3} \,\,\,\,\, {\rm for} \,\,\,\, u_r\gg1\,,
\label{xx big u}
\eeq
where the constant  $B^{pp}_{xx}$  is given by
\bew
\beq
B^{pp}_{xx} \nonumber\\
=\int\frac{d\Omega'}{2\pi}\frac{d^2Q'}{(2\pi)^2} \exp [i(\Omega'-Q'_x)]H_{pp}\bigg({\Omega'\over Q'}, \theta_{{\bf Q}'}; \bigg\{{v_\rho\over v_0},{v_p\over v_0},{\gamma\over v_0},{c_0\over v_0}\bigg\}\bigg){Q_x^{'2}Q_y^{'2}\over Q^{'4}}\,,
\label{Bppxx}
\eeq
\ew
and,
like $A^{pp}_{xx}$,
is
again a function of all of the ratios ${v_\rho\over v_0}$, ${v_p\over v_0}$, ${\gamma\over v_0}$, and ${c_0\over v_0}$, and will again be of $O(1)$ when all of these ratios are of $O(1)$.

In summary,  the behavior of the scaling function $F^{ pp}_{xx}(u_r)$ is given by:
\beqn
F^{pp}_{xx}(u_r)
=\left\{
\begin{array}{ll}
A^{pp}_{xx}\sep&u_r\ll1\\\\
{B^{pp}_{xx}\over u_r^3}\sep&u_r\gg1
\end{array}
\right.
\sep
\label{fppxx}
\eeqn

We now evaluate $C^{\rho\rho}_{xx}(v_0(t-t')\hx, z, z, t-t')$, which from (\ref{vcorrneutral}) is expressed as
\bew 
\beq
C^{\rho\rho}_{xx}(v_0(t-t')\hx, z, z, t-t') = v_a'^2\int\frac{d\omega}{2\pi}\frac{d^2q}{(2\pi)^2} \exp [i(\omega-v_0q_x)(t-t')
-2qz] C_{\rho\rho}({\bf q},\omega)\bigg[1-{2q_x^2z\over q}+ {z^2q_x^4\over q^2}\bigg]
\eeq
\ew

Making  the same change of variables (\ref{varchange})  as before, we obtain an scaling form for $C^{\rho\rho}_{ij}(v_0(t-t')\hx, z, z, t-t')$, given by
\beq
C^{\rho\rho}_{xx}(v_0(t-t')\hx, z, z, t-t') = \left({D_p\rho_c^2 v_a'^2\over v_0 z}\right)F^{\rho\rho}_{xx}\left({v_0|t-t'|\over z}\right)
\label{Crho2xx}
\eeq
where $F^{\rho\rho}_{xx}\left({v_0|t-t'|\over z}\right)$ is written as
\bew
\beq
F^{\rho\rho}_{xx}(u_r) \nonumber\\
=\int\frac{d\Omega}{2\pi}\frac{d^2Q}{(2\pi)^2} \exp [i(\Omega-Q_x)u_r-2Q]H_{\rho\rho}\bigg({\Omega\over Q}, \theta_{\bf Q}; \bigg\{{v_\rho\over v_0},{v_p\over v_0},{\gamma\over v_0},{c_0\over v_0}\bigg\}\bigg)
\bigg[{1 \over Q^2} - {2 Q_x^2 \over Q^3} + {Q_x^4 \over Q^4}\bigg]\,.
\label{rho2xxscalefn}
\eeq
\ew
where $H_{\rho\rho}\bigg({\Omega\over Q}, \theta_{\bf Q}; \bigg\{{v_\rho\over v_0},{v_p\over v_0},{\gamma\over v_0},{c_0\over v_0}\bigg\}\bigg)$ is defined as
\bew
\beq
H_{\rho\rho}\bigg({\Omega\over Q}, \theta_{\bf Q}; \bigg\{{v_\rho\over v_0},{v_p\over v_0},{\gamma\over v_0},{c_0\over v_0}\bigg\}\bigg) = {2\bigg({v_\rho\over v_0}\bigg)^2(\sin\theta_{\bf Q})^2 \over \bigg({\Omega\over Q}-{c_+(\theta_{\bf Q})\over v_0}\bigg)^2\bigg( {\Omega\over Q}-{c_-(\theta_{\bf Q})\over v_0}\bigg)^2+\bigg({\Omega\over Q} {\psi(1,\theta_{\bf Q}) \over v_0} -{v_\rho \over v_0}{\psi(\varphi,\theta_{\bf Q}) \over 
v_0} \cos\theta_{\bf Q}\bigg)^2} \,. 
\label{Hrho2scale}
\eeq
\ew

The limiting behaviors of the scaling function $F^{\rho\rho}_{xx}(u)$ can be obtained by an almost identical analysis to that used for $F^{pp}_{xx}(u_r)$, giving:
\beqn
F^{\rho\rho}_{xx}(u_r)
=\left\{
\begin{array}{ll}
A^{\rho\rho}_{xx}\sep&u\ll1\\\\
{(B^{\rho\rho}_{xx})_1\over u_r} + {(B^{\rho\rho}_{xx})_2\over u_r^2} + {(B^{\rho\rho}_{xx})_3\over u_r^3}\sep&u_r\gg1
\end{array}
\right.
\nonumber\\
\label{frho2xx}
\eeqn
with
\bew
\beq
A^{\rho\rho}_{xx} \nonumber\\
=\int\frac{d\Omega}{2\pi}\frac{d^2Q}{(2\pi)^2} e^{-2Q}H_{\rho\rho}\bigg({\Omega\over Q}, \theta_{\bf Q}; \bigg\{{v_\rho\over v_0},{v_p\over v_0},{\gamma\over v_0},{c_0\over v_0}\bigg\}\bigg)\bigg[{1 \over Q^2} - {2 Q_x^2 \over Q^3} + {Q_x^4 \over Q^4}\bigg]\,.
\label{Arho2xx}
\eeq
\ew

and
\newpage

\bew

\bea
(B^{\rho\rho}_{xx})_1 &=& \int\frac{d\Omega'}{2\pi}\frac{d^2Q'}{(2\pi)^2} \exp [i(\Omega'-Q'_x)]H_{\rho\rho}\bigg({\Omega'\over Q'}, \theta_{{\bf Q}'}; \bigg\{{v_\rho\over v_0},{v_p\over v_0},{\gamma\over v_0},{c_0\over v_0}\bigg\}\bigg)
{1 \over Q^{'2}}\,, 
\eea

\bea
(B^{\rho\rho}_{xx})_2 &=& -2\int\frac{d\Omega'}{2\pi}\frac{d^2Q'}{(2\pi)^2} \exp [i(\Omega'-Q'_x)]H_{\rho\rho}\bigg({\Omega'\over Q'}, \theta_{{\bf Q}'}; \bigg\{{v_\rho\over v_0},{v_p\over v_0},{\gamma\over v_0},{c_0\over v_0}\bigg\}\bigg){Q_x^{'2}\over Q^{'3}}\,, 
\eea
\bea
(B^{\rho\rho}_{xx})_3 &=& \int\frac{d\Omega'}{2\pi}\frac{d^2Q'}{(2\pi)^2} \exp [i(\Omega'-Q'_x)]H_{\rho\rho}\bigg({\Omega'\over Q'}, \theta_{{\bf Q}'}; \bigg\{{v_\rho\over v_0},{v_p\over v_0},{\gamma\over v_0},{c_0\over v_0}\bigg\}\bigg)
{Q_x^{'4}\over Q^{'4}}\,, 
\label{Brho2xx}
\eea
\ew


  Like $A^{pp}_{xx}$ (\ref{Appxx}) and $B^{pp}_{xx}$ (\ref{Bppxx}),  $A^{\rho\rho}_{xx}$ and $(B^{\rho\rho}_{xx})_1$, $(B^{\rho\rho}_{xx})_2$, and $(B^{\rho\rho}_{xx})_3$ 
are functions of all of the ratios ${v_\rho\over v_0}$, ${v_p\over v_0}$, ${\gamma\over v_0}$, and ${c_0\over v_0}$, and  will  all be of $O(1)$ when all of these ratios are of $O(1)$.

Taking (\ref{Cppxx}) and (\ref{Crho2xx}) together, we find that $C^{v}_{xx}(v_0(t-t')\hx, z, z, t-t')$ is  given by
\bew
\beq
C^{v}_{xx}(v_0(t-t')\hx, z, z, t-t') = \left({D_pv_0\over z}\right)F^{ pp}_{xx}\left({v_0|t-t'|\over z}\right) + \left({D_p\rho_c^2 v_a'^2\over v_0 z}\right)F^{\rho\rho}_{xx}\left({v_0|t-t'|\over z}\right)
\label{xscale}
\eeq
\ew
with $F^{ pp}_{xx}\left({v_0|t-t'|\over z}\right)$, and $F^{\rho\rho}_{xx}\left({v_0|t-t'|\over z}\right)$ given by (\ref{fppxx}), and (\ref{frho2xx}) respectively.

 Because of the ${1\over u_r}$ piece of the scaling function $F^{\rho\rho}_{xx}\left(u_r\right)$, the integral of this correlation function over $u$ for fixed $z$ diverges logarithmically at large $u_r$.
 We will show  in section (\ref{dif})
    below that this implies logarithmic superdiffusion (equations (\ref{anomdifshort}) and (\ref{anomdiflongx})  in the $x$ direction.

The derivation of similar results for the remain components of the velocity correlator tensor is given in Appendix~\ref{vel-corr-app}.

\section{Anomalous diffusion of tracers}\label{dif}

\subsection{Neutrally buoyant tracers}

A neutrally buoyant tracer is a particle that is simply carried around passively by the flows in the bulk fluid. That is, if the particle is at position $r_p(t)$ at time $t$, its  instantaneous velocity $\dot\br_p(t)$ is
\beq
\dot\br_p(t)=\bv(\br_p(t), t) \,,
\label{v tracer}
\eeq
where $\bv(\br,t)$ is the velocity field of the passive bulk fluid. Therefore, the instantaneous position of a tracer particle $\br_p(t)$ that starts at $\br_p(t=0)$ at some later time $t$ is given by
\begin{eqnarray}
\br_p(t)-\br_p(0) = \int^t_0
\bv(\br_p(t'), t') dt' \,.
\label{r tracer}
\end{eqnarray}

Since diffusion, and even the superdiffusion than we eventually will find, is a much slower process that ballistic transport, we can take, for a neutrally buoyant particle (i.e., one which has no net speed in the $z$-direction), 
\beq
\br_p(t')=(v_0t'+x(t=0))\hx+y(0)\hy+z\hz
\label{r ballsitic tracer}
\eeq
in (\ref{r tracer}), where we have taken into account the fact that the passive fluid is, on average, flowing along the $\hx$ direction at a speed $v_0$. We have also simply written $z(t=0)$ as $z$. Doing so gives
\begin{eqnarray}
\br_p(t)-\br_p(0) = \int^t_0
\bv((v_0t'+x(t=0))\hx+y(0)\hy+z\hz, t') dt' \,.\nonumber\\
\label{r tracer 2}
\end{eqnarray}
Autocorrelating the $x$, $y$, and $z$ components of this equation with themselves gives
\beq
\langle(x(t)-x(0))^2\rangle=\int_0^t dt'\int_0^t dt'' \, C^v_{xx}(t'-t'') \,,
\label{x diff 1}
\eeq
\beq
\langle(y(t)-y(0))^2\rangle=\int_0^t dt'\int_0^t dt'' \, C^v_{yy}(t'-t'') \,,
\label{y diff 0}
\eeq
\beq
\langle(x(t)-x(0))(y(t)-y(0))\rangle=\int_0^t dt'\int_0^t dt'' \, C^v_{xy}(t'-t'')=0 \,,
\label{xy diff 0}
\eeq
\beq
\langle(z(t)-z(0))^2\rangle=\int_0^t dt'\int_0^t dt'' \, C^v_{zz}(t'-t'') \,,
\label{z diff 0}
\eeq
where the correlation functions $C^v_{xx}(t'-t'')$, $C^v_{yy}(t'-t'')$, $C^v_{xy}(t'-t'')$,and $C^v_{zz}(t'-t'')$ are precisely those we calculated in equations (\ref{xscale}), (\ref{yscale}), and (\ref{zscale}) respectively of 
the preceding section.
We will now use these expressions to obtain the diffusive motion - that is, the small departure from the mean ballistic motion at velocity $v_0\hx$ of tracers in the $\hx$ direction - for all three directions.

We begin with the $x$-direction. Using our scaling form (\ref{xscale})    for $C^v_{xx}(t'-t'')$, we can rewrite this as

\bew
\beq
\langle(x(t)-x(0))^2\rangle=\left({D_pv_0\over z}\right)\int_0^t dt'\int_0^t dt'' \, F^{pp}_{xx}\left({v_0|t'-t''|\over z}\right) + \left({D_p\rho_c^2v_a'^2\over v_0z}\right)\int_0^t dt'\int_0^t dt'' \, 
F^{\rho\rho}_{xx}\left({v_0|t'-t''|\over z}\right)  \,.
\label{x diff 2}
\eeq
\ew

Now, changing variables of integration in the integrals above from $t''$ to $\delta t\equiv t''-t'$, we can rewrite this as

\newpage

\bew

\beq
\langle(x(t)-x(0))^2\rangle=\left({D_pv_0\over z}\right)\int_0^t dt'\int_{-t'}^{t-t'} d\delta t \, F^{pp}_{xx}\left({v_0|\delta t |\over z}\right) + \left({D_p\rho_c^2v_a'^2\over v_0z}\right)\int_0^t dt'\int_{-t'}^{t-t'} d\delta t \, F^{\rho\rho}_{xx}\left({v_0|\delta t |\over z}\right) \,.
\label{x diff 3}
\eeq

\ew

Now  recalling, as shown by equations (\ref{xx big u})     and (\ref{frho2xx}),  that $F^{pp}_{xx}(u_r) \sim 1/u_r^3$, and $F^{\rho\rho}_{xx}(u_r)\sim 1/u_r$ for $u_r\gg 1$ (that is, for  
$t\gg\left({z\over v_0}\right)$), we expect the first integral over $\delta t$ on the right hand side of (\ref{x diff 3}) to converge as its limits go to $\infty$, and the second integral to diverge. Thus 
the second integral clearly dominates at long times and (\ref{x diff 3}) can be approximated by

\beq
\langle(x(t)-x(0))^2\rangle\approx\left({D_p\rho_c^2v_a'^2\over v_0z}\right)\int_0^t dt'\int_{-t'}^{t-t'} d\delta t \, F^{\rho\rho}_{xx}\left({v_0|\delta t |\over z}\right) \,.
\label{x diff 4}
\eeq

Changing variables of integration from $\delta t$ to $u\equiv {v_0\delta t\over z}$, we obtain for the integral over $\delta t$:
\beq
\int_{-t'}^{t-t'} d\delta t \, F^{\rho\rho}_{xx}\left({v_0|\delta t |\over z}\right)=\left({z\over v_0}\right) \int_{-{v_0t'\over z}}^{{v_0(t-t')\over z}} du_r \, F^{\rho\rho}_{xx}(u_r)\,.
\label{x diff 5}
\eeq
Using the fact that $F^{\rho\rho}_{xx}(u_r)$ is an even function of $u_r$, we can rewrite this as:
\bew
\beq
\int_{-t'}^{t-t'} d\delta t \, F^{\rho\rho}_{xx}\left({v_0|\delta t |\over z}\right)=\left({z\over v_0}\right) \left\{\int_0^{v_0t'\over z} du_r \, F^{\rho\rho}_{xx}(u_r)+\int_0^{{v_0(t-t')\over z}} du_r \, F^{\rho\rho}_{xx}(u_r)\right\}\,.
\label{x diff 6}
\eeq
\ew
It is convenient to break up the first integral in this expression into a part coming from $u_r<1$ and another part from $u_r>1$:
\beq
\int_0^{v_0t'\over z} du_r \, F^{\rho\rho}_{xx}(u_r)=\int_0^1 du_r \, F^{\rho\rho}_{xx}(u_r)+\int_1^{v_0t'\over z} du_r \, F^{\rho\rho}_{xx}(u_r)
\label{split1a}
\eeq
The integral from $0$ to $1$ in this expression is clearly $O(1)$, since, as we showed earlier, the integrand is. The second integral can be rewritten:
\bea
\int_1^{v_0t'\over z} du_r \, F^{\rho\rho}_{xx}(u_r)
 &=& (B^{\rho\rho}_{xx})_1\int_1^{v_0t'\over z} {du_r\over u_r} \nonumber \\
&+&\int_1^{v_0t'\over z} du_r \, \left(F^{\rho\rho}_{xx}(u_r)-{ (B^{\rho\rho}_{xx})_1 \over u_r}\right) \,.
\nonumber\\
\label{pick log x}
\eea
The second integral  on the right hand side of  this equation (\ref{pick log x})  is also ${\cal O}(1)$, as can be seen from the fact that the integrand is ${\cal O}(1)$, and the expression in parentheses falls off like ${1\over u_r^2}$ for large $u_r$, so the integral converges, even if the upper limit is taken to $\infty$. The first integral in (\ref{pick log x}) is, of course, elementary (about as elementary as they come, in fact!). We thereby obtain:
\beq
\int_0^{v_0t'\over z} du_r \, F^{\rho\rho}_{xx}(u_r)= (B^{\rho\rho}_{xx})_1\ln\left({v_0t'\over z}\right)+{\cal O}(1) \,.
\label{t'int x}
\eeq
An almost identical analysis shows that
\beq
\int_0^{{v_0(t-t')\over z}} du_r \, F^{\rho\rho}_{xx}(u_r)= (B^{\rho\rho}_{xx})_1\ln\left({v_0(t-t')\over z}\right)+{\cal O}(1) \,.
\label{other t'int x}
\eeq
Inserting (\ref{t'int x}) and (\ref{other t'int x}) into  (\ref{x diff 6}) gives
\bew
\beq
\int_{-t'}^{t-t'} d\delta t \, F^{\rho\rho}_{xx}\left({v_0|\delta t |\over z}\right)=\left({z\over v_0}\right) \left\{
 (B^{\rho\rho}_{xx})_1 \left[\ln\left({v_0t'\over z}\right)+\ln\left({v_0(t-t')\over z}\right)\right]+O(1)\right\}\,.
\label{x diff 7}
\eeq
\ew
Inserting this in turn into (\ref{x diff 4}), we obtain:
\bew
\beq
\langle(x(t)-x(0))^2\rangle={D_p\rho_c^2v_a'^2 \over v_0^2}\int_0^t dt'\left\{ (B^{\rho\rho}_{xx})_1 \left[\ln\left({v_0t'\over z}\right)+\ln\left({v_0(t-t')\over z}\right)\right]+O(1)\right\}\,.
\label{x diff 8}
\eeq
\ew
All of the integrals in this expression are elementary, yielding our final superdiffusive expression for the mean squared displacement in the $x$ direction:
\bew
\beq
\langle(x(t)-x(0))^2\rangle= 2(B^{\rho\rho}_{xx})_1 {D_p\rho_c^2v_a'^2 \over v_0^2} t \left(\ln\left({v_0t\over z}\right)+O(1)\right)\,,
\label{x diff final}
\eeq
\ew
which the alert reader will recognize as equation (\ref{anomdifshort}) of the introduction.

We have derived this result assuming that the particle has not moved appreciably in the $z$ direction from its original height $z$. This will  be true only for  $t\ll{z^2\over D_z}$, where $D_z$ is the ($z$-independent) diffusion constant we calculated  below in Eq.~(\ref{z diff}). For much longer times, the particle will typically be a distance 
$z(t)\sim\sqrt{D_z t}$ above the surface. Since the logarithm is quite insensitive to the precise position (i.e., to factors of $O(1)$ in this estimate of $z$, it will suffice, to leading logarithmic order, to replace 
$z$ in (\ref{x diff final}) with $z(t)\sim\sqrt{D_z t}$ in (\ref{x diff final}). Doing so we obtain

\bew
\beq
\langle(x(t)-x(0))^2\rangle= (B^{\rho\rho}_{xx})_1 {D_p\rho_c^2v_a'^2 \over v_0^2} t \left(\ln\left({v_0^2t\over  D_z}\right)+{\cal O}(1)\right)\,,
\label{x diff really big time}
\eeq
\ew
which that same  alert reader will recognize as equation (\ref{anomdiflongx}) of the introduction.




For short times $t\ll\left({z\over v_0}\right)$ (which, it should be noted, can actually get arbitrarily {\it long} as $z\to\infty$), the arguments of  $F^{pp}_{xx}$ and $F^{\rho\rho}_{xx}$ are always much less than $1$ throughout the region of integration over $t'$ and $t''$ in (\ref{x diff 2}). Therefore,  
$F^{pp}_{xx}$ and $F^{\rho\rho}_{xx}$  can be replaced in that integral by their small $u_r$ limits 
$A^{pp}_{xx}$ and $A^{\rho\rho}_{xx}$ respectively. And because they are constants, we obtain {\it ballistic} scaling in this regime:

\bew
\beq
\langle(x(t)-x(0))^2\rangle= \left[A^{pp}_{xx}\left({D_pv_0\over z}\right) + A^{\rho\rho}_{xx}
\left({D_p\rho_c^2v_a'^2\over v_0z}\right) \right]t^2   \sep t\ll\left({z\over v_0}\right)\,.
\label{x ballistic}
\eeq
\ew
The interested reader can easily verify that this ballistic behavior matches smoothly onto the $z$-independent diffusive behavior (\ref{x diff 5}) for $t\sim\left({z\over v_0}\right)$, up to logarithmic factors, as it should.

Virtually identical reasoning applies to  diffusion in the $y$ direction, with the result:
\bew
\beq
\langle(y(t)-y(0))^2\rangle= 2(B^{pp}_{yy})_1 D_p t \left(\ln\left({v_0t\over z}\right)+O(1)\right)\,,
\label{y diff final}
\eeq
\ew
and at very long times, 
\bew
\beq
\langle(y(t)-y(0))^2\rangle= (B^{pp}_{yy})_1 D_p t \left(\ln\left({v_0^2t\over  D_z}\right)+O(1)\right)\,.
\label{y diff really big time}
\eeq
\ew
The above equation is easy to recognize as (\ref{anomdiflongy}) of the introduction.

As for $x$-direction, the motion in the $y$-direction is also ballistic for short times $t\ll\left({z\over v_0}\right)$:
\bew
\beq
\langle(y(t)-y(0))^2\rangle= \left[A^{pp}_{yy}\left({D_pv_0\over z}\right)+ A^{\rho\rho}_{yy}\left({D_p\rho_c^2v_a'^2\over v_0z}\right) \right] t^2 \sep t\ll\left({z\over v_0}\right)\,.
\label{y ballistic}
\eeq
\ew
Once again, the interested reader  can easily verify that this ballistic behavior matches smoothly onto the $z$-independent diffusive behavior (\ref{y diff final}) for $t\sim\left({z\over v_0}\right)$, as it should.

In the $z$-direction, we see that both $F^{pp}_{zz}(u_r) \sim 1/u_r^3$, and $F^{\rho\rho}_{zz}(u_r)\sim 1/u_r^3$ for 
$u\gg 1$ or $t\gg\left({z\over v_0}\right)$. This implies  diffusive behavior at long times; i.e.

\beq
\langle(z(t)-z(0))^2\rangle = 2D_z t \, ,
\label{z diff1}
\eeq
with
\bew
\beq
D_z=D_p\int_{0}^{\infty} du_r \, F^{pp}_{zz}(u) + {D_p\rho_c^2 v_a'^2 \over v_0^2}\int_{0}^{\infty} du_r \, 
F^{\rho\rho}_{zz}(u_r)\,,
\label{z diff}
\eeq
\ew
where $D_z$ converges to a finite value.

As for motion in the $x$ and $y$ direction,  at short times $t\ll\left({z\over v_0}\right)$  we obtain {\it ballistic} scaling for $z$:
\bew
\beq
\langle(z(t)-z(0))^2\rangle=\left[A^{pp}_{zz}\left({D_pv_0\over z}\right)+ A^{\rho\rho}_{zz}\left({D_p\rho_c^2v_a'^2\over v_0z}\right) \right] t^2 \sep t\ll\left({z\over v_0}\right)\,.
\label{z ballistic}
\eeq
\ew
As for $x$,  this ballistic behavior for $z$ matches smoothly onto the $z$-independent diffusive behavior (\ref{z diff1}) for $t\sim\left({z\over v_0}\right)$, as it should.

We can also calculate the off diagonal correlation 
\beq
\langle(x(t)-x(0))(y(t)-y(0))\rangle=\int_0^t dt'\int_0^t dt'' \, C^v_{xy}(t'-t'') \,,
\label{xy diff 1}
\eeq
which turns out to be zero as $C^v_{xy}(t'-t'')$ consists of an integral over q of $C_{p\rho}(\bq)$ times an even function of $q_y$. As we have seen from Eq. () that $C_{p\rho}(\bq)$ is an odd in $q_y$, this integral turns out to be zero. So 
\beq
\langle(x(t)-x(0))(y(t)-y(0))\rangle=0 \,,
\label{xy diff final}
\eeq

The above equation is easy to recognize as (\ref{anomdiflongxy}) of the introduction.

\subsection{Sedimenting tracers}

For sedimenting particles, we first note that, since we just showed that diffusion in the $z$  direction is normal, and homogeneous, we need to consider diffusive motion in the $x$, and $y$ direction as well as the 
$x-y$ plane. We'll start by considering sedimenting particles whose sedimentation speed $v_{\rm sed}<<v_0$. Such particles will spend a time of $O({z_0\over v_{\rm sed}})$ at a height of order $z_0$. Due to the aforementioned insensitivity of the logarithmic factor in (\ref{x diff final}), (\ref{y diff final}), and the 
(\ref{xy diff final}), we can therefore accurately estimate the total mean squared $y$ displacement 
$\langle(x(z=z_0)-x(z=0))^2\rangle$, $\langle(y(z=z_0)-y(z=0))^2\rangle$, $\langle(x(z=z_0)-x(z=0))
(y(z=z_0)-y(z=0))\rangle$ of a sedimenting particle that starts at $z=z_0$ and sinks at speed $v_{\rm sed}$ all the way down to the surface by simply replacing $t$ in (\ref{x diff final}), (\ref{y diff final}), and (\ref{xy diff final}) equations by ${z_0\over v_{\rm sed}}$, which is the time it takes the sedimenting tracer to sink to the bottom. Doing so gives
\bew
\bea
\langle(x(z=z_0)-x(z=0))^2\rangle &=& 2(B^{\rho\rho}_{xx})_1 {D_p\rho_c^2v_a'^2 \over v_0^2} \left({z_0\over v_{\rm sed}}\right) \left(\ln\left({v_0\over v_{\rm sed}}\right)+O(1)\right)\,,
\label{x diff sed} \\
\langle(y(z=z_0)-y(z=0))^2\rangle &=& 2 B^{pp}_{yy} D_p \left({z_0\over v_{\rm sed}}\right) \left(\ln\left({v_0\over v_{\rm sed}}\right)+O(1)\right)\,,
\label{y diff sed} \\
\langle(x(z=z_0)-x(z=0))(y(z=z_0)-y(z=0))\rangle &=&  0
\label{xy diff sed}
\eea
\ew
which are just the equations (\ref{anomsedx}-\ref{anomsedxy}) of the introduction.

For more rapidly sedimenting particles (i.e., denser ones), for which $v_{\rm sed}\gg v_0$, the time scale $t\sim{z_0\over v_{\rm sed}}$ of the sinking of the sedimenting particle is in the ballistic regime $t\ll\left({z\over v_0}\right)$, and so we need to use 
(\ref{x ballistic}) and(\ref{y ballistic}) with $t$ replaced with the sedimenting time ${z_0\over v_{\rm sed}}$. This gives 
\bew
\bea
\langle(x(z=z_0)-x(z=0))^2\rangle &=& A^{pp}_{xx}\left({D_pv_0 z_0\over v_{\rm sed}^2}\right) + 
A^{\rho\rho}_{xx}\left({D_p\rho_c^2v_a'^2z_0\over v_0v_{\rm sed}^2}\right)  \,,
\label{x ball sed} \\
\langle(y(z=z_0)-y(z=0))^2\rangle &=& A^{pp}_{yy}\left({D_pv_0 z_0\over v_{\rm sed}^2}\right) + 
A^{\rho\rho}_{yy}\left({D_p\rho_c^2v_a'^2z_0\over v_0v_{\rm sed}^2}\right)\,,
\label{y ball sed} \\
\eea
\ew
These are equations (\ref{anomsedx}) and (\ref{anomsedy}) of the introduction.

We close this Section by noting that, for both neutrally buoyant particles and sedimenting tracers, the aspect ratio $\sqrt{(y(t)-y(0))^2/(x(t)-x(0))^2}$ is not exactly 1, but is a model parameter dependent ${\cal O}(1)$ number, reflecting the geometric anisotropy due to the $x$-direction being the preferred direction of orientation for the polar particles. Nonetheless, the aspect ratio is independent of time $t$. This means an anisotropy exponent $\zeta_{ani}$ that describes the relative scaling between distances measured along the $x$- and $y$-directions is unity (i.e., $\zeta_{ani}=1$),  unlike in the Toner-Tu model of flocking, for which 
$\zeta_{ani}\ne1$~\cite{tonertu95,toner98} due to strongly relevant anharmonic effects. 

The anomalous diffusion we find here  has some features common with the phenomenon of 
``Taylor diffusion''\cite{taylor}, which occurs in parallel plate shear flow. One could obtain the 
Taylor diffusion geometry from ours by replacing the active fluid layer at the bottom \ by a flat 
surface, moving at constant velocity, and our free top surface   with a 
stationary rigid wall with no-slip boundary conditions. However, our anomalous diffusion  is 
fundamentally different  from Taylor diffusion. One principal difference is that, in our case, 
anomalous diffusion arises purely due to the fluctuations away from uniform alignment
and motion in the active layer. There is obviously no analog of this if one replaces the active layer in our problem with a uniformly
moving rigid body, as in Taylor diffusion.

\section{Renormalization group argument and irrelevance of non-linearities}\label{RG}

So far, we have worked strictly with the linear theory. For dry active matter, it is well-known\cite{tonertu95, toner98, toner05}  that non-linear effects radically change the long-wavelength behavior (indeed, it is {\it only} the effects of non-linearities that even make the ordered state in two dimensions possible). It therefore clearly behooves us to ask whether non-linearities have such important effects in our problem. 

In this section, we will use a simple renormalization group power counting argument to show that they do not. In fact, the linear theory presented earlier is asymptotically exact at long distances. 

There are several sources of nonlinearities. For instance, treating the various speeds and  pressures in the problem as functions of $\rho$, and then expanding in powers of $\delta\rho$ produces these nonlinear terms. In addition there are nonlinear terms which originate from the fixed length constant on the polarization $\bf p$.

In order to ascertain the relevance or irrelevance of the nonlinear effects, we need to consider the lowest order nonlinear terms in the equations for $p_y$ and $\rho$,  and in the ``active partial slip'' boundary condition (\ref{activebc}). The RG analysis we are about to present will make it clear that the most important terms at long distances are those with the fewest possible spatial derivatives, and the smallest number of fields.

{ There are two types of additional terms. The first type arises very straightforwardly from expanding, e.g, the pressures $P_{p,s}(\rho)$ and the velocity $v_a(\rho)$, to higher order in $\rho$. The second come from including the $p_kp_i$ piece of the transverse projection operator $T_{ki}=\delta_{ki}-p_kp_i$ in the full equation of motion \eqref{pi}. The latter gives rise to the following extra  quadratic  order in 
the fields $p_y$ and $\delta\rho$ 
contribution to $\pp_tp_y$:
\beq
\pp_tp_y(\brp,t)_{\rm quad}=-\alpha p_yv_{sx}(\brp,t) \,.
\label{quadproj}
\eeq
Since this term already has one power of $p_y$ multiplying $v_{sx}$, it is  obviously sufficient, to quadratic  order in the fields, to use our linear solution  \eqref{vsxFTclosed} for $v_{sx}$ in \eqref{quadproj}. Since it is most convenient for our RG analysis to work in real space, we rewrite \eqref{quadproj} in real space, where it reads
\bew
\beq
\delta v_{sx}(\brp,t)=(v'_a+(\bar\zeta-\sigma) \pp_x)\delta\rho(\brp,t)-\mu\eta v_a'\int d^2 r'_\perp  {\cal K}_{pp\rho}({\brp-\brp'}) \delta\rho (\brp',t)+\zeta_{20}\pp_yp_y(\brp,t)-\mu\eta v_0\int d^2 r'_\perp  {\cal K}_{p\rho}({\brp-\brp'}) p_y (\brp',t) \,,
\eeq
\ew
where the kernels ${\cal K}_{pp\rho}(\brp)$ and ${\cal K}_{p\rho}(\brp)$ are the inverse  two-dimensional Fourier transforms of 
$\left(\frac{q^2+q_x^2}{q}\right)$ and $\left(\frac{q_xq_y}{q}\right)$ respectively\cite{UV}, and are given, at large distances~\cite{UV},  by
\beq
{\cal K}_{pp\rho}(\brp) \approx -{3 x^2\over 2\pi r_\perp^5}
\label{pprhokernel}
\eeq
 and
\beq
{\cal K}_{p\rho}(\brp) \approx-{3 xy\over 2\pi r_\perp^5}
\label{prhokernel}
\eeq
 as we demonstrate in appendix \ref{appc}.
}
Including such terms, the equations of motion take the form
\bew
\begin{eqnarray}
 \frac{\partial p_y}{\partial t} &=& -v_p \partial_x p_y - \gamma\int d^2 r'_\perp {\cal K}_{pp}({\brp-\brp'}) p_y (\brp')  - \frac{\gamma_\rho}{\rho_c}\int d^2 r'_\perp {\cal K}_{p\rho}({\brp-\brp'}) \delta\rho (\brp')- 
 \sigma_t \partial_y\delta\rho+ f_y \nonumber \\
  &-& { \gamma_{_{NL}}}\int d^2 r'_\perp p_y (\brp) {\cal K}_{p\rho}({\brp-\brp'}) p_y (\brp')-{ g_{pp\rho1}}\int d^2 r'_\perp p_y (\brp) {\cal K}_{pp\rho}({\brp-\brp'}) \delta\rho (\brp')
  \nonumber \\
  &-& { g_{pp\rho2}}\int d^2 r'_\perp p_y (\brp') {\cal K}_{pp}({\brp-\brp'}) \delta\rho (\brp') - { \sigma_{_{NL}}} p_y \partial_x\delta\rho \nonumber\\ &+&  \frac{\lambda_1}{2}\partial_y p_y^2  +
 \lambda_2 \delta\rho\partial_x p_y+{\lambda_3} p_y f_x,
 \label{py-real}
\end{eqnarray}
\ew

 and
 \newpage
 
\bew
\begin{eqnarray}
 \frac{\partial \delta\rho}{\partial t} = -v_\rho\partial_x\delta\rho - v_p\rho_c\partial_y p_y - \lambda_4\partial_y (p_y\delta \rho) + \lambda_5 \partial_x(\delta\rho)^2 + {\boldsymbol\nabla}\cdot {f}_\rho \,.\label{rho-gen}
\end{eqnarray}
\ew

{ In equation \eqref{py-real}, the non-linear terms appear on the last three lines. The ``bare'' values of the parameters in these equations of motion (hereafter denoted by a superscript ``$0''$) are related as follows:
\begin{eqnarray} 
&&\gamma_{_{NL}}^0=\gamma^0\,,
\label{rr0a}\\ 
&&g_{pp\rho1}^0= \gamma^0_\rho/\rho^0_c \,,
\\ 
&&g_{pp\rho2}^0= \alpha  v'_a\mu\eta\,,
\\ 
&&\sigma^0_{_{NL}}= \sigma^0_t \,,
\\ 
&&\lambda_1^0=(\nu^0_1 - \lambda^0_{\rho v})v_0 - \lambda^0 \,,
\\
&&\lambda^0_2=\bigg(\frac{\nu^0_1+1}{2} \bigg) v'^0_a- \lambda^0_{\rho v}(v_a')^0 \,,
\\
&&\lambda^0_3=-1 \,,
\\
&&\lambda^0_4=v_\rho \,,
\\
 &&\lambda^0_5=-\rho'^0_e(v_a')^0 \,.
 \label{bare}
\end{eqnarray}
However, none of these parameters  will  continue to maintain these relations to the other parameters upon renormalization, which is why we have introduced them as independent parameters in the equations of motion \eqref{py-real} and \eqref{rho-gen}.
}

Here, the kernel ${\cal K}_{pp}(\brp)$ is the inverse  two-dimensional Fourier transform of $\left(\frac{q^2+q_y^2}{q}\right)$, 
and is given, at large distances~\cite{UV},  by
\beq
{\cal K}_{pp}(\brp) \approx -{3 y^2\over 2\pi r_\perp^5} \,,
\label{ppkernel}
\eeq
as we demonstrate in appendix \ref{appc}. Notice that we have restored  the { number-conserving}  noise ${\bf f}_\rho$ in (\ref{rho-gen}). That noise  is assumed to be of zero-mean and Gaussian-distributed with  variance
\begin{equation}
 \langle f_{\rho\,i}({\bf x},t) f_{\rho\,j}(({ {\bf x}',t'})\rangle = 2D_\rho { \delta_{ij}}\delta({\bf x}{ -{\bf x}'})\delta(t{ -t'}).
 \label{frhocorr}
\end{equation}

We will now assess the importance of the non-linear terms in 
these equations of motion using the dynamical renormalization group (DRG). Readers interested in a more complete and pedagogical  discussion of the DRG are referred to \cite{FNS} for the details of this general approach.

This approach begins by decomposing the Fourier modes of the fields $p_y({\bf q},\omega)$ and $\rho({\bf q},\omega)$, and the noises $f_p$ and $f_\rho$ into a rapidly varying parts $p_y^>({\bf q},\omega)$,  $\rho^>({\bf q},\omega)$, $f_p^>({\bf q},\omega)$  and  $f_\rho^>({\bf q},\omega)$, and   slowly varying parts $p_y^<({\bf q},\omega)$ and  $\rho^><{\bf q},\omega)$ $f_p^<({\bf q},\omega)$  and  $f_\rho^<({\bf q},\omega)$. The rapidly varying parts are supported in the momentum shell $\Lambda\ee^{-\dd\ell}<q<\Lambda$, where $\dd\ell$ is an infinitesimal
and $\Lambda$ is the ultraviolet cutoff. The slowly varying part is supported in $q<\Lambda\ee^{-\dd\ell}$.

The  DRG procedure then consists of two steps. In step 1, we eliminate the rapidly varying parts $p_y^>({\bf q},\omega)$  and  $\rho^>({\bf q},\omega)$ from the equations of motion.
We do this by solving them iteratively for  $p_y^>({\bf q},\omega)$  and  $\rho^>({\bf q},\omega)$. This solution is a series in the non-linearities 
which depends on the slow fields $p_y^<({\bf q},\omega)$ and  $\rho^<{\bf q},\omega)$. We substitute this solution into the equations of motion for the slow fields, and average over  the short wavelength components   $f^>_y({\bf q},\omega)$ and $\bff^>_\rho({\bf q},\omega)$ of the noises $f_y$ and $\bff_\rho$, which gives a closed EOM  for the slow fields  $p_y^<({\bf q},\omega)$ and  $\rho^<{\bf q},\omega)$.

Step 2 consists of rescaling  space and time as follows: ${\bf r}_{_\perp}'={\bf r}_\perp\ee^{-\dd\ell},\,t'=t\ee^{-z\dd\ell}$, where we will choose the ``dynamical exponent'' $z$ for our convenience. The rescaling of $\br$ has the effect of restoring the ultraviolet cutoff $\Lambda$ to its original value.

We simultaneously rescale the fields $p_y({\bf r}_{_\perp},t)$ and $\rho({\bf r}_{_\perp},t)$ according to  
\begin{equation}
 p_y({\bf r}_{_\perp},t)=e^{\chi_pd\ell}p_y({\bf x'},t'),\,\rho({\bf r}_{_\perp},t)=e^{\chi_\rho d\ell}\rho({\bf x'},t') \,
\end{equation}
where we will also choose the field rescaling exponents $\chi_p$ and $\chi_\rho$ for our convenience.
We then reorganize the resultant EOM so that it has the same form as  our original equations of motion, but with all of the coefficients changed, or, to use the standard jargon, ``renormalized''.
This process is then repeated. The result is a set of differential recursion relations for the various parameters in the equations of motion (\ref{py-real}) and (\ref{rho-gen}), which are:
{
\bew
\begin{eqnarray} 
&&{dD_p\over d\ell}=(z-2-2\chi_p)D_p +{\rm nonlinear \, corrections}\,,
\label{rr0}\\ 
&&{dD_\rho\over d\ell}= (z-4-2\chi_\rho)D_\rho+{\rm nonlinear \, corrections} \,,
\\ 
&&{d\sigma_t\over d\ell}= (z-1+\chi_\rho-\chi_p)\sigma_t+{\rm nonlinear \, corrections} \,,
\\ 
&&{d\gamma\over d\ell}=(z-1)\gamma +{\rm nonlinear  \, corrections} \,,
\\ 
&&{d\over d\ell}\left({\gamma_\rho\over\rho_c}\right)=(z+\chi_\rho-\chi_p-1)\left({\gamma_\rho\over\rho_c}\right)+{\rm nonlinear  \, corrections} \,,
\\
&&{d v_p\over d\ell}=(z-1)v_p+{\rm nonlinear  \, corrections} \,,
 \\
 &&{d v_\rho\over d\ell}=(z-1)v_\rho+{\rm nonlinear  \, corrections} \,,
  \\
&& {d (v_p\rho_c)\over d\ell}=(\chi_p-\chi_\rho+z-1)v_p\rho_c+{\rm nonlinear  \, corrections} \,,
\\ 
&&{ {d\gamma_{_{NL}}\over d\ell}=(z+\chi_p-1)\gamma_{_{NL}}+{\rm nonlinear  \, corrections}} \,,
\\
&&{ {dg_{pp\rho1}\over d\ell}=(z+\chi_\rho-1)g_{pp\rho1}+{\rm nonlinear  \, corrections}} \,,
\\
&&{ {dg_{pp\rho2}\over d\ell}=(z+\chi_\rho-1)g_{pp\rho2}+{\rm nonlinear  \, corrections}} \,,
\\
&&{ {d\sigma_{_{NL}}\over d\ell}=(z+\chi_\rho-1)\sigma_{_{NL}}+{\rm nonlinear  \, corrections}} \,,
\\
&&{d\lambda_1\over d\ell}=(z+\chi_p-1)\lambda_1+{\rm nonlinear  \, corrections} \,,
 \label{rr}
\end{eqnarray}
\ew
\bew
\beqn
&&{d\lambda_2\over d\ell}=(z+\chi_\rho-1)\lambda_2+{\rm nonlinear  \, corrections} \,,
\\
&&{d\lambda_3\over d\ell}=(\chi_p)\lambda_3+{\rm nonlinear  \, corrections} \,,
\\
&&{d\lambda_4\over d\ell}=(z+\chi_p-1)\lambda_4+{\rm nonlinear  \, corrections} \,,
\\
 &&{d\lambda_5\over d\ell}=(z+\chi_\rho-1)\lambda_4+{\rm nonlinear  \, corrections} \,.
 \label{rr2}
\end{eqnarray}
\ew
The terms we have explicitly displayed in the above equations all come trivially from the rescaling of length, time, and fields. The ``non-linear corrections'' denote corrections arising due to the couplings between the ``fast'' and ``slow'' modes that arise in the equations of motion  (\ref{py-real}) and (\ref{rho-gen}) as a result of the non-linear terms in those equations. Since they arise from those non-linear terms, these terms must vanish when those non-linear terms do, and must, by continuity, be small when the coefficients $\lambda_i$ of those non-linear terms are small. 

Of course, how small the $\lambda_i$ have to be to ensure that the non-linear corrections in the above recursion relations are negligible depends on the linear parameters in the equations of motion, since those determine the size of the fluctuations in the fields. This suggests a very simple, and very standard, way  to decide if the non-linear terms are important at long distances and times: simply choose the rescaling exponents $z$ and $\chi_{p,\rho}$ to keep the linear parameters that control the fluctuations fixed. If, {\it with this choice of  $z$ and $\chi_{p,\rho}$}, the $\lambda_i$ all then flow to zero as $\ell\to\infty$, then the non-linear terms are guaranteed to be unimportant at long distances. This is the approach we will now take.

We begin by noting that the linear parameters which control the size of the fluctuations in the linear theory are $D_p$, $v_{p,\rho}$, $\sigma_t$,$\gamma$, ${\gamma_\rho\over\rho_c}$, and $v_p\rho_c$, as we showed in section (\ref{corr})      above, where we calculated the fluctuations in the linear theory.

If we assume that all of the $\lambda_i$ are initially small enough that the non-linear corrections in the recursion relations (\ref{rr0})-(\ref{rr}) are negligible, then we can keep the above parameters fixed by choosing
\beq
\chi_p=\chi_\rho=-{1\over2},\;\;   z=1.
\label{rescaling choice}
\eeq
The choice $z=1$ implies that the scaling}  argument of the scaling functions $F_{pp}, F_{\rho\rho}$ and $F_{p\rho}$ should be $\omega/q$  (since for general $z$, it would be $\omega/q^z$).

Using these choices \eqref{rescaling choice} in the recursion relation for $D_\rho$, and again neglecting the non-linear terms, we see that
\beq
{dD_\rho\over d\ell}=-2D_\rho \,,
\label{Drholin}
\eeq
which clearly shows that $D_\rho(\ell\to\infty)\to0$. Thus, $D_\rho$ is irrelevant at long distances and times, which justifies our neglect of it in our earlier, linear analysis.

We also find that all $\lambda_i$, $i=1\to5$ have same recursion relation:

\beq
{d\lambda_i\over d\ell}=-\left({1\over2}\right)\lambda_i+{\rm nonlinear  \, corrections} \,,
\eeq
and that the other three non-linear coefficients $\gamma_{_{NL}}$, $g_{pp\rho}$, and $\sigma_{_{NL}}$ have the same RG eigenvalues:
\begin{eqnarray} 
&&{ {d\gamma_{_{NL}}\over d\ell}=-\left({1\over2}\right)\gamma_{_{NL}}+{\rm nonlinear  \, corrections}} \,,
\nonumber\\
&&{ {dg_{pp\rho1}\over d\ell}=-\left({1\over2}\right)g_{pp\rho1}+{\rm nonlinear  \, corrections}} \,,
\nonumber\\
&&{ {dg_{pp\rho2}\over d\ell}=-\left({1\over2}\right)g_{pp\rho2}+{\rm nonlinear  \, corrections}} \,,
\nonumber\\
&&{ {d\sigma_{_{NL}}\over d\ell}=-\left({1\over2}\right)\sigma_{_{NL}}+{\rm nonlinear  \, corrections}}  \,.
\nonumber\\
 \label{NLrg}
\end{eqnarray}
Therefore, all of these nonlinearities are also irrelevant at long distances, at least if they are initially small.

Since $\chi_{p, \rho}$ are both $<0$, any terms with more fields are less relevant. Likewise, any fields with more gradients are also less relevant. This includes all possible other nonlinearities. Therefore, all non-linearities are irrelevant. This implies that our linear results are asymptotically exact at long length and time scales.

\section{Summary and conclusions}

In this paper, we have studied the  the stability and fluctuations of a large polar-ordered flock at a solid-liquid interface, which is a natural intermediate case between the two previously studied  cases of  wet and dry polar  active  fluids. Such a flock is affected by {\em both} the friction force from the solid substrate underneath and the long range hydrodynamic interaction mediated by the overlying passive, isotropic bulk fluid.   As a result, such a flock is simultaneously momentum nonconserving, but affected by the hydrodynamic interactions of the bulk surrounding fluid. Friction with the substrate also breaks {\em Galilean invariance}. 

These  features lead to novel  behavior at  long length and time  scales,   radically different from both dry  and  wet polar flocks. First of all, a flock of arbitrarily large size with long range polar  order is {\em stable} for a range of the model parameters. Although this prediction is qualitatively same as that of the original Toner-Tu model for a dry polar-ordered flock~\cite{tonertu95,toner98}, there are significant differences. For instance, the effective damping in the present theory is ${\cal O}(q)$ in contrast to the ${\cal O}(q^2)$ damping in the  linearized Toner-Tu model~\cite{tonertu95,toner98} in the long wavelength limit.   As a result, fluctuations are significantly smaller at a solid-fluid interface than in the Toner-Tu model. { Therefore, } the predictions   of the linear theory  are asymptotically exact in the long wavelength limit. This is contrast to the larger fluctuations in the Toner-Tu model,  in which  the predictions from the linear theory break down at long length and time scales due to the fluctuations.  Furthermore,  although both the solid-fluid interface problem treated here and the Toner-Tu model are anisotropic, we find isotropic {\it scaling} in our problem, while the Toner-Tu model exhibits anisotropic scaling~\cite{tonertu95,toner98}.

 We have shown the existence of giant number fluctuation in our model, with the variance of the number scaling as the 3/4th power of the mean.  In addition
 to this unusual scaling, we also find that the number fluctuations in a given area depend on the {\it shape}, as well as the size, of the area.

In addition, we find that the bulk fluid is ``stirred'' by the active particles on the interface, giving rise to long ranged fluctuations in the bulk fluid velocity. These in turn lead to anomalous diffusion of 
 tagged particles in the bulk fluid. 
  Specifically, we find that 
the displacement variances $\langle(x(t)-x(0))^2\rangle$ and $\langle(y(t)-y(0))^2\rangle$ in the plane become anomalous,  scaling as $t\ln t$ in the large time limit. In contrast,  diffusion in the  $z$-direction remains normal, i.e., $\langle(z(t)-z(0))^2\rangle$  scales linearly with $t$. 
 
 Non-interacting particles sedimenting through the bulk fluid, as a result, will land on the solid substrate in a region whose typical dimensions exhibit an anomalous logarithmic dependence on the sedimenting speed, as summarized in equations (\ref{x diff sed})-(\ref{xy diff sed}).

Finally, we have shown that non-linearities are irrelevant to the long-distance, long-time behavior of these systems, in contrast to dry active matter.

 There are many possible extensions of the work reported here. One could, for example, consider replacing either the bulk fluid or the bulk solid of our problem with a liquid crystal (e.g., nematic or smectic).
 One could  also ask how the presence of multiple species, instead of one as here, might affect the macroscopic properties. 

It would also be interesting to study the order-disorder transition in the present model. Will a linear theory suffice,  as  we have found it does for the ordered phase?

 It would also be interesting to study a  variant of our model, in which the passive fluid layer has finite height $h$. This is precisely the experimental geometry of Ref.~\cite{bricard2013}. In the limit of lateral length scales $L_\perp\gg h$, such a system  reduces to the system of polar-ordered flocks { suspended in a fluid on a substrate  studied in Ref.~\cite{maitra2020}}.
  Lastly, one might consider another variant of our system, in which there is a bulk fluid of finite thickness resting on a solid substrate, and is covered by a fluid membrane at the top surface containing self-propelled 
particles attached to the membrane.  How the order of the polar flock couples with the membrane undulations,  and how this depends on the bulk fluid thickness,  is an interesting, and completely open, question. 

Another interesting extension would be to consider a variant of our system,  in which the bulk fluid above has additional properties, e.g., if it is anisotropic or has long ranged correlations. This could be achieved if the passive bulk fluid is replaced, e.g., by a nematic liquid crystal~\cite{guillamat}.
The special ordering direction of the bulk nematic  will be system specific (i.e., it will depend on the active particles, the nematic material used, and properties of the substrate, as well as (possibly) temperature and pressure. It will not, in general, be parallel to the direction of self-propulsion in the active fluid layer. Indeed, in some cases, it may even  be along the direction orthogonal to the active fluid layer (  the $z$-direction in our geometry). Similarly, one might also extend this study by replacing the isotropic bulk fluid by a bulk smectic A or C liquid crystal. With a bulk smectic too, there could be competition between the in-plane self-propulsion direction in the active fluid layer and the alignment direction of the smectic molecules. How all these different possibilities conspire with the activity to produce a steady state of some type is an unresolved but open question.

Acknowledgements:  One of us (AB) thanks 
the SERB, DST (India) for partial financial support through the MATRICS scheme [file no.: MTR/2020/000406]. NS is partially supported by Netherlands Organization for Scientific Research (NWO), through the Vidi grant No. 2016/N/00075794. We thank S. Ramaswamy for sharing reference \cite{maitra2018} with us.  NS thanks Institut Curie and MPIPKS for their   support through postdoctoral fellowships while some of this work was being done.  AB thanks the MPIPKS, Dresden for their hospitality, and their support through their Visitors' Program, while a portion of this work was underway. JT likewise thanks the MPIPKS for their hospitality, and their support through the Martin Gutzwiller Fellowship, and the Higgs Center of the University of Edinburgh for their support with a Higgs Fellowship.

\begin{appendix}

\section{Relating the 3D bulk fluid flow to the surface velocity field}
\label{appa}

\setcounter{equation}{0}
\setcounter{figure}{0}
\setcounter{table}{0}
\setcounter{page}{1}
\makeatletter
\renewcommand{\theequation}{A\arabic{equation}}
\renewcommand{\thefigure}{A\arabic{figure}}
\renewcommand{\bibnumfmt}[1]{[S#1]}
\renewcommand{\citenumfont}[1]{S#1}

The 3D bulk passive fluid satisfies the Stoke's equation
\bea
\eta\nabla^2_3{\bf v}={\boldsymbol\nabla}_3 \Pi, \label{stokespass}
\eea
where ${\boldsymbol\nabla}_3$ is the 3D gradient operator, and $\Pi$ is the
hydrostatic pressure associated with the bulk fluid in the
$z>0$ region. We impose 3D incompressibility for this fluid, so that ${\boldsymbol\nabla}_3\cdot
{\bf v}=0$. Therefore, taking the divergence of (\ref{stokespass}) implies
\bea
\nabla_3^2\Pi=0\,. \label{incompass}
\eea
Taking the Laplacian of (\ref{stokespass}) and using (\ref{incompass}) then implies
\beq
\nabla_3^4v_\alpha=0\,.
\label{del4v}
\eeq
We now look for plane wave solutions of this equation; that is, solutions of the form
\beq
\bv(\br_\perp,z)={\bf V}(z)\exp(i\bq\cdot\br_{_\perp})\,,
\label{bulkFTdef}
\eeq
where $\bf q$ is a two-dimensional in-plane Fourier wavevector and $\brp\equiv(x,y,0)$ is the projection of $\br$ onto the plane of the surface.

We likewise assume that the velocity $\bv_s(x,y)$ on the surface (which by assumption is parallel to the surface) also takes a plane wave form:
\beq
\bv_s(\br_\perp)={\bf v}_s(\bq)\exp(i\bq\cdot\br_{_\perp})\,,
\label{surfFTdef}
\eeq

The general solution of this equation (\ref{del4v}) of the form (\ref{bulkFTdef}) that vanishes as $z\to\infty$ is
\bea
\bv(\br_\perp,z)=(\bw_1(\bq)+\bw_2(\bq)z)\exp(-qz)\exp(i\bq\cdot\br_{_\perp})\,. 
\label{vpass}
\eea

The boundary condition we impose on the 3D passive fluid at the fluid-solid interface is
$v_z(z=0)=0$ and $\bv_\perp(x,y,z=0)=\bv_s(x,y)$, where $\perp$ denotes components of the velocity parallel to the fluid-solid interface, and 

These boundary conditions entirely determine $\bw_1$ in equation(\ref{vpass}):
\beq
\bw_1(\bq)=\bv_s(\bq) \,,
\label{w1sol}
\eeq
which implicitly fixes ${\rm w}_{1z}=0$.

To determine $\bw_2$, we take the curl of (\ref{stokespass}) to eliminate the pressure $\Pi_3$.  This gives:
\beq
{\boldsymbol\nabla}_3\times\nabla_3^2\bv={\bf 0}.
\label{curl}
\eeq
The Laplacian of our velocity field (\ref{vpass}) is 
\bew
\beq
\nabla^2_3{\bf v}=\bw_1\nabla^2_3\bigg[\exp(-qz)\exp(i\bq\cdot\br_{_\perp})\bigg]+\bw_2\nabla^2_3\bigg[z\exp(-qz)\exp(i\bq\cdot\br_{_\perp})\bigg]=-2q\bw_2\exp(-qz)\exp(i\bq\cdot\br_{_\perp}) \,.
\eeq
\ew
Inserting this into (\ref{curl}) gives
\beq
(-q\hz+i\bq)\times\bw_2={\bf 0} \,.
\label{w2cond}
\eeq
Taking the dot product of this with $\bq$ (which we remind the reader lies in the plane of the surface, and, hence, perpendicular to $\hz$) implies
\beq
\bq\cdot(\hz\times\bw_2)=0 \,.
\label{triple1}
\eeq
Using the cyclic properties of triple products, this in turn implies
\beq
\hz\cdot(\bq\times\bw_2)=0 \,.
\label{triple2}
\eeq
Since $\hz\cdot\bq\times\hz=0$, this only imposes a condition on the in-plane components $\bw_{2\perp}$ of $\bw$:
\beq
\bq\times\bw_{2\perp}={\bf 0} \,,
\label{triple3}
\eeq
the most general solution of which is 
\beq
\bw_{2\perp}=\bq\phi \,.
\label{phidef}
\eeq
To determine the remaining unknown quantities, which are $\phi$ and ${\rm w}_{2z}$, we insert our expression (\ref{vpass}) for $\bv$  into the incompressibility condition:
\bew
\beq
{\boldsymbol\nabla}_3\cdot
{\bf v}=[i(\bq\cdot\bv_s)+{\rm w}_{2z}]\exp(-qz)\exp(i\bq\cdot\br_{_\perp})+(i\bq\cdot\bw_2-q{\rm w}_{2z})z\exp(-qz)\exp(i\bq\cdot\br_{_\perp})=0 \,,
\label{div}
\eeq
\ew
where we have replaced $\bw_1$ with $v_s$ everywhere it appears.

To satisfy this equation, the coefficients of both $\exp(-qz)\exp(i\bq\cdot\br_{_\perp})$ and $z\exp(-qz)\exp(i\bq\cdot\br_{_\perp})$ must vanish.
The former condition  implies
\beq
{\rm w}_{2z}=-i\bq\cdot\bv_s \,,
\label{w2z}
\eeq
which can be used in the latter to give
\beq
i\bq\cdot\bw_2=-iq(\bq\cdot\bv_s) \,.
\label{wperpcond}
\eeq
Using (\ref{phidef}) to rewrite this equation in terms of $\phi$, and solving for $\phi$, gives
\beq
\phi=-{\bq\cdot\bv_s\over q} \,,
\label{phisol}
\eeq
which in turn implies via (\ref{phidef}) that
\beq
\bw_\perp=-{\hat q}\bq\cdot\bv_s \,.
\label{wpsol}
\eeq
Now using (\ref{w1sol}),  (\ref{w2z}), and  (\ref{wpsol}) in  (\ref{vpass}), we obtain:
\beq
\bv(\br_\perp, z, t)=[\bv_s-z(\bq\cdot\bv_s)({\hat \bq}+i{\hat \bz})]e^{-qz+i\bq\cdot\br_\perp} \,.\label{vbulk-plane}
\eeq
\newline

We have so far focused on solutions to the Stokes' equation (\ref{del4v}) that are single plane waves. The most general solution, of course, can be obtained by summing up plane waves with all possible values of the wavevector $\bq$. This gives
\beq
\bv(\br_\perp, z, t)=\sum_{\bf q}[\bv_s(\bq)-z(\bq\cdot\bv_s(\bq))({\hat \bq}+i{\hat \bz})]e^{-qz+i\bq\cdot\br_\perp} \,.\label{vbulk}
\eeq



\section{Derivation of the stability condition}
\label{appb}

\setcounter{equation}{0}
\setcounter{figure}{0}
\setcounter{table}{0}
\setcounter{page}{1}
\makeatletter
\renewcommand{\theequation}{B\arabic{equation}}
\renewcommand{\thefigure}{B\arabic{figure}}
\renewcommand{\bibnumfmt}[1]{[S#1]}
\renewcommand{\citenumfont}[1]{S#1}

We will now prove,  in two steps, that the  conditions $c_0^2>0$, $\gamma>0$, and $\phi$ sufficiently small  imply that ${\rm Im}(\Upsilon(\theta))<0$ for all $\theta$. First, we will show that ${\rm Im}(\Upsilon(\theta=\pi/2))<0$. Then  we will show that ${\rm Im}(\Upsilon(\theta))$ cannot {\it equal} $0$ for any value of $\theta$. Since ${\rm Im}(\Upsilon(\theta))$ is obviously a continuous function of $\theta$, this implies that it can never become positive, since to get from its negative value at $\theta=\pi/2$ to a positive value, it would first have to pass through $0$.

We begin by showing ${\rm Im}(\Upsilon(\theta=\pi/2))<0$. Setting $\theta=\pi/2$ in (\ref{ev2}) gives
\beq
\Upsilon(\pi/2)^2+2i\gamma\Upsilon(\pi/2)-c_0^2=0 \,,
\label{evpi/2}
\eeq
which is readily solved to give
\beq
\Upsilon(\pi/2)=-i\gamma\pm\sqrt{c_0^2-\gamma^2} \,.
\label{evpi/2sol}
\eeq

If $c_0^2$ and $\gamma$ are both $>0$, then the magnitude of the argument of the square root in this expression is clearly less than $\gamma$. Hence, even if $\gamma^2>c_0^2$, so that the argument of the square root is negative, making the square root itself purely imaginary, it can only add to the imaginary part of $\Upsilon$ a contribution smaller in magnitude than $-i\gamma$; hence, it cannot make the imaginary part positive. If $\gamma<c_0^2$, things are even simpler: the square root in (\ref{evpi/2sol}) is real, and 
${\rm Im}(\Upsilon(\theta=\pi/2))=-\gamma$, which is $<0$ if $\gamma>0$.

So our stability conditions $c_0^2>0$, $\gamma>0$ imply that ${\rm Im}(\Upsilon(\theta=\pi/2))<0$. We now complete our proof that ${\rm Im}(\Upsilon(\theta))<0$ for {\it all} $\theta$ by showing that ${\rm Im}(\Upsilon(\theta))\ne0$ for any value of $\theta$. We will prove this by contradiction: assume that 
${\rm Im}(\Upsilon(\theta))=0$ at some value of $\theta$. Then at that value of $\theta$, $\Upsilon(\theta))$
is real. Therefore, the imaginary part of (\ref{ev2}) reads

\beq
\Upsilon\gamma(1+\sin^2\theta)-\gamma v_\rho\cos\theta(1+\varphi\sin^2\theta)=0 \,.
\label{Imups}
\eeq
which can be solved for $\Upsilon$:
\beq
\Upsilon=v_\rho\cos\theta\left({1+\varphi\sin^2\theta\over1+\sin^2\theta}\right) \,,
\label{Upssol}
\eeq

Inserting (\ref{Upssol}) into the real part of (\ref{ev2}) then implies

\bew
\beq
v_\rho^2\cos^2\theta\left({1+\varphi\sin^2\theta\over1+\sin^2\theta}\right)^2-v_\rho(v_\rho+v_p)\cos^2\theta\left({1+\varphi\sin^2\theta\over1+\sin^2\theta}\right)+v_pv_\rho \cos^2\theta-c_0^2\sin^2\theta=0 \,.
\label{proof}
\eeq
\ew
After a bit of algebra, and using $\varphi=1-{\gamma_\rho\over\gamma}$, this can be rewritten
\bew
\beq
\bigg[\left({v_\rho\gamma_\rho\over\gamma}\right)\cos^2\theta\left[v_p-v_\rho+(v_p-\varphi v_\rho)\sin^2\theta\right]-c_0^2(1+\sin^2\theta)^2\bigg]\sin^2\theta=0 \,.
\label{proof2}
\eeq
\ew

Our original assumption that ${\rm Im}(\Upsilon(\theta))=0$ can therefore only be satisfied if (\ref{proof2}) has a solution for some real $\theta$ between $-\pi$ and $\pi$. To say this another way, our system will be {\it stable} if (\ref{proof2}) has {\it no} solution for {\it any} real $\theta$. 

Since, in that range of $\theta$, $\sin^2\theta\ne0$, we can divide (\ref{proof2}) by $\sin^2\theta$. Doing so, and reorganizing a bit further, we can rewrite (\ref{proof2}) in dimensionless form as
\bew
\beq
\left({m^2\gamma_\rho\over\gamma}\right)\cos^2\theta\left[\varpi-1+(\varpi-\varphi)\sin^2\theta\right]=(1+\sin^2\theta)^2 \,,
\label{dimless cond}
\eeq
\ew
where we have defined the ``Mach number''
\beq
m\equiv {v_\rho\over c_0} \,,
\label{mach}
\eeq
and the speed ratio 
\beq
\varpi\equiv
{v_p\over v_\rho} \,.
\label{ratio}
\eeq

This equation clearly has no solution for real $\theta$, implying that the system is stable, if the left hand side of the equation is bounded {\it above} by $1$, since the right hand side is clearly bounded 
{\it below} by $1$.

We can derive such a bound by using $\cos^2\theta\le1$,  $\sin^2\theta\le1$, $\varpi-1\le |\varpi-1|$, and $\varphi=1-{\gamma_\rho\over\gamma}$ to show that the left hand side (LHS)  obeys
\beq
{\rm LHS}<\left({m^2\gamma_\rho\over\gamma}\right)\left[2|\varpi-1|+{\gamma_\rho\over\gamma}\right] \,.
\label{LHS bound}
\eeq

The right hand side of equation (\ref{LHS bound}) is a quadratic function of the ratio ${\gamma_\rho\over\gamma}$. Requiring that it be less than $1$ leads to the bounds on that ratio:
\bew
\beq
-|\varpi-1|-\sqrt{(\varpi-1)^2+{1\over m^2}}<{\gamma_\rho\over\gamma}<-|\varpi-1|+\sqrt{(\varpi-1)^2+{1\over m^2}} \,.
\label{stabapp}
\eeq
\ew
It is easy to see that this condition can always be satisfied for sufficiently small $\gamma_\rho$; in particular, the allowed region always includes $\gamma_\rho=0$. 

Note that we have shown that the condition (\ref{stabapp}) is {\it sufficient} for stability; however, it is by no means necessary.



 Nonetheless, having such a sufficient condition proves that the ordered state in this system, unlike that of ``wet'' systems, {\it can} be stable. More specifically, as long as $c_0^2$ and $\gamma$ are positive, there is always a window of stability at sufficiently small  $\gamma_\rho$.

  At $\theta=0,\,\pi$, $Im\Upsilon(\theta)=0$; however, there are stabilizing ${\cal O}(q^2)$ damping terms that have been neglected in the previous discussion. For these two special directions of propagation, the dynamics of the density fluctuations $\delta\rho$ decouple, up to ${\cal O}(q)$,  from those of $p_y$. Hence the density fluctuations are {\em not damped} at this order. This is not really a problem, as there are diffusive damping terms  at ${\cal O}(q^2)$ which provide damping to the density fluctuations. These damping terms are the $[D_\rho \nabla_s^2 + D_{\rho 2}\partial_x^2]\delta\rho$ terms, which, for $\theta=0,\,\pi$ in Fourier space, give $-(D_\rho+ D_{\rho 2})q^2\delta \rho$ and  ultimately damp the density fluctuations at $\theta=0,\,\pi$. In fact, this decoupling of $\delta\rho$ and $p_y$ at ${\cal O}(q)$ at $\theta=0,\,\pi$ is  the origin of the special behaviors of the scaling functions at those points.

 Q.E.D.
\beq
\nonumber\\
\eeq

\section{Evaluation of the integral for the equal time correlation function}
\label{appc}

\setcounter{equation}{0}
\setcounter{figure}{0}
\setcounter{table}{0}
\setcounter{page}{1}
\makeatletter
\renewcommand{\theequation}{C\arabic{equation}}
\renewcommand{\thefigure}{C\arabic{figure}}
\renewcommand{\bibnumfmt}[1]{[S#1]}
\renewcommand{\citenumfont}[1]{S#1}
 In this appendix, we  evaluate the integral
\begin{eqnarray}
I\equiv \int_{-\infty}^\infty 
{{ (s+\psi(\varphi, \theta_\bq)}^2ds\over(s-\psi_+( \varphi,  \theta_\bq))^2(s-\psi_-({ \varphi}, \theta_\bq))^2+s^2}\,,
\nonumber\\
\label{ETint}
\end{eqnarray}
in equation (\ref{pet2}), which arises
in the calculation
of the equal time correlation function $C_{pp}(\bq)\equiv\langle |p_y(\bq,t)|^2\rangle$, is given by $I=\pi$, independent of all parameters, and $\theta_\bq$.

This proves to be true if the two quantities 
\beq
\psi_\pm(\theta_\bq)\equiv{c_\pm(\theta_\bq)-v_\rho \cos\theta_\bq \,  \Xi(\varphi, \theta_\bq)\over{ \Psi(1, \theta_\bq)}} \,
\label{psidefa}
\eeq

\noindent obey $\psi_+(\theta_\bq)>0$, and $\psi_-(\theta_\bq)<0$. We will now show that $\psi_+(\theta_\bq)$, and $\psi_-(\theta_\bq)$ do indeed obey these inequalities  whenever the system is stable. We will then show that when these inequalities are satisfied, $I=\pi$.

We begin by showing that $\psi_+(\theta_\bq)>0$, and $\psi_-(\theta_\bq)<0$.
This can be shown by first rewriting (\ref{psidefa}) using our expression (\ref{cplusminus}) for the sound speeds $c_\pm(\theta_\bq)$:
\bew
\begin{eqnarray}
\psi_{\pm}\left(\theta_\bq \right) =
\left\{\left({v_\rho(1-2\Xi) + v_p \over 2}\right)\cos\theta_\bq
\pm \sqrt{{1 \over 4}\left(v_\rho -v_p\right)^2 \cos^2
\theta_\bq + c^2_0
\sin^2 \theta_\bq}\right\}/ \Psi(1, \theta_\bq)\quad .
\label{psiplusminus}
\end{eqnarray}
\ew

Since $\Psi(1, \theta_\bq)=\gamma(1+\sin^2\theta)$ is always positive if $\gamma>0$ (as it must be for stability), we will clearly have
$\psi_+(\theta_\bq)>0$, and $\psi_-(\theta_\bq)<0$ if the magnitude of the square root in (\ref{psiplusminus}) is greater than the magnitude of the term $\left({v_\rho(1-2\Xi) + v_p \over 2}\right)\cos\theta_\bq$ in that equation.
This is equivalent to the argument of that square root being  bigger than the {\it square} of $\left({v_\rho(1-2\Xi) + v_p \over 2}\right)\cos\theta_\bq$. This leads to the condition
\beq
{1 \over 4}\left(v_\rho -v_p\right)^2 \cos^2
\theta_\bq + c^2_0
\sin^2 \theta_\bq >\left[\left({v_\rho(1-2\Xi) + v_p \over 2}\right)\cos\theta_\bq
\right]^2
\label{psicond}
\eeq
as a necessary and sufficient condition for making $\psi_+(\theta_\bq)>0$, and $\psi_-(\theta_\bq)<0$.

After a little (!) algebra, and using the fact that 
\beq
\Xi(\Phi, \theta)\equiv{\Psi(\Phi, \theta)\over\Psi(1, \theta)}={1+\Phi\sin^2\theta\over1+\sin^2\theta} \,,
\label{Xidefa}
\eeq
this can be rewritten as 
\bew
\beq
\left({v_\rho\gamma_\rho\over\gamma}\right)\cos^2\theta\left[v_p-v_\rho+(v_p-\varphi v_\rho)\right]<c_0^2(1+\sin^2\theta)^2 \,,
\label{psicond2}
\eeq
\ew
which the alert reader will note is precisely the same as our stability condition (\ref{stability}). So, if our system is stable, $\psi_+(\theta_\bq)>0$, and $\psi_-(\theta_\bq)<0$.

Now we turn to the evaluation of the integral $I$ equation (\ref{ETint}).
This can clearly be rewritten as
\beq
I=I_1+2\psi I_2+ \psi^2 I_3 \,,
\label{I123def}
\eeq
with
\begin{eqnarray}
I_1=\int_{-\infty}^\infty 
{s^2ds\over(s-\psi_+(\theta_\bq))^2(s-\psi_-(\theta_\bq))^2+s^2}\,,\nn\\
\label{I1}
\end{eqnarray}
\begin{eqnarray}
I_2=\int_{-\infty}^\infty 
{sds\over(s-\psi_+(\theta_\bq))^2(s-\psi_-(\theta_\bq))^2+s^2}\,,\nn\\
\label{I2}
\end{eqnarray}
and 
\begin{eqnarray}
I_3=\int_{-\infty}^\infty 
{ds\over(s-\psi_+(\theta_\bq))^2(s-\psi_-(\theta_\bq))^2+s^2}\,.\nn\\
\label{I3}
\end{eqnarray}

We will now evaluate each of these using the method of partial fractions. This begins with the identity
\bew
\beq
{1\over(s-\psi_+(\theta_\bq))^2(s-\psi_-(\theta_\bq))^2+s^2}={1\over2is}\left[{1\over(s-\psi_+(\theta_\bq))(s-\psi_-(\theta_\bq))-is}-{1\over(s-\psi_+(\theta_\bq))(s-\psi_-(\theta_\bq))+is}\right] \,.
\label{factor}
\eeq
\ew
Using this, we can
rewrite $I_1$ (\ref{I1}) as

\newpage

\bew

\begin{eqnarray}
I_1=\int_{-\infty}^\infty 
{s^2ds\over(s-\psi_+(\theta_\bq))^2(s-\psi_-(\theta_\bq))^2+s^2} ={1\over 2i}(I_{1-}-I_{1+})\,,\nn\\
\label{intfac}
\end{eqnarray}
\ew

where we have defined
\begin{eqnarray}
 &&I_{1\pm}\equiv \int_{-\infty}^\infty 
{sds\over(s-\psi_+(\theta_\bq))(s-\psi_-(\theta_\bq))\pm is} \,.
\label{ipm}
\end{eqnarray}
We will now evaluate $I_{1\pm}$; in particular, we will show that both are independent of $\psi_\pm$.

Consider first $I_{1+}$, and  the poles of its integrand in the complex $s$ plane. There are obviously two of these, which we can write as 
\beq
s_{1,2}=r_{1,2}e^{i\phi_{1,2}} \,,
\label{s12}
\eeq
with $r_{1,2}$ both real and positive, and $\phi_{1,2}$ real. 
Inspection of the denominator in (\ref{ipm}) shows that these
must obey
\beq
r_1e^{i\phi_1}+r_2e^{i\phi_2}=\psi_+(\theta_\bq)+\psi_-(\theta_\bq)-i 
\label{rootsum}
\eeq
and
\beq
r_1r_2e^{i(\phi_1+\phi_2)}=\psi_+(\theta_\bq)\psi_-(\theta_\bq)\,.
\label{rootprod}
\eeq
Since $\psi_+>0$ and $\psi_-<0$, $\psi_+\psi_-<0$. Using this fact in (\ref{rootprod}), and recalling that $r_1$ and $r_2$ are both real and positive, we see that 
\beq
\phi_1+\phi_2=\pi  \,,
\label{phicon}
\eeq
from which it immediately follows that  
\beq
s_2=r_2e^{i(\pi-\phi_1)}=-r_2e^{-i\phi_1} \,.
\label{s2}
\eeq
Note that (\ref{s2}) implies that $s_1$ and $s_2$ lie on the same side of the real axis.  Using (\ref{s2}) in  (\ref{rootsum}) gives 
\beq
r_1e^{i\phi_1}-r_2e^{-i\phi_1}=\psi_+(\theta_\bq)+\psi_-(\theta_\bq)-i \,.
\label{rootsum2}
\eeq
Taking the imaginary part of this expression gives
\beq
(r_1+r_2)\sin\phi_1=-1 \,,
\label{phi1con}
\eeq
which implies that $\sin\phi_1<0$, which in turn implies that $s_1$, and, hence, $s_2$, both lie in the lower half plane.

 Then if we consider the contour $\mathcal{C}$ in the complex $s$ plane consisting of the real axis plus the infinite semicircle {\it above} the real axis connecting its ends, it follows from the absence of poles in the upper half plane that 
\beq
\oint_\mathcal{C}{sds\over(s-\psi_+(\theta_\bq))(s-\psi_-(\theta_\bq))+ is}=0 \,.
\label{contour1}
\eeq
The integral in this expression can be written as 
\bew
\beq
\oint_\mathcal{C}{sds\over(s-\psi_+(\theta_\bq))(s-\psi_-(\theta_\bq))+ is}=I_{1+}+\int_\mathcal{S}{sds\over(s-\psi_+(\theta_\bq))(s-\psi_-(\theta_\bq))+ is} \,,
\label{contour2}
\eeq
\ew
where the  contour $\mathcal{S}$ is the aforementioned  infinite semicircle. Since that semicircle is infinite, $\psi_{\pm}$ and $is$ can be neglected in the denominator of the integrand relative to $s^2$. Thus we have
\beq
\int_\mathcal{S}{sds\over(s-\psi_+(\theta_\bq))(s-\psi_-(\theta_\bq))+ is}=\int_\mathcal{S}{sds\over s^2}=\int_\mathcal{S}{ds\over s}=i\pi \,.
\label{semi}
\eeq
Using this in (\ref{contour2}) and using the result in (\ref{contour1}) gives 
\beq
I_+=-i\pi \,.
\label{I_+}
\eeq
It is straightforward to repeat this reasoning for $I_-$; in that case, both poles lie in the upper half plane, and the semicircle must lie in the lower half plane. We obtain
\beq
I_-=i\pi \,.
\label{I_-}
\eeq
Using (\ref{I_+}) and (\ref{I_-}) in (\ref{intfac}), we obtain
\begin{eqnarray}
I_1=\int_{-\infty}^\infty 
{s^2ds\over(s-\psi_+(\theta_\bq))^2(s-\psi_-(\theta_\bq))^2+s^2} =\pi \,.\nn\\
\label{intans}
\end{eqnarray}
as claimed earlier.

Similar reasoning can be applied to $I2$ and $I3$. For $I2$, we get
\bew
\begin{eqnarray}
I_2=\int_{-\infty}^\infty 
{sds\over(s-\psi_+(\theta_\bq))^2(s-\psi_-(\theta_\bq))^2+s^2} ={1\over 2i}(I_{2-}-I_{2+})\,,\nn\\
\label{intfac2}
\end{eqnarray}
\ew
where we have defined
\begin{eqnarray}
 &&I_{2\pm}\equiv \int_{-\infty}^\infty 
{ds\over(s-\psi_+(\theta_\bq))(s-\psi_-(\theta_\bq))\pm is} \,.
\label{ipm1}
\end{eqnarray}
Proceeding as we did above for $I_1$, we find
\bew
\beq
\oint_\mathcal{C}{ds\over(s-\psi_+(\theta_\bq))(s-\psi_-(\theta_\bq))+ is}=I_{2+}+\int_\mathcal{S}{ds\over(s-\psi_+(\theta_\bq))(s-\psi_-(\theta_\bq))+ is} =0\,.
\label{contour2.2}
\eeq
\ew
Again using the fact that semicircle is infinite, we can again neglect
$\psi_{\pm}$ and $is$ in the denominator of the integrand relative to $s^2$. Thus we have
\bew
\beq
\int_\mathcal{S}{ds\over(s-\psi_+(\theta_\bq))(s-\psi_-(\theta_\bq))+ is}=\int_\mathcal{S}{ds\over s^2}=\lim_{R\to\infty}\int_\mathcal{S}{ds\over s^2}=\lim_{R\to\infty}(1-e^{-i\pi})/R=0 \,.
\label{semi2}
\eeq
\ew

{
Thus we find 
\beq
I_{2+}=0 \,.
\label{I2+res}
\eeq
Similar reasoning shows that 
\beq
I_{2-}=0 \,,
\label{I2-res}
\eeq
as well.
Taking these two results \ref{I2+res} and \ref{I2+res} together in \ref{intfac2} gives
\beq
I_{2}=0 \,.
\label{I2res}
\eeq

Finally, applying this partial fraction approach to $I_3$, we obtain
\beq
I_3=\left({1\over2i}\right)(I_{3-}-I_{3+})\,,\nn\\
\label{I3.1}
\eeq
where we have defined
\beq
I_{3\pm}\equiv\int_{-\infty}^\infty{ds\over (s-i\epsilon)\left[(s-\psi_+(\theta_\bq))(s-\psi_-(\theta_\bq))\pm is\right]} \,,
\label{I3pm}
\eeq
and we have shifted the pole of ${1\over s}$ above the real axis by a small amount $\epsilon$, which we will take to zero at the end of our calculation. It is straightforward to check that choosing to move the pole {\it below} the real axis by a small amount $\epsilon$ leads to exactly the same final answer for $I_3$. 

Proceeding with the choice of moving the pole above the axis, we note that, for this choice, all of the poles in the integrand for $I_{3-}$ lie in the upper half plane. Therefore, evaluating the integral by closing the contour in the lower half plane (which we can do with impunity, since the integrand vanishes like ${1\over s^2}$ as $|s|\to\infty$, which implies that the semi-infinite semi-circle with we close the contour contributes nothing to the integral), we find that $I_{3-}=0$. 

On the other hand, for $I_{3+}$, there is a single pole at $s=i\epsilon$ and the upper half plane, and two poles (as discussed earlier) in the lower half plane. Making the easy choice of closing the contour in the upper half plane, and taking the limit $\epsilon\to0$, we easily find
\beq
I_{3+}={2\pi i\over\psi_+\psi_-} \,.
\label{I3+}
\eeq
Using this in \eqref{I3.1}, we obtain our final result for $I_3$:
\beq
I_3=-{\pi\over\psi_+\psi_-} \,.
\label{I3final}
\eeq

Using this result \ref{I3final} and our earlier results $I_1=\pi$ and  $I_2=0$ in our expression \ref{I123def} for the integral $I$ gives
\beq
I=\pi\left(1-{\psi^2\over\psi_+\psi_-}\right) \,.
\eeq

Using our expressions (\ref{psiplusminus})      and (\ref{psi n def})       for $\psi_{\pm}$ and $\psi$ in this result, we obtain, after a little (!) algebra, our final expression

}
 \bew
 \beq
 I=\pi\bigg(1+\left({\gamma_\rho\over\gamma}\right)^2\frac{m^2\sin^2\theta\cos^2\theta}{\left\{\left(1+\sin^2\theta\right)^2-m^2[\varpi-1+(\varpi-\varphi)\sin^2\theta]\cos^2\theta\right\}}\bigg) \,.
 \label{I final}
\eeq
\ew

\section{Evaluation of the non-local kernels}
\label{appd}

Introducing an exponential UV cutoff, we can write
\bew
\beq
{\cal K}_{pp}(\brp) =\int {d^2q\over(2\pi)^2}
\left(\frac{q^2+q_y^2}{q}\right)e^{i\bq\cdot\brp-aq}=-(\pp_x^2+2\pp_y^2)\phi(\brp)
\label{ppkernelapp1}
\eeq
\ew
where $a\equiv2\pi/\Lambda$, with $\Lambda$ being the ultraviolet cutoff, and
\bew
\bea
\phi(\brp)&=&\int {d^2q\over(2\pi)^2}
\left(\frac{1}{q}\right)e^{i\bq\cdot\brp-aq}={1\over4\pi^2}\int_0^{2\pi}d\theta\int_0^\infty \exp([ir_\perp\cos\theta-a]q)dq={1\over4\pi^2}\int_0^{2\pi}{d\theta\over a-ir_\perp\cos\theta}\nonumber\\&=&{1\over4\pi^2}\int_0^{2\pi}{(a+ir_\perp\cos\theta)d\theta\over a^2+r^2_\perp\cos^2\theta} \,.
\label{phidef2}
\eea
\ew
The $\cos\theta$ term in the numerator is odd under $\theta\to\pi-\theta$, so its integral vanishes. The remainder of the integral is even under $\theta\to\pi-\theta$, so we can replace its integral with twice the integral from $-\pi/2$ to $\pi/2$. Doing so, we get
\beq
\phi(\brp)={a\over2\pi^2}\int_{-\pi/2}^{\pi\over2}{d\theta\over a^2+r^2_\perp\cos^2\theta} \,.
\label{phi2}
\eeq
The integral can be done straightforwardly with the substitution $u=\tan\theta$, with the result
\beq
\phi(\brp)={1\over2\pi\sqrt{a^2+r_\perp^2}} \,.
\label{phiUV}
\eeq

Inserting \eqref{phiUV} into the last equality of \eqref{ppkernelapp1} and taking the derivatives gives
\beq
{\cal K}_{pp}(\brp) =\left({3 a^2-3 y^2\over 2\pi(a^2+ r_\perp^2)^{5/2}}\right) \,.
\label{ppkernelapp}
\eeq
Likewise, 
\beq
{\cal K}_{pp\rho}(\brp) =\left({3 a^2-3 x^2\over 2\pi(a^2+ r_\perp^2)^{5/2}}\right) \,.
\label{pprhokernelapp}
\eeq
and,
\beq
{\cal K}_{p\rho}(\brp) =-\pp_x\pp_y\phi(\brp)=-{3 xy\over2\pi (a^2+ r_\perp^2)^{5/2}} \,.
\label{prhokernelapp}
\eeq

The limiting behaviors of these for $r\gg a$ are given by equations \eqref{ppkernel} and \eqref{prhokernel} respectively. Note also that both are well-behaved as $r\to0$.

\section{Giant number fluctuations}\label{giant-app}

We here revisit giant number fluctuations in this model and consider the case $L_y  \gg L_x$, i.e.,  aspect ratio $\alpha_A\ll 1$. We  again begin with Eq.~(\ref{Nvar1}). Now consider the integral over $y'$ for fixed $r = (x, y)$ in this limit.
Note first that in this limit, for most of the range of integration over $|y- y|\gg  L_x$. For such values of $y$,  in analogy with our approach for $\alpha_A\gg 1$, we can split
the integral over $y'$ into three parts

\bew
\begin{eqnarray}
&&\int_0^{L_y}dy'\frac{|{\bf r-r'}|}{({\bf r-r'})^2 + (x-x')^2}\nonumber \\
 &&\times \left[1-\frac{(v_p-v_\rho)^2(y-y')}{({\bf r-r'})^2 + (x-x')^2}\frac{\gamma_\rho v_\rho}{\gamma c_0^2} - \left(\frac{v_\rho\gamma_\rho}{c_0\gamma}\right)^2 \frac{(y-y')^2(x-x')^2} {\{({\bf r-r'})^2 + (x-x')^2\}^2}\right]^{-1}=I_4+I_5+I_6 \,,
\label{split}
\end{eqnarray}
\ew

where we have defined
\bew
\begin{eqnarray}
I_4&\equiv&\int_0^{y-\tilde CL_x}dy'\frac{|{\bf r-r'}|}{({\bf r-r'})^2 + (x-x')^2}\nonumber \\
 &&\times \left[1-\frac{(v_p-v_\rho)(y-y')^2}{({\bf r-r'})^2 + (x-x')^2}\frac{\gamma_\rho v_\rho}{\gamma c_0^2} - \left(\frac{v_\rho\gamma_\rho}{c_0\gamma}\right)^2 \frac{(y-y')^2(x-x')^2} {\{({\bf r-r'})^2 + (x-x')^2\}^2}\right]^{-1} \,,
\label{I4def}
\end{eqnarray}

\begin{eqnarray}
I_5&\equiv&\int_{y-\tilde CL_x}^{y+\tilde CL_x}dy'\frac{|{\bf r-r'}|}{({\bf r-r'})^2 + (x-x')^2}\nonumber \\
 &&\times \left[1-\frac{(v_p-v_\rho)(y-y')^2}{({\bf r-r'})^2 + (x-x')^2}\frac{\gamma_\rho v_\rho}{\gamma c_0^2} - \left(\frac{v_\rho\gamma_\rho}{c_0\gamma}\right)^2 \frac{(y-y')^2(x-x')^2} {\{({\bf r-r'})^2 + (x-x')^2\}^2}\right]^{-1} \,,
\label{I5def}
\end{eqnarray}
and 
\begin{eqnarray}
I_6&\equiv&\int_{y+\tilde CL_x}^{L_y}dy'\frac{|{\bf r-r'}|}{({\bf r-r'})^2 + (x-x')^2}\nonumber \\
 &&\times \left[1-\frac{(v_p-v_\rho)(y-y')^2}{({\bf r-r'})^2 + (x-x')^2}\frac{\gamma_\rho v_\rho}{\gamma c_0^2} - \left(\frac{v_\rho\gamma_\rho}{c_0\gamma}\right)^2 \frac{(y-y')^2(x-x')^2} {\{({\bf r-r'})^2 + (x-x')^2\}^2}\right]^{-1} \,.
\label{I6def}
\end{eqnarray}
\ew

Once again, we have chosen the constant $\tilde C$ to be sufficiently large compared to 1 that, throughout the regions of integration of $I_4$ and $I_6$, 
$|y- y'|\gg L_x$. Hence, in those regions of integration,
\begin{eqnarray}
&&\frac{|{\bf r-r'}|}{({\bf r-r'})^2 + (x-x')^2}\approx{1\over y-y'} \,,\\
 &&\frac{(y-y')^2}{(x-x')^2 + ({\bf r-r'})^2}\approx 1\,,\\
 &&\frac{(y-y')^2(x-x')^2}{[(x-x')^2 + ({\bf r-r'})^2]^2}\ll 1\,.
\end{eqnarray}
This implies that
\bew
\beq
I_4\approx\left[1-\frac{(v_p-v_\rho)\gamma_\rho v_\rho}{\gamma c_0^2}\right]^{-1}\int_0^{y-\tilde CL_x}{dy'\over y-y'}= \ln\left({y\over \tilde CL_x}\right)\left[1-\frac{(v_p-v_\rho)\gamma_\rho v_\rho}{\gamma c_0^2}\right]^{-1} = 
\bigg\{\ln\left({y\over L_x}\right)-\ln \tilde C\bigg\}\left[1-\frac{(v_p-v_\rho)\gamma_\rho v_\rho}{\gamma c_0^2}\right]^{-1} \,.
\label{I4result}
\eeq
\ew

Virtually identical reasoning can be applied to $I_6$, giving the result
\bew
\beq
I_6= \bigg\{\ln\left({L_y-y \over L_x}\right)-\ln \tilde C \bigg\}\left[1-\frac{(v_p-v_\rho)\gamma_\rho v_\rho}{\gamma c_0^2}\right]^{-1} \,.
\label{I6result}
\eeq
\ew

 For $I_5$, a simple shift of variables of integration $y'=y+y''$ shows that $I_5$ is independent of $y$:
\bew
\begin{eqnarray}
I_5&=&\int_{-\tilde CL_x}^{\tilde CL_x}dy''\frac{\sqrt{y''^2 + (x-x')^2}}{2(x-x')^2 + y''^2}\nonumber \\
&&\times \left[1-\frac{\gamma_\rho v_\rho}{\gamma c_0^2}\frac{(v_p-v_\rho)y''^2}{ {2(x-x')^2 + y''^2}} - \left(\frac{v_\rho\gamma_\rho}{c_0\gamma}\right)^2 \frac{(x-x')^2{ y''}^2}{[ {2(x-x')^2 + y''^2}]^2}\right]^{-1}\,.
\label{I5result}
\end{eqnarray}
\ew

{
Inserting these results (\ref{I4result}),  (\ref{I5result}), and  (\ref{I6result}) into our expression (\ref{split}), and using that result in our expression (\ref{Nvar1}) for $\langle (\delta N)^2\rangle$, we obtain

\bew
\begin{eqnarray}
 &&\langle (\delta N)^2\rangle  
\approx \frac{D_p\rho_c^2 m^2}{2\pi \gamma } \left\{ \int_0^{L_x} dx\int_0^{L_x} dx' \int_0^{L_y}dy\left(\ln\left({y\over L_x}\right)+\ln\left({L_y-y\over L_x}\right)-2\ln\tilde C \right) \left[1-\frac{(v_p-v_\rho)\gamma_\rho v_\rho}{\gamma c_0^2}\right]^{-1} \right. \nonumber\\
&& \left.
+\int_0^{L_x} dx\int_0^{L_x} dx' \int_0^{L_y}dy\int_{-\tilde CL_x}^{\tilde CL_x}dy''\frac{\sqrt{y''^2 + (x-x')^2}}{2(x-x')^2 + y''^2}
\times \left[1-\frac{\gamma_\rho v_\rho}{\gamma c_0^2}\frac{(v_p-v_\rho)y''^2} {[2(x-x')^2 + y''^2]} - \left(\frac{v_\rho\gamma_\rho}{c_0\gamma}\right)^2 \frac{(x-x')^2 y''^2}{[2(x-x')^2 + y''^2]^2}\right]^{-1} \right\}
\nonumber\\
\label{delNfinal2}
\end{eqnarray}
\ew
The integrals over $x$ and $x'$ in the (triple) integral in the first line of this expression trivially give a factor of $L_x^2$, 
since the integrand is independent of $x$ and $x'$. The remaining integral over $y$ is elementary. The net result is
\bew
\begin{eqnarray}
 &&\int_0^{L_x} dx\int_0^{L_x} dx' \int_0^{L_y}dy\left(\ln\left({y\over L_x}\right)+\ln\left({L_y-y\over L_x}\right)-2\ln\tilde C \right)   \left[1-\frac{(v_p-v_\rho)\gamma_\rho v_\rho}{\gamma c_0^2}\right]^{-1}\nonumber\\
 && 
= 2L_x^2L_y\left(\ln\left({L_y\over \tilde CL_x}\right)-1\right) \left[1-\frac{(v_p-v_\rho)\gamma_\rho v_\rho}{\gamma c_0^2}\right]^{-1}\,.
\label{1st line}
\end{eqnarray}
\ew
 
The $y$ integral in the remaining (quadruple) integral which appears on the second line in equation (\ref{delNfinal2}) can be done immediately, since the integrand is independent of $y$, yielding a factor of $L_y$. The remaining triple integral over $x$, $x'$, and $y''$ can be done by changing variables of integration to new, rescaled variables $u_x$, $u_y$, and $u'_y$ via

\beq
x\equiv L_xu_x
\sep
x'\equiv L_xu'_x
\sep
y''\equiv L_xu_y \,.
\label{varchange xyy'2}
\eeq
This gives
\bew
\bea
\int_0^{L_x} dx\int_0^{L_x} dx' \int_0^{L_y}dy\int_{-\tilde CL_x}^{\tilde CL_x}dy''\frac{\sqrt{y''^2 + (x-x')^2}}{2(x-x')^2 + y''^2}
&&
\times \left[1-\frac{\gamma_\rho v_\rho}{\gamma c_0^2}\frac{(v_p-v_\rho)y''^2}{[2(x-x')^2 + y''^2]} - \left(\frac{v_\rho\gamma_\rho}{c_0\gamma}\right)^2 \frac{(x-x')^2 y''^2}{[2(x-x')^2 + y''^2]^2}\right]^{-1}\nonumber \\
&&=\tilde C'L_x^2 L_y \,,\nonumber\\
\label{2nd line}
\eea
\ew
where
\bew
\bea
\tilde C'&\equiv& \int_0^1 du_x\int_0^1 du_x'\int_{-C}^C\,du_y  \frac{\sqrt{(u_x-u_x')^2 + u_y^2}}{u_y^2 + 
2 (u_x-u_x')^2}\nonumber \\&&
\left[1-\frac{\gamma_\rho v_\rho}{\gamma c_0^2}\frac{(v_p-v_\rho)u_y^2}{2(u_x-u_x')^2 + u_y^2} - \left(\frac{v_\rho\gamma_\rho}{c_0\gamma}\right)^2 \frac{u_y^2 (u_x-u_x')^2}{[2(u_x-u_x')^2 + u_y^2]^2}\right]^{-1} 
\label{tildeC'def}
\eea
\ew
is an ${\cal O}(1)$ constant.

Comparing (\ref{1st line})  and (\ref{2nd line}), we see that the first line of 
(\ref{delNfinal2}) actually dominates the second in the small aspect ratio limit $L_x\ll L_y$ that we are considering here. Therefore, we obtain, in the limit of small aspect ratio ($\alpha_A\ll1$):
\bew
\begin{equation}
 \sqrt{\langle (\delta N)^2\rangle} \approx \frac{\rho_c m\sqrt{\tilde C'' D_p}}{\sqrt{2\pi \gamma}} 
 \sqrt{L_x^2L_y\ln\left({L_y\over L_x}\right)}\approx\frac{\rho_c m\sqrt{\tilde C'' D_p}}{\sqrt{2\pi \gamma}} L_y^{3/2}\alpha_A
 \sqrt{\ln\left({1\over\alpha_A}\right)}\,,
 \label{Nvarlar2}
\end{equation}
\end{widetext}
where we have defined 
\beq
\tilde C''\equiv 2\left[1-\frac{(v_p-v_\rho)\gamma_\rho v_\rho}{\gamma c_0^2}\right]^{-1}
\label{tildeC''def}
\eeq
Note  that $C'$ is an $O(1)$, parameter-dependent but aspect ratio and box size independent constant.

This can be rewritten in terms of 
the mean particle number $\overline N$ in the same area $A$using (\ref{LyNb}), which gives 
\begin{equation}
 \Delta N= \sqrt{\langle(\delta N)^2\rangle} = \left({\rho_c m\over \rho_0^{3/4}}\right)\left(\sqrt {C'D_p\over
 2\pi \gamma}\right){\overline N}^{3/4}\alpha_A^{1/4}  \sqrt{\ln\left({1\over\alpha_A}\right)}   \,.
\end{equation}
}

\section{Bulk velocity correlations in the $y$- and $z$-directions}\label{vel-corr-app}

{

In this section we calculate the bulk velocity correlations in the $y$ and $z$-directions. We first work out
the velocity correlations in the $y$-direction.

Using (\ref{cvij}) and (\ref{vcorrneutral}),  we see that the velocity correlation in the $y$ direction is given by
\bew
\bea
 C^v_{yy}(v_0(t-t')\hx, z, z, t-t') &=& C^{pp}_{yy}(v_0(t-t')\hx, z, z, t-t') + C^{\rho\rho}_{yy}(v_0(t-t')\hx, z, z, t-t')
 \label{cvyy} \\ 
 &=& v_0^2\int\frac{d\omega}{2\pi}\frac{d^2q}{(2\pi)^2} \exp [i(\omega-v_0q_x)(t-t')
-2qz] C_{pp}({\bf q},\omega)\left(1-{2q_y^2z\over q}+{q_y^4z^2\over q^2}\right) \nonumber \\ 
&& + v_a'^2\int\frac{d\omega}{2\pi}\frac{d^2q}{(2\pi)^2} \exp [i(\omega-v_0q_x)(t-t')
-2qz] C_{\rho\rho}({\bf q},\omega){z^2 q_x^2q_y^2 \over q^2}.
\label{vyycorrneutral}
\eea
\ew

Making the change of variables (\ref{varchange}) as before, we  find that $C^v_{yy}(v_0(t-t')\hx, z, z, t-t')$ also obeys a scaling law:
\bew
\beq
C^v_{yy}(v_0(t-t')\hx, z, z, t-t')=\left({D_pv_0\over z}\right)F^{pp}_{yy}\left({v_0|t-t'|\over z}\right) + 
\left({D_p\rho_c^2 v_a'^2\over v_0 z}\right)F^{\rho\rho}_{yy}\left({v_0|t-t'|\over z}\right) \,,
\label{yscale}
\eeq
\ew
where now the dimensionless scaling functions are given by 


\bew
\bea
F^{pp}_{yy}(u_r) &=& \int\frac{d\Omega}{2\pi}\frac{d^2Q}{(2\pi)^2} \exp [i(\Omega-Q_x)u_r-2Q]H_{pp}\bigg({\Omega\over Q}, \theta_{\bf Q}; \bigg\{{v_\rho\over v_0},{v_p\over v_0},{\gamma\over v_0},{c_0\over v_0}\bigg\}\bigg)\left(1-2{Q_y^2\over Q}+{Q_y^4\over Q^2}\right)\left({1\over Q^2}\right)\,.
\label{fppyyscale} \\ 
F^{\rho\rho}_{yy}(u_r) &=& \int\frac{d\Omega}{2\pi}\frac{d^2Q}{(2\pi)^2} \exp [i(\Omega-Q_x)u_r-2Q]H_{\rho\rho}\bigg({\Omega\over Q}, \theta_{\bf Q}; \bigg\{{v_\rho\over v_0},{v_p\over v_0},{\gamma\over v_0},{c_0\over v_0}\bigg\}\bigg){Q_x^2Q_y^2\over Q^4}\,.
\label{frho2yyscale} 
\eea
\ew
The limiting behaviors of these scaling functions can be obtained by an almost identical analysis to that used for $F^{pp}_{xx}(u_r)$, with the result:
\beqn
F^{pp}_{yy}(u_r)
=\left\{
\begin{array}{ll}
A^{pp}_{yy}\sep&u_r\ll1\\\\
{ (B^{pp}_{yy})_1\over u_r}+{(B^{pp}_{yy})_2\over u_r^2}+{(B^{pp}_{yy})_3\over u_r^3}\sep&u\gg1
\end{array}
\right.
\nonumber\\
\label{fppyy}
\eeqn
and
\beqn
F^{\rho\rho}_{yy}(u_r)
=\left\{
\begin{array}{ll}
A^{\rho\rho}_{yy}\sep&u_r\ll1\\\\
{B^{\rho\rho}_{yy}\over u_r^3}\sep&u_r\gg1
\end{array}
\right.
\nonumber\\
\label{frho2yy}
\eeqn
where the constants $A^{pp}_{yy}$, $A^{\rho\rho}_{yy}$, $(B^{pp}_{yy})_1$, $(B^{pp}_{yy})_2$, 
$(B^{pp}_{yy})_3$, and $B^{\rho\rho}_{yy}$ are given by
\bew
\bea
A^{pp}_{yy} &=& \int\frac{d\Omega}{2\pi}\frac{d^2Q}{(2\pi)^2} e^{-2Q}H_{pp}\bigg({\Omega\over Q}, \theta_{\bf Q}; \bigg\{{v_\rho\over v_0},{v_p\over v_0},{\gamma\over v_0},{c_0\over v_0}\bigg\}\bigg)\left(1-2{Q_y^2\over Q}+{Q_y^4\over Q^2}\right)\left({1\over Q^2}\right)\,,
\label{Appyy} \\ 
A^{\rho\rho}_{yy} &=& \int\frac{d\Omega}{2\pi}\frac{d^2Q}{(2\pi)^2} e^{-2Q}H_{\rho\rho}\bigg({\Omega\over Q}, \theta_{\bf Q}; \bigg\{{v_\rho\over v_0},{v_p\over v_0},{\gamma\over v_0},{c_0\over v_0}\bigg\}\bigg)
{Q_x^2Q_y^2\over Q^4}\,.
\label{Arho2yy}
\eea
\ew

and 

\newpage

\bew

\bea
(B^{pp}_{yy})_1 &=& \int\frac{d\Omega'}{2\pi}\frac{d^2Q'}{(2\pi)^2} \exp [i(\Omega'-Q'_x)]H_{pp}\bigg({\Omega'\over Q'}, \theta_{{\bf Q}'}; \bigg\{{v_\rho\over v_0},{v_p\over v_0},{\gamma\over v_0},{c_0\over v_0}\bigg\}\bigg){1\over Q^{'2}}\,, \\ 
(B^{pp}_{yy})_2 &=&  -2 \int\frac{d\Omega'}{2\pi}\frac{d^2Q'}{(2\pi)^2} \exp [i(\Omega'-Q'_x)]H_{pp}\bigg({\Omega'\over Q'}, \theta_{{\bf Q}'}; \bigg\{{v_\rho\over v_0},{v_p\over v_0},{\gamma\over v_0},{c_0\over v_0}\bigg\}\bigg){Q_y^{'2}\over Q^{'3}} \,, \\ 
(B^{pp}_{yy})_3 &=& \int\frac{d\Omega'}{2\pi}\frac{d^2Q'}{(2\pi)^2} \exp [i(\Omega'-Q'_x)]H_{pp}\bigg({\Omega'\over Q'}, \theta_{{\bf Q}'}; \bigg\{{v_\rho\over v_0},{v_p\over v_0},{\gamma\over v_0},{c_0\over v_0}\bigg\}\bigg){Q_y^{'4}\over Q^{'4}}.
\label{Bppyy} \\ 
B^{\rho\rho}_{yy} &=& \int\frac{d\Omega'}{2\pi}\frac{d^2Q'}{(2\pi)^2} \exp [i(\Omega'-Q'_x)]H_{\rho\rho}\bigg({\Omega'\over Q'}, \theta_{{\bf Q}'}; \bigg\{{v_\rho\over v_0},{v_p\over v_0},{\gamma\over v_0},{c_0\over v_0}\bigg\}\bigg){Q_x^{'2}Q_y^{'2}\over Q^{'4}}. \label{Brho2yy}
\eea
\ew
As we found earlier, all of these constants are again  functions of all of the ratios ${v_\rho\over v_0}$, ${v_p\over v_0}$, ${\gamma\over v_0}$, and ${c_0\over v_0}$, and will again be of $O(1)$ when all of these ratios are of ${\cal O}(1)$.

However, the important point is that the dominant term in (\ref{Bppyy}) at large $u$, which is obviously the 
$(B^{pp}_{yy})_1$ term, is clearly not integrable over $u$ all the way out to $u=\infty$; its integral diverges logarithmically.
We will show in the next subsection that this leads to superdiffusive motion in the $y$ direction.  Thus, taking this together with our result for velocity fluctuations in the $x$ direction, we see that there will be superdiffusion for both directions parallel to the surface.

Interestingly, the superdiffusion in the $x$-direction is {\em not} correlated with that in the $y$-direction. 
To see this, we calculate the cross-correlation function $ C^v_{xy}$.  Using (\ref{cvij}) and (\ref{vcorrneutral}), we see that this is given by
\bew
\bea
 C^v_{xy}(v_0(t-t')\hx, z, z, t-t') &=& C^{p\rho}_{yy}(v_0(t-t')\hx, z, z, t-t')\nonumber\\
 &=& v_0v_a'\int\frac{d\omega}{2\pi}\frac{d^2q}{(2\pi)^2} \exp [i(\omega-v_0q_x)(t-t')
-2qz] C_{p\rho}({\bf q},\omega)\left(1-qz+{2q_x^2q_y^2z^2\over q^2}\right) \,.\nonumber\\
\label{vxycorrneutral}
\eea
\ew
This vanishes due to the fact that $C_{p\rho}({\bf q},\omega)$ is odd in $q_y$;  see (\ref{crossprhoFT}) above. Thus, there is {\em no} cross-correlation between the superdiffusion in the $x$-direction and that in the $y$-direction.

We now turn to the motion of the tracer particles in the $z$-direction. The relevant velocity correlation using (\ref{cvij}), can be expressed as 
\bew
\begin{eqnarray}
 C^v_{zz}(v_0(t-t')\hx, z, z, t-t') = C^{pp}_{zz}(v_0(t-t')\hx, z, z, t-t')
 + C^{\rho\rho}_{zz}(v_0(t-t')\hx, z, z, t-t') 
 \label{cvzz}
\end{eqnarray}
\ew

We evaluate each of the two terms in (\ref{cvzz}). From (\ref{vcorrneutral}), we see that 
$C^v_{zz}(v_0(t-t')\hx, z, z, t-t')$ is given by
\bew
\begin{eqnarray}
 &&C^v_{zz}(v_0(t-t')\hx, z, z, t-t') \nonumber\\
 &&=v_0^2\int\frac{d\omega}{2\pi}\frac{d^2q}{(2\pi)^2} \exp [i(\omega-v_0q_x)(t-t')
-2qz] C_{pp}({\bf q},\omega)q_y^2z^2 + v_a'^2\int\frac{d\omega}{2\pi}\frac{d^2q}{(2\pi)^2} 
\exp [i(\omega-v_0q_x)(t-t')-2qz] C_{\rho\rho}({\bf q},\omega)q_x^2z^2. \nonumber \\
\label{vzzcorrneutral}
\end{eqnarray}
\ew

Making the same change of variables (\ref{varchange}) that we made when we were analyzing $ C^v_{xx}(v_0(t-t')\hx, z, z, t-t')$, we find that $C^v_{zz}(v_0(t-t')\hx,z,z,t-t')$ obeys an almost identical scaling law:
\bew
\beq
C^v_{zz}(v_0(t-t')\hx, z, z, t-t')= \left({D_pv_0\over z}\right)F^{pp}_{zz}\left({v_0|t-t'|\over z}\right) + 
\left({D_p\rho_c^2 v_a'^2\over v_0 z}\right)F^{\rho\rho}_{zz}\left({v_0|t-t'|\over z}\right)
\label{zscale}
\eeq
\ew
where now the dimensionless scaling functions are given by 

\newpage

\bew

\bea
F^{pp}_{zz}(u_r) &=& \int\frac{d\Omega}{2\pi}\frac{d^2Q}{(2\pi)^2} \exp [i(\Omega-Q_x)u_r-2Q]H_{pp}\bigg({\Omega\over Q}, \theta_{\bf Q}; \bigg\{{v_\rho\over v_0},{v_p\over v_0},{\gamma\over v_0},{c_0\over v_0}\bigg\}\bigg){Q_y^2\over Q^2}\,, \\ 
F^{\rho\rho}_{zz}(u_r) &=& \int\frac{d\Omega}{2\pi}\frac{d^2Q}{(2\pi)^2} \exp [i(\Omega-Q_x)u_r-2Q]H_{\rho\rho}\bigg({\Omega\over Q}, \theta_{\bf Q}; \bigg\{{v_\rho\over v_0},{v_p\over v_0},{\gamma\over v_0},{c_0\over v_0}\bigg\}\bigg)
{Q_x^2\over Q^2}\,.
\label{fzz2}
\eea
\ew

The limiting behaviors of these scaling functions can be obtained by an almost identical analysis to that used for $F^{pp}_{xx}(u)$, with almost identical results:
\beqn
F^{pp}_{zz}(u_r)
=\left\{
\begin{array}{ll}
A^{pp}_{zz}\sep&u_r\ll1\\\\
{B^{pp}_{zz}\over u_r^3}\sep&u_r\gg1
\end{array}
\right.
\nonumber\\
\label{fppzz}
\eeqn

and
\beqn
F^{\rho\rho}_{zz}(u_r)
=\left\{
\begin{array}{ll}
A^{\rho\rho}_{zz}\sep&u_r\ll1\\\\
{B^{\rho\rho}_{zz}\over u_r^3}\sep&u_r\gg1
\end{array}
\right.
\nonumber\\
\label{frho2zz}
\eeqn
with
\bew
\bea
A^{pp}_{zz} &=& \int\frac{d\Omega}{2\pi}\frac{d^2Q}{(2\pi)^2} e^{-2Q}H_{pp}\bigg({\Omega\over Q}, \theta_{\bf Q}; \bigg\{{v_\rho\over v_0},{v_p\over v_0},{\gamma\over v_0},{c_0\over v_0}\bigg\}\bigg){Q_y^2\over Q^2}\,,
\label{Appzz}  \\ 
A^{\rho\rho}_{zz} &=& \int\frac{d\Omega}{2\pi}\frac{d^2Q}{(2\pi)^2} e^{-2Q}H_{\rho\rho}\bigg({\Omega\over Q}, \theta_{\bf Q}; \bigg\{{v_\rho\over v_0},{v_p\over v_0},{\gamma\over v_0},{c_0\over v_0}\bigg\}\bigg){Q_x^2\over Q^2}\,,
\label{Arho2zz} 
\eea
\ew
and
\newpage

\bew

\bea
B^{pp}_{zz} &=&\int\frac{d\Omega'}{2\pi}\frac{d^2Q'}{(2\pi)^2} \exp [i(\Omega'-Q'_x)]H_{pp}\bigg({\Omega'\over Q'}, \theta_{{\bf Q}'}; \bigg\{{v_\rho\over v_0},{v_p\over v_0},{\gamma\over v_0},{c_0\over v_0}\bigg\}\bigg){Q_y^{'2}\over Q^{'2}}\,,
\label{Bppzz} \\ 
B^{\rho\rho}_{zz} &=& \int\frac{d\Omega'}{2\pi}\frac{d^2Q'}{(2\pi)^2} \exp [i(\Omega'-Q'_x)]H_{\rho\rho}\bigg({\Omega'\over Q'}, \theta_{{\bf Q}'}; \bigg\{{v_\rho\over v_0},{v_p\over v_0},{\gamma\over v_0},{c_0\over v_0}\bigg\}\bigg){Q_x^{'2}\over Q^{'2}}\,,
\label{Brho2zz}
\eea

\ew

Note that, as we found for the $xx$ correlations, $A^{pp}_{zz}$, $A^{\rho\rho}_{zz}$, $B^{pp}_{zz}$ and 
$B^{\rho\rho}_{zz}$
are functions of all of the ratios ${v_\rho\over v_0}$, ${v_p\over v_0}$, ${\gamma\over v_0}$, and ${c_0\over v_0}$, and will be of $O(1)$ when all of these ratios are of $O(1)$.

 Note also that the scaling function for the $z$ direction falls off sufficiently rapidly as a function of its argument $u$ that their integral over all $u$ converges. We  show  in Section (VII)}  that this implies that the motion of tracer particles in the $z$ direction is simply conventional diffusion. Furthermore, we also showed there  that the diffusion constant for the $z$-direction is independent of the height $z$ of the particles above the surface.

\end{appendix}


\begin{thebibliography}{7}
\bibitem{xtalfoot} Exceptions to this result are two dimensional crystals (see, e.g., \cite{2dxtal})
and fluctuating tethered membranes (see, e.g., \cite{teth}).
\bibitem{smecticfoot} The only example of a finite $Q$ as $q\to0$ we know of is smectic liquid crystals; see \cite{smectic}.
\bibitem{funny speed} It is a peculiarity of our problem, and in particular of the fact that damping in this system scales as $q$, rather than the usual $q^2$,  that a damping coefficient can have the dimensions of a speed, rather than a diffusion constant.
\bibitem{maitra comment} In most of the literature, the $\zeta$ terms in this expression are taken to be the divergence of a tensor: that is, instead of those terms, one writes  $\partial_j(\zeta p_ip_j)=\zeta(\rho)(\hp\cdot\nabla_s)p_i+\zeta(\rho)p_i(\nabla_s\cdot\hp)+p_i(\hp\cdot\nabla_s)\zeta(\rho)$, which is a special case (specifically, $\zeta_1=\zeta_2=\zeta$) of the more general expression we have used here. That our more general expression is actually allowed - i.e., that these terms need {\it not} add up to the divergence of a tensor in non-equilibrium systems-   was first recognized by~\cite{maitra2018}.
\bibitem{SC} The only equilibrium systems we know of that shows giant number fluctuations away from a critical point are superfluids. See, e.g., Ref.~\cite{chaikin} above.
\end{thebibliography}
\end{document}